\documentclass[aps,twocolumn,showpacs,prl,amsmath,amssymb,floatfix,superscriptaddress]{revtex4-2}
\usepackage[colorlinks=true,citecolor=blue,linkcolor=magenta,urlcolor=blue]{hyperref}
\usepackage{amsmath,amssymb,graphicx,color}
\usepackage{amsmath}
\usepackage{mathrsfs}
\usepackage{times}
\usepackage[table]{xcolor}
\usepackage{color,cancel}
\usepackage{float}
\usepackage[caption=false]{subfig}
\usepackage{empheq}
\usepackage{enumitem}
\usepackage{cancel}

\usepackage{pifont}
\usepackage{bm}

\usepackage{mathptmx} 
\DeclareMathAlphabet{\mathcal}{OMS}{cmsy}{m}{n}
\DeclareSymbolFont{largesymbols}{OMX}{cmex}{m}{n}

\begin{document}
\title{
Nonreciprocal and Geometric Frustration in Dissipative Quantum Spins
}

\author{Guitao Lyu}
\affiliation{Division of Natural and Applied Sciences, Duke Kunshan University, Kunshan, Jiangsu 215300, China}

\author{Myung-Joong Hwang}	\email{myungjoong.hwang@duke.edu}
\affiliation{Division of Natural and Applied Sciences, Duke Kunshan University, Kunshan, Jiangsu 215300, China}

\date{\today}

\begin{abstract}
Nonreciprocal interactions often create conflicting dynamical objectives that cannot be simultaneously satisfied, leading to nonreciprocal frustration. On the other hand, geometric frustration arises when conflicting static objectives in energy minimization cannot be satisfied. In this work, we show that nonreciprocal interaction among three collective quantum spins, mediated by a damped cavity, induces not only nonreciprocal frustration, intrinsic to nonreciprocity, but also geometric frustration with a remarkable robustness against disorder. It therefore ensures that the accidental degeneracy for steady states remains intact even when the system is perturbed away from a fine-tuned point of enhanced symmetry, in sharp contrast to the equilibrium case. Leveraging this finding, we identify a nonreciprocal phase transition driven by both geometric and nonreciprocal frustration. It gives rise to a time-dependent state, which shows a chiral dynamics along a geometry shaped by the geometric frustration and dynamically restores the broken discrete symmetries. Moreover, it constitutes a time-crystalline order, with multiple harmonics set by an emergent time scale that exhibits critical slowing down. Our predictions have important physical implications for a three-component spinor BEC-cavity system, which manifest as a geometric frustration in the structural phase transition and chiral dynamics of the frustrated self-organized BECs. We demonstrate the feasibility of experimental observation despite the presence of disorder in the spin-cavity coupling strengths.
\end{abstract}

\maketitle

\textit{Introduction.---}
Reciprocity is a foundational principle governing microscopic interactions under equilibrium conditions~\cite{Ivlev2015PRX}. However, nonreciprocal interaction violating this principle arises ubiquitously in nonequilibrium systems, including classical systems~\cite{Ivlev2015PRX, Brandenbourger2019NatComm, Nagulu2020NatElec, Baconnier2022NatPhy, Nassar2020NatRevMat} and quantum devices~\cite{estep2014NatPhy,ruesink2016NatComm,fang2017NatPhy,Peterson2017PRX,xu2019Nature,wanjura2023NatPhy,Reisenbauer2024NatPhy,Metelmann2015PRX,Metelmann2017PRA,Huang2021Light,Wang2023PRXQuantum,Manipatruni2009PRL,Lin2011PRL,fan2012Science,Wang2013PRL,shen2016NatPhot,Cao2017PRL,Xia2018PRL,Tang2022PRL}. A recent seminal work~\cite{Fruchart2021Nature} on nonreciprocal phase transitions, resulting in time-dependent nonstationary phases through exceptional points, has sparked growing interest in the role of nonreciprocal interactions in collective properties of many-body systems~\cite{Hanai2024,Hanai2024PRL,Avni2025PRL, Avni2025PRE,Zhu2024PRL,Weiderpass2025PE,Brauns2024PRX, Nadolny2025PRX,Gu2025Nature}. In particular, a nonreciprocal variant~\cite{Nunnenkamp2019PRL,Nunnenkamp2023PRL,Jachinowski2025} of the open Dicke model—a paradigmatic model in quantum optics exhibiting  superradiant phase transitions~\cite{Baumann2010nature,Baumann2011PRL, Kirton2018AdvQuanTech,Mivehvar2021AdvPhy, Klinder2015PNAS,Klinder2015PRL,Donner2017Nature, Benjamin2018PRL,Guo2021Nature}—has emerged as a minimal model for exploring nonreciprocal criticality in open quantum systems.

Geometric frustration, on the other hand, arises when competing constraints prevent the simultaneous minimization of all interaction energies~\cite{ Diep2005book, Moessner2006PhysicsToday}. The resulting extensive degeneracy in ground states gives rise to novel phases of matter, including spin ices~\cite{Steven2001Science, Ross2011PRL, Lee2012PRB,Wang2006Nature}, spin liquids~\cite{Balents2010nature, Savary2016RPP, Broholm2020science}, and spin glasses~\cite{Stein2013book,Sherrington1975PRL,Benjamin2023arXiv,Benjamin2024PRX}. In quantum optical systems, spin frustration can be optically mediated~\cite{Chiocchetta2021NatComm, Kim2010Nature, Benjamin2009NatPhy,Benjamin2021PRX,Benjamin2024PRX}; moreover, spins can also induces geometric frustration of cavity fields through frustrated superradiant phase transitions~\cite{Zhao2022PRL}, which exhibits anomalous critical properties~\cite{Zhao2022PRL, Zhao2023PRL, Zhang2022PRL,  Zhang2024PRR, Luo2025}. Nonreciprocal interaction also create competing dynamical constraints among constituents~\cite{Fruchart2021Nature}, leading to a dynamic frustration dubbed as nonreciprocal frustration~\cite{Hanai2024}. While a direct analogy between geometric and nonreciprocal frustration has recently been established in terms of marginal orbits~\cite{Hanai2024}, the role of their coexistence and interplay in many-body systems and nonreciprocal phase transitions remains largely unexplored~\cite{Avni2025PRL,Weiderpass2025PE}.

In this Letter, we demonstrate that nonreciprocal interaction among three spin ensembles, mediated by a damped cavity field, induces geometric frustration, in addition to the nonreciprocal frustration. In contrast to its coherent and reciprocal counterpart, where the accidental degeneracy of ground states requires fine-tuning, the geometrically frustrated phases arising from nonreciprocal interaction shows robustness against disorder. This finding sets the stage for identifying and investigating a nonreciprocal phase transition where both the geometric and nonreciprocal frustration coexists. Upon modulating the spin-cavity coupling, thus the degree of nonreciprocity, a nonstationary state emerges as a limit cycle that traverses all frustrated metastable states, thereby dynamically restoring the broken translational symmetry due to geometric frustration. This dynamical phase spontaneously breaks time-translational symmetry—akin to continuous time crystals~\cite{Kongkhambut2022Science, Iemini2018PRL, Sacha2018Reports}—as a generic consequence of nonreciprocal frustration~\cite{Fruchart2021Nature,Hanai2024}, and is characterized by an emergent time scale that exhibits critical slowing down and its multiple harmonics. In a weak coupling regime, a limit cycle whose period is set by the bare spin frequency appears, similar to the ones observed in previous studies~\cite{Nunnenkamp2019PRL,Nunnenkamp2023PRL,Jachinowski2025}. These two dynamical phases with distinct time-crystalline orders are separated by a chaotic phase.

While our predictions are broadly applicable for any physical realizations of nonreciprocally coupled dissipative quantum spins, here we focus their physical implications in the context of a three-component spinor Bose–Einstein condensate (BEC) inside an optical cavity~\cite{Donner2019Science, Donner2018PRL}. In this setting, we find that the frustrated phases manifest as geometric frustration in a structural phase transition, where the self-organization of the BEC into a checkerboard pattern is constrained: each species disfavors occupying the same half of the pattern, a condition that cannot be simultaneously satisfied by all three species. We demonstrate the feasibility of experimental observation by showing that the extra degeneracy resulting from the frustrated self-organization as well as the associated limit-cycle dynamics i) can be directly probed via the phase-space trajectories of the cavity field, and ii) remain robust against disorder arising from the inhomogeneous spin-cavity coupling strengths in spinor BEC–cavity systems.

\textit{Model.---}
We consider a three-species nonreciprocal Dicke model where three collective spins, labelled $m \in \{-1,  0,  1\}$, are coupled to a single-mode cavity field $\hat a$ with detuning frequency $\omega_c$. The collective spins are described by the angular momentum operator $\hat J_{x(y,z),m}=\sum_{j=1}^{N}\hat \sigma_{x(y,z),m}^{j}$ where $\hat \sigma_{x(y,z),m}^{j}$ is the Pauli matrix for the j-th spin in ensemble $m$. The dynamics is governed by the master equation
$\dot{\hat\rho}=-i[\hat H, \hat \rho]+\kappa D[\hat a] \hat \rho+ \frac{\gamma}{N} \sum_m D[\hat J_{-,m}]\hat \rho$
with the Lindblad dissipator $D[\hat o]\hat \rho \equiv 2 \hat o\hat \rho \hat o^\dagger - \hat o^\dagger \hat o \hat \rho - \hat \rho \hat o^\dagger \hat o$, where $\kappa$ and $\gamma$ denote the cavity decay and the spin damping with $\hat J_{-, m}=(\hat J_{x, m} - i\hat J_{y, m})/2$, respectively. The coherent part of the system is described by the Hamiltonian
\begin{align}
\label{eq: H}
\hat H=\omega_c \hat a^\dagger \hat a+\sum_{m=-1}^{1} \left[ \frac{\Omega}{2}\hat J_{z,m}+\frac{\lambda}{2\sqrt{N}} (e^{-im\phi} \hat a + e^{im\phi} \hat a^\dagger ) \hat J_{x,m}\right].
\end{align}
For simplicity, we assume the identical frequency $\Omega$, population $N$, and spin-boson interaction strength $\lambda$ for all species, except when we examine the effect of disorder. A key feature of the model is the $m$-dependent phase of the cavity quadrature operator, $e^{-im\phi} \hat a + e^{im\phi} \hat a^\dagger$, that couples to spins. Such complex interaction can be readily realized using  three magnetic sublevels ($m_F=\pm1,0$) within the $F=1$ hyperfine manifold of a $^{87}\mathrm{Rb}$ BEC trapped in a single-mode cavity~\cite{Donner2018PRL, Donner2019Science,Nunnenkamp2019PRL}, where the spin-dependent phase is tunable by adjusting the polization angle of the pump field~\cite{Donner2018PRL}. See \textit{implementation} at the end for more details.

The mean-field equation of motions (EOMs), which becomes exact in the thermodynamic limit $N\rightarrow\infty$, read
\begin{align}
\label{eq: eoms1}
  \dot S_{x,m}=&-\Omega  S_{y,m} +  \gamma  \frac{S_{x,m} S_{z,m}}{N},\\
\label{eq: eoms2}
  \dot S_{y,m}=&\Omega  S_{x,m}-\frac{ \lambda \left(e^{-i m\phi} \alpha  + e^{i m\phi }\alpha^* \right) S_{z,m} }{\sqrt{N}} + \gamma  \frac{S_{y,m} S_{z,m}}{N},\\
\label{eq: eoms3}
  \dot S_{z,m}=&\frac{ \lambda \left( e^{-i m\phi} \alpha + e^{i m\phi }\alpha^* \right)  S_{y,m} }{\sqrt{N}} -\gamma \frac{N^2- S_{z,m}^2}{N},\\
\label{eq: eoms4}
  \dot \alpha=& - (i\omega _c+\kappa)\alpha  -i  \frac{ \lambda }{2\sqrt{N}} \sum_{m=-1}^1 e^{i m \phi } S_{x,m},
\end{align}
where $\alpha \equiv \langle \hat a \rangle$, $S_{x(y,z), m}\equiv \langle \hat J_{x(y,z),m} \rangle$. The EOMs~(\ref{eq: eoms1}-\ref{eq: eoms4}) possess a $Z_2$ parity symmetry, $ (\alpha, S_{x (y),m}) \leftrightarrow (-\alpha, -S_{x (y),m}) $, whose spontaneous breaking leads to a superradiant phase transition. In addition, they are invariant under $ (\phi, S_{x (y,z),1}) \leftrightarrow (-\phi , S_{x (y,z),-1})$. Combining the two symmetries, the mean-field phase diagram is mirror symmetric around $\phi=\pi/2$ upon spin transformation $S_{x(y),\pm1} \leftrightarrow -S_{x(y),\mp 1}$ and $S_{z,\pm1}\leftrightarrow S_{z,\mp1}$. For a particular value $\phi = 2\pi/3$ (similarly, for $\pi/3$), an additional $Z_3$ symmetry exits. It originates from a cyclic permutation of three spin species, or equivalently, a translational symmetry on a triangle, accompanied by a $2\pi/3$ phase rotation of the cavity field: $(\alpha, S_{x(y,z),m}) \leftrightarrow (\alpha e^{i 2\pi/3}, S_{x(y,z),m-1})$. This enhanced symmetry at the fine-tuned point therefore becomes $Z_6$.


In an experimentally relevant parameter regime, $\omega _c>\kappa \gg \Omega$~\cite{Donner2018PRL, Donner2019Science}, the cavity field adiabatically follows the dynamics of the spins. Consequently, its steady-state expectation value is determined by the instantaneous spin configurations,
\begin{align}
	\label{eq: alpha}
	\alpha=-\frac{\lambda  \left(\omega _c + i \kappa \right)}{2 \sqrt{N} \left(\omega _c^2 + \kappa^2 \right)} \sum_{m=-1}^1 e^{im \phi } S_{x,m}.
\end{align}
By eliminating the cavity field, Eq.~(\ref{eq: eoms2}) becomes
\begin{align}
\label{eq: eoms2new}
	\dot S_{y,m}&=\Omega  S_{x,m}+ \frac{\lambda ^2  S_{z,m} F_m }{N(\omega _c^2+\kappa ^2)}+ \gamma \frac{ S_{y,m} S_{z,m}}{N},
\end{align}
where $F_m =\sum_{m'=-1}^1 D[(m-m')\phi] S_{x,m'}$ is a spin-dependent function
with $ D[\Phi] \equiv \omega _c\cos\Phi +\kappa  \sin\Phi $.
The asymmetry of  $ D[(m-m')\phi]\neq  D[(m'-m)\phi]$ constitutes the nonreciprocal coupling  \cite{Fruchart2021Nature,Hanai2024,Nunnenkamp2023PRL} among the spins, mediated by the damped cavity field.

\begin{figure}[tpb!]
	\includegraphics[width=0.99\linewidth]{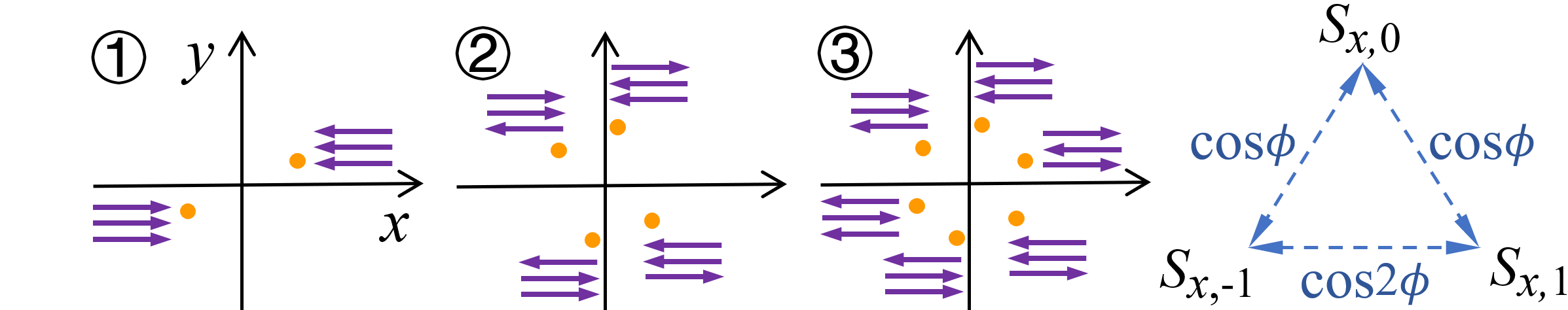}
	\includegraphics[width=0.99\linewidth]{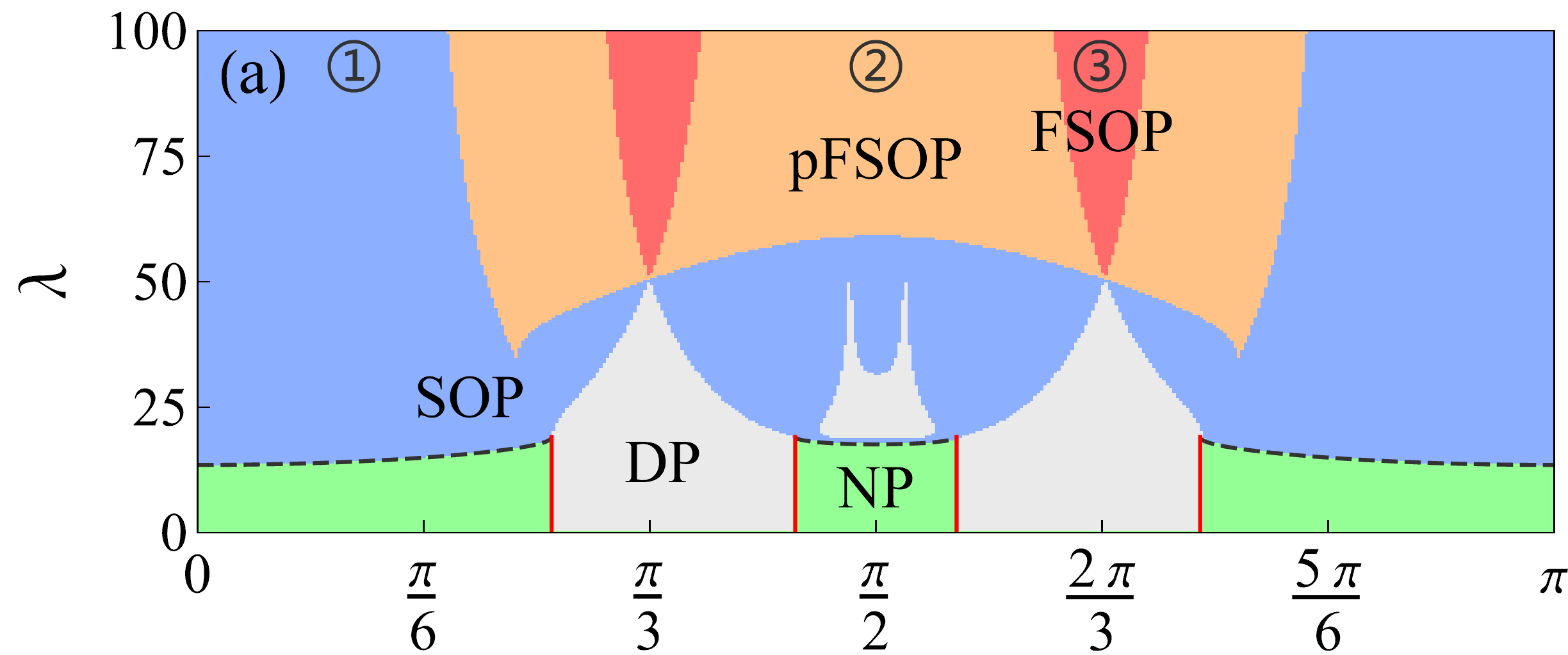}
	\includegraphics[width=0.99\linewidth]{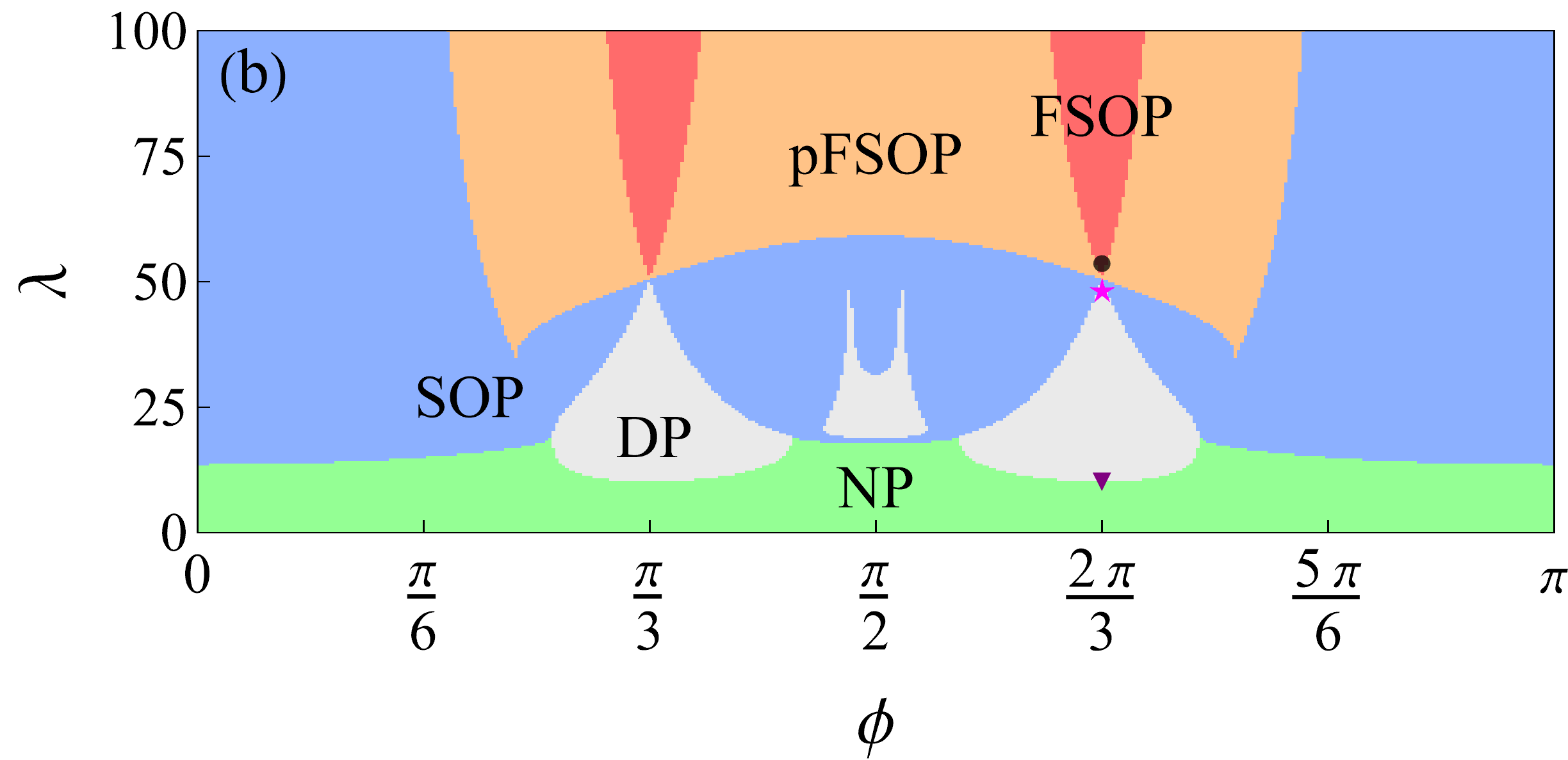}
	\caption{Steady-state phase diagram as a function of spin-boson coupling strength $\lambda$ and phase $\phi$ for (a) no spin damping $\gamma=0$ and (b) $\gamma=0.05$. We set $\omega_c = 500$ and $\kappa=150$ with $\Omega=1$ being the unit of energy throughout the paper.  The top-left panel illustrates representative spin configurations corresponding to the regions  \ding{172}–\ding{174} of the phase diagram, where $S_{x,-1}$, $S_{x,0}$, and $S_{x,1}$ are represented by purple arrows from top to bottom with rightward (leftward) direction indicating $+N$ ($-N$). The orange dots denote the corresponding values of $\alpha=x+iy$ in the cavity phase space. The top-right panel illustrates the cavity-mediated coherent interaction strength among $S_{x,m}$.}
	\label{fig:phase diagram}
\end{figure}

\textit{Phase diagram.---}
By solving spin-only EOMs~(\ref{eq: eoms1}) and (\ref{eq: eoms2new}), along with a stability analysis~\cite{SM_link}, we obtain the phase diagram in the $\phi-\lambda$ plane in Fig.~\ref{fig:phase diagram}(a) for $\gamma=0$. The rich phase diagram is categorized into the following distinct phases. First, the normal phase (NP) features $S_{x(y),m}=\alpha=0$ as a stable fixed point. Second, the dynamical phase (DP), characterized by the onset of nonstationary states, emerges from the NP at $
\phi_c = \arccos( \pm \frac{1}{2} (1 \pm 3\kappa/\sqrt{\kappa^2 + \omega_c^2})^{1/2})$.
The NP-DP phase boundary corresponds to a nonreciprocal phase transition, marked by the appearance of exceptional points \cite{SM_link}. Third, for each $\phi\in\textrm{NP}$, there is a threshold value $\lambda_c =\sqrt{\Omega (-3 \omega_c + \sqrt{P})/(2 \cos 2 \phi +\cos 4 \phi -3)}$ with $P=\omega _c^2 (2\cos 2 \phi +1)^2-8 \kappa ^2 \sin^2\phi (\cos 2 \phi +2)$, above which the $Z_2$-symmetry-breaking stationary solutions, $S_{x(y),m},\alpha\neq0$, emerge. We refer them as self-organized phases (SOPs), to emphasize that the spontaneous magnetization corresponds to the self-organization of the BEC into a checkerboard pattern, where the sign of $S_{x,m}$ determines which half of the checkerboard lattice is predominantly occupied by atoms in the magnetic sublevel $m_F = m$ \cite{Donner2019Science}. Remarkably, the SOPs are divided further into conventional (SOP), partially-frustrated (pFSOP), and frustrated (FSOP) phases with varying degeneracies. These additional degeneracies are attributed to geometric frustration caused by nonreciprocal interactions, as we discuss below. For a nonzero spin damping $\gamma\neq0$, the overall phase diagram remains largely unchanged, except that a portion of the DP is restored to the NP [Fig.~\ref{fig:phase diagram}(b)].

\begin{figure}[tpb!]
	\centering
	\includegraphics[width=0.5\linewidth]{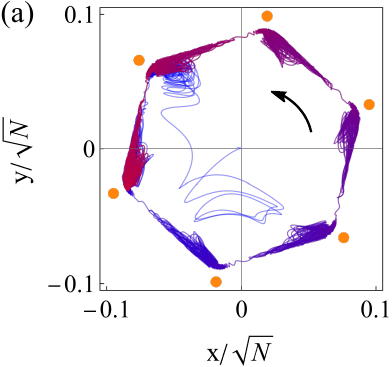}
	\hspace*{0.2cm} \raisebox{0.4em}{\includegraphics[width=0.42\linewidth]{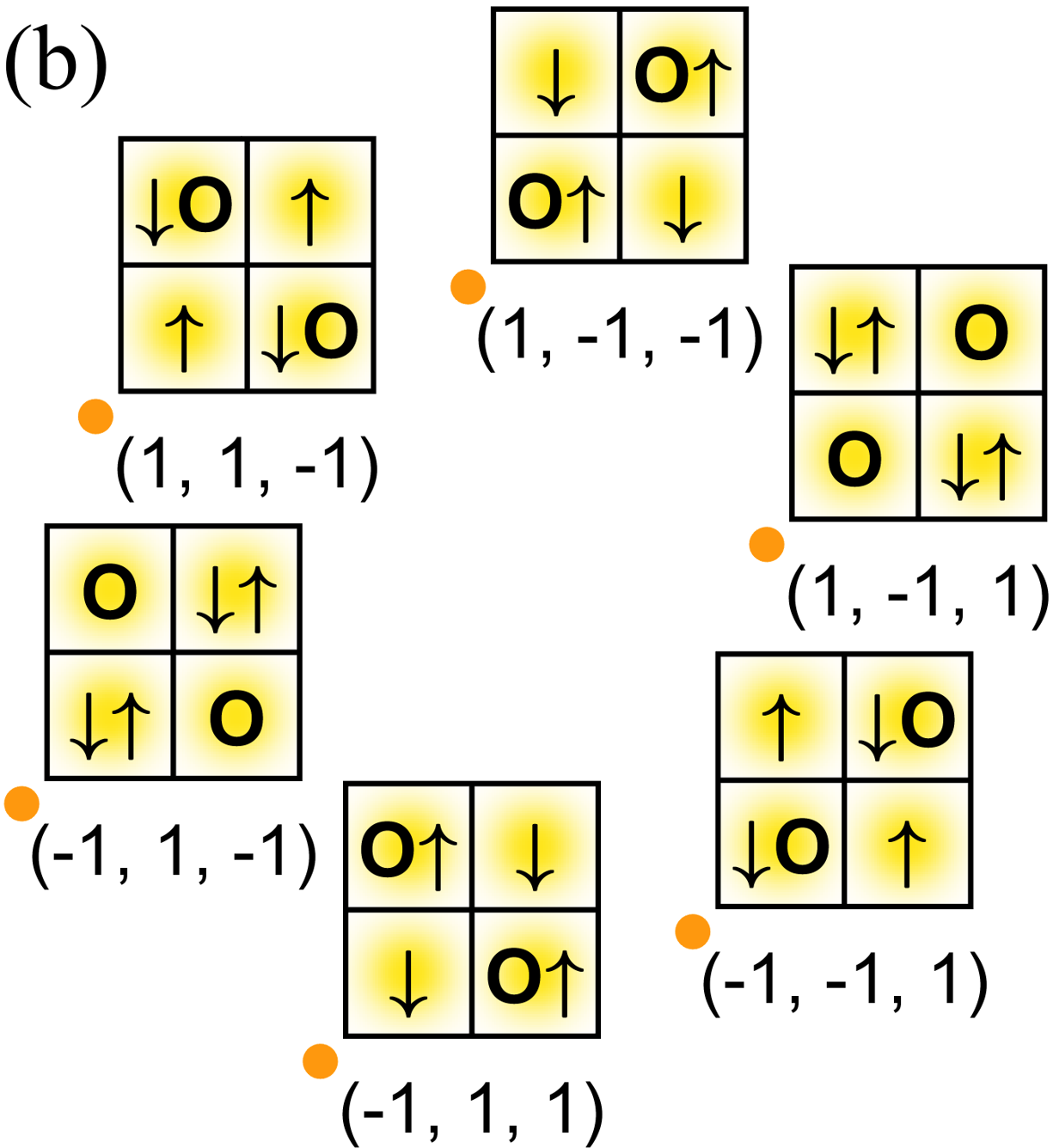}}
	\includegraphics[width=1\linewidth]{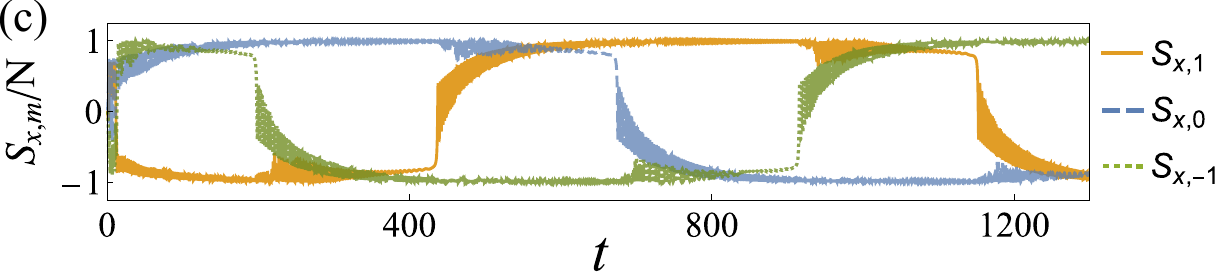}
	\includegraphics[width=0.48\linewidth]{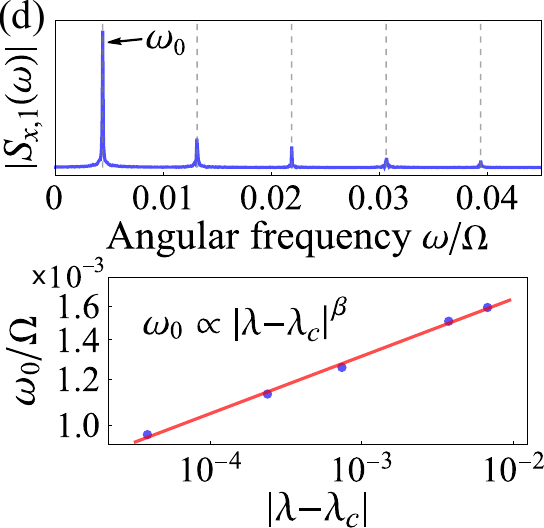} 
	\hspace*{0.02cm}
	\includegraphics[width=0.50\linewidth]{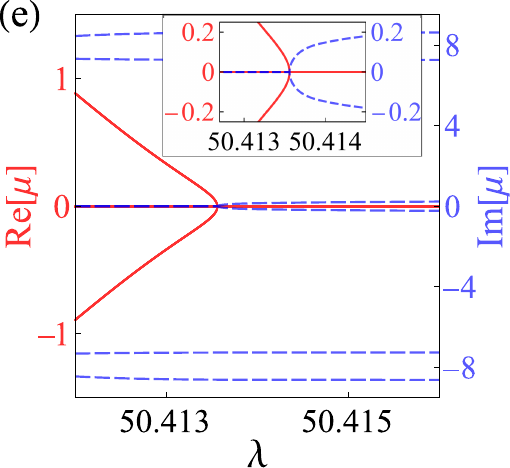}

\caption{Swap phase dynamics
(a) The cavity field dynamics in the phase space at $\phi=2\pi/3$ and $\lambda=49$, marked by a star in Fig.~\ref{fig:phase diagram}(b). The trajectory starts in blue and gradually turn to red as they evolve over time $t\in[0,1800]$. The orange dots represent the stable solutions in FSOP nearby, marked by a dark dot (at $\lambda=54$) in Fig.\ref{fig:phase diagram}(b).
(b) Schematics for geometrically frustrated self-organization. The diagonal (off-diagonal) blocks denote the even (odd) sublattice of the checkerboard pattern. The symbols $\bm{\downarrow}$, $\textbf{O}$, and  $\bm{\uparrow}$ denote the spinor BEC in hyperfine states $m_F=-1$, $0$, and $1$, respectively. Each pattern corresponds to a spin state $(S_{x,-1},S_{x,0},S_{x,1})/N$ indicated below and a cavity state (orange dot). (c) Spin dynamics of $S_{x,m}$ as a function of time. (d) Fourier spectrum (in arbitrary units) of $S_{x,1}(t)$. The gray dashed lines indicate the odd multiple harmonics of a fundamental frequency $\omega_0$. The Fourier spectrum of $\alpha(t)$ (not shown) exhibits a similar structure, except that certain odd harmonics are suppressed due to destructive interference in the collective modes. $\omega_0$ decreases with a power-law scaling with an exponent $\beta\approx 0.0985$ near the transition point. Note that the resolution of Fourier frequency is about $5\times10^{-6}/\Omega$, due to the finite evolution time.
(e) Spectrum of the dynamical matrix of FSOP at $\phi=2\pi/3$. The insert shows the magnified view of the exceptional point at which the nonreciprocal phase transition occurs.
}
	\label{fig: dynamics hexagon}
\end{figure}

\textit{Nonreciprocity-induced frustrated self-organization.---}
For sufficiently large $\lambda$, as $\phi$ increases from zero, the system undergoes a sequence of first-order phase transitions: from a SOP with degeneracy $d=2$, where all three spins are ferro-magnetically aligned, to a pFSOP with $d=4$, where $m=\pm1$ spins are anti-aligned and $m=0$ spin is frustrated, and finally to a FSOP with $d=6$, where all three spins are frustrated. Schematic configurations for the spins and the corresponding cavity field in the phase space are presented in Fig.~\ref{fig:phase diagram}, while the numerical solutions are presented in \cite{SM_link}. Moreover, the correspondence between the frustrated spins and frustrated self-organization is illustrated in Fig.~\ref{fig: dynamics hexagon}(d).

Identifying geometric frustration in terms of conflicting constraints on energy minimization is not applicable for open quantum systems with nonreciprocal interactions. Therefore, we consider the reciprocal limit $\kappa\rightarrow0$ where coherent interactions mediated by a cavity may induce geometric frustration among spins, which provides a useful reference for highlighting the distinctive effects of nonreciprocal interactions. After integrating out the cavity field from Eq.~(\ref{eq: H}), the spin-only mean-field energy reads
\begin{align}
	\label{eq: E_eff}
	E_\text{eff}=& \sum_{m=-1}^1 -\frac{\Omega }{2} \sqrt{N^2 -S_{x,m}^2}
	-\frac{\lambda ^2}{4 N \omega_c} S_{x,m}^2 +E_\text{int},
\end{align}
with the interaction energy $E_\text{int}= -\frac{\lambda ^2 }{2N \omega_c} (S_{x,-1} S_{x,0} \cos\phi + S_{x,1} S_{x,0}\cos\phi  + S_{x,-1}S_{x,1}\cos2\phi ) $. The local mean-field energy for each spin (the first two terms) undergoes a single- to double-well potential transition upon increasing $\lambda$ \cite{SM_link}, leading to two degenerate spin solutions $S_{x,m}/N \approx \pm1$ for $\lambda\gg\sqrt{\Omega\omega_c}$. In addition, the spins needs to minimize the interaction energy $E_\text{int}= -\frac{\lambda ^2 }{2N \omega_c} (S_{x,-1} S_{x,0} \cos\phi + S_{x,1} S_{x,0}\cos\phi  + S_{x,-1}S_{x,1}\cos2\phi )$. Note that the phase $\phi$ modulates the spin-dependent interaction strength. For $2\pi/3$, all interaction strengths become identical and anti-ferromagnetic, which cannot be simultaneously minimized on a triangle, leading to FSOP with the six degenerate ground states (similarly, for $\pi/3$ through mirror symmetry around $\pi/2$). For $\pi/3<\phi <2\pi/3$, where $|\cos2\phi |>|\cos\phi|$ and $\cos2\phi <0$, while the anti-alignment between $m=\pm1$ is favored, the spin $m=0$ cannot simultaneously minimize the interaction with both $m=1$ and $-1$, leading to the pFSOP with the partially lifted degeneracy of four. For other values of $\phi$, it is straightforward to show that the frustration is completely lifted, leading to the conventional SOP. See~\cite{SM_link} for the phase diagram for $\kappa=0$ case.

We emphasize that in the reciprocal limit the FSOP occurs at the points of extended $Z_6$ symmetry, $\phi=\pi/3$ and $2\pi/3$, where the interaction strength for all bonds are identical. Any disorder in interaction strengths would partially or completely lift the degeneracy. In stark contrast, nonreciprocal interaction stabilizes the FSOP over a \textit{finite} range around these $Z_6$ symmetric points, where interaction strength among bonds are not identical (see Fig.~\ref{fig:phase diagram}). Moreover, the pFSOP also extends beyond the reciprocal range $\pi/3 < \phi < 2\pi/3$, with the additional degeneracy increasing or decreasing by two through a sequence of first-order phase transitions. From Eq.~(\ref{eq: alpha}), the non-, partially-, and fully-frustrated spin configurations leave distinct imprints on the cavity field $\alpha$, forming a line, rectangle, and hexagon, respectively (see the top panels in Fig.~\ref{fig:phase diagram}). The global phase rotation of $\alpha$ arises from dissipation, a typical feature of dissipative superradiant phase transitions~\cite{TorrePRA2013,Hwang2018PRA}.

\textit{Chiral phase.---} The DPs are characterized by the emergence of nonstationary dynamics, which we investigate by numerically solving the full EOMs~(\ref{eq: eoms1}–\ref{eq: eoms4}) with spin damping ($\gamma/\Omega=0.05$), focusing on the symmetric point $\phi=2\pi/3$. Near the NP-DP boundary for small $\lambda$, we observe a limit cycle where the phase $\theta_m\equiv \arctan(S_{y,m}/S_{x,m})$ of each species linearly increases in time (counter-clockwise rotation) with the relative phase being locked at $2\pi/3$. This limit cycle preserves the $Z_3$ cyclic symmetry at any time $t$ in the long-time limit. We identify this type of limit cycle, a chiral phase, characterized by a chiral rotation at a constant angular velocity (set by the bare spin frequency $\Omega$) and synchronized relative phases~\cite{Fruchart2021Nature}. We note that the chiral phase here emerges without the Goldstone mode, unlike the scenario considered in Ref.~\cite{Fruchart2021Nature}; instead, it arises via a nonreciprocal phase transition from a symmetric phase.

\begin{figure}[tpb!]
	\includegraphics[width=0.99\linewidth]{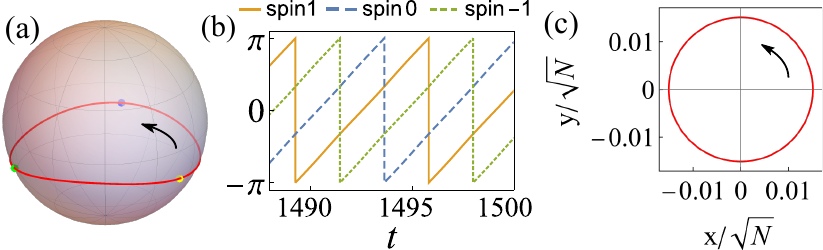}
	\caption{Chiral phase dynamics. The dynamics of the three spins near the NP-DP boundary at $\phi=2\pi/3$ and $\lambda=10.8$ [indicated by a triangular marker in Fig.~\ref{fig:phase diagram}(b)] are shown  (a) on Bloch sphere, where the dots represent individual spins, and (b) with their azimuthal angles. The limit cycles of each spins are identical showing a chiral rotation with a fixed frequency $\Omega$ and relative phase $2\pi/3$. The cavity field $\alpha=x+iy$ on the phase space forms a circular limit cycle.
	}
	\label{fig: limit circle}
\end{figure}

\textit{Swap phase.---} Near the critical point of the FSOP-to-DP transition, we find that geometric frustration strongly influences the character of the resulting time-dependent states. The spins in the long-time limit show a periodic dynamics [see Fig.~\ref{fig: dynamics hexagon} (c)]. To identify limit cycles, we consider the collective coordinates of spins, whose $x$-components are $X_c=(S_{x,1}+S_{x,0}+S_{x,-1})/\sqrt{3}$, $X_{d1}=(S_{x,1}-2S_{x,0}+S_{x,-1})/\sqrt{6}$, and $X_{d2}=(-S_{x,1}+S_{x,-1})/\sqrt{2}$. Moreover, from Eq.~\eqref{eq: alpha}, we note that $X_{d1}$ and $X_{d2}$ correspond to the real and imaginary part of $\alpha$, up to a common factor, for $\phi=2\pi/3$. In Fig.~\ref{fig: dynamics hexagon}(a), the limit cycle of the cavity field in phase space, which coincides with the collective spin dynamics projected onto the $X_{d1}$–$X_{d2}$ plane, forms a hexagon. Its vertices are set by the six frustrated stable solutions in the FSOP near the critical point (indicated by orange dots). The spins oscillate around one of the six (meta-)stable states for a period of time before transitioning to a neighboring state in the counter-clockwise direction. Therefore, on average, the broken $Z_6$ symmetry is dynamically restored. Moreover, the Fourier spectrum  reveals that the limit-cycle dynamics are governed by multiple odd harmonics of an emergent fundamental frequency $\omega_0$ [Fig.~\ref{fig: dynamics hexagon}(d)]. This emergent time-scale exhibits critical slowing down; namely, the time required to complete one full cycle diverges as the system approaches the critical point. We identify this limit-cycle as a swap phase, characterized by the dynamically restored broken discrete symmetries and the time-crystalline order with multiple harmonics~\cite{Fruchart2021Nature}.

The transition from FSOP to swap phase is driven by the dual roles of nonreciprocal interactions: inducing nonreciprocal frustration that breaks time-translational symmetry and introducing geometric frustration that dictates the shape of the resulting limit cycle. We demonstrate that this transition constitutes a nonreciprocal phase transition through the emergence of exceptional points (EPs) [Fig.~\ref{fig: dynamics hexagon} (e)]. The dynamical matrix, obtained by linearizing the equations of motion around each stationary solution in the FSOP, exhibits an exceptional point (EP) where the imaginary parts of the eigenvalues coalesce and an anti-damped mode emerges. The simultaneous emergence of six exceptional points (EPs) provides an intuitive understanding for the observed hexagonal limit cycle. When the system becomes unstable around one of the stationary solutions, it initially exhibits oscillatory dynamics in its vicinity, reminiscent of chiral phase around normal solution. However, before a closed limit cycle is established within the metastable manifold, anti-damping drives the system out of this manifold, leading to a transition into a neighboring one. While it is natural to expect that the dynamics within a metastable manifold and transitions between manifolds may be governed by different time-scales, it is noteworthy that they are related through harmonics of a single emergent frequency.

\textit{Chiral-chaos-swap transition.---}The chiral and swap phases, in the weak and strong coupling regimes, respectively, represent qualitatively distinct limit cycles that are not be smoothly connected. In the nonreciprocal flocking model, for example, these two time-crystal phases are separated by a time quasicrystal, arising from a mixture of chiral and swap dynamics~\cite{Fruchart2021Nature}. In stark contrast, we find that the two phases here are separated by a chaotic phase \cite{SM_link}. Notably, a similar transition from the chiral to chaotic phase has been observed in two-species models~\cite{Nunnenkamp2023PRL, Jachinowski2025}, and we observe the further transition into the swap phase (not shown). Based on this, we expect that the sequence of chiral–chaotic–swap transition is a generic feature of multi-species nonreciprocal Dicke models, and we leave a detailed investigation of the nature of transition for future work. Finally, we identify another isolated region of the DP embedded within the SOP, a feature absent in the two-species case. This phase arises via a nonreciprocal phase transition from the SOP and exhibits a swap phase dynamics with two metastable manifolds on a line \cite{SM_link}.

\textit{Implementation.---} The three-species nonreciprocal Dicke model can be implemented by utilizing the hyperfine states $|F=1, m=0,\pm1\rangle $ of $^{87}\text{Rb}$ BEC loaded in a single-mode cavity. By modulating the angle between the pump laser polarization vector and the cavity field, a tunable $m$-dependent phase $\phi_{m}=m \phi$ can be realized~ \cite{Donner2018PRL}. In turn, coupling strength also becomes $m$-dependent, leading to inhomogeneous coupling strength with $\lambda_{\pm1}/\lambda_0=\sqrt{N_{\pm1}/N_0}\sqrt{1+\tan\phi^2}$. We find that, by choosing $N_{m}$ to make $\Lambda\equiv\lambda_{m} \sqrt{N_m}$ uniform, the effect of disorder, though still present, can be significantly reduced. As a result, the phase diagram recovers all distinct phases predicted for the case of homogeneous coupling. At $\phi=2\pi/3$ with $N_0=2N_{\pm1}$, the hexagonal limit-cycle in the swap phase emerges with a slight distortion. An additional modulation of spin frequencies $\Omega_0=\Omega_{\pm1}/2$ further reduces the distortion. We present the details of the compensation scheme and the numerical results for the phase diagram and limit-cycle dynamics in Sec.~F of SM~\cite{SM_link}, which demonstrates the feasibility of experimental observation of the geometrically frustrated self-organization and the associated swap phase dynamics.

\textit{Conclusions.---} In conclusion, we have demonstrated that nonreciprocal interactions among dissipative quantum spins stabilize geometrically frustrated stationary phases that are robust against disorder. Furthermore, we have identified a nonreciprocal phase transition from the frustrated phase, which leads to a limit cycle whose geometry is governed by the associated accidental degeneracy, manifested as a swap phase. Our predictions are amenable to experimental verification in existing spinor BEC–cavity systems. An intriguing direction for future work would be to extend our study to many dissipative spins with nonreciprocal interactions, in order to explore the potential emergence of spin glasses with fundamentally distinct properties from their reciprocal counterparts owing to their robustness against disorder.

\textit{Acknowledgments.}
This work was supported by the Innovation Program for Quantum Science and Technology Grant No. 2021ZD0301602.

\bibliography{reference}

\begin{thebibliography}{76}%
\makeatletter
\providecommand \@ifxundefined [1]{%
 \@ifx{#1\undefined}
}%
\providecommand \@ifnum [1]{%
 \ifnum #1\expandafter \@firstoftwo
 \else \expandafter \@secondoftwo
 \fi
}%
\providecommand \@ifx [1]{%
 \ifx #1\expandafter \@firstoftwo
 \else \expandafter \@secondoftwo
 \fi
}%
\providecommand \natexlab [1]{#1}%
\providecommand \enquote  [1]{``#1''}%
\providecommand \bibnamefont  [1]{#1}%
\providecommand \bibfnamefont [1]{#1}%
\providecommand \citenamefont [1]{#1}%
\providecommand \href@noop [0]{\@secondoftwo}%
\providecommand \href [0]{\begingroup \@sanitize@url \@href}%
\providecommand \@href[1]{\@@startlink{#1}\@@href}%
\providecommand \@@href[1]{\endgroup#1\@@endlink}%
\providecommand \@sanitize@url [0]{\catcode `\\12\catcode `\$12\catcode
  `\&12\catcode `\#12\catcode `\^12\catcode `\_12\catcode `\%12\relax}%
\providecommand \@@startlink[1]{}%
\providecommand \@@endlink[0]{}%
\providecommand \url  [0]{\begingroup\@sanitize@url \@url }%
\providecommand \@url [1]{\endgroup\@href {#1}{\urlprefix }}%
\providecommand \urlprefix  [0]{URL }%
\providecommand \Eprint [0]{\href }%
\providecommand \doibase [0]{https://doi.org/}%
\providecommand \selectlanguage [0]{\@gobble}%
\providecommand \bibinfo  [0]{\@secondoftwo}%
\providecommand \bibfield  [0]{\@secondoftwo}%
\providecommand \translation [1]{[#1]}%
\providecommand \BibitemOpen [0]{}%
\providecommand \bibitemStop [0]{}%
\providecommand \bibitemNoStop [0]{.\EOS\space}%
\providecommand \EOS [0]{\spacefactor3000\relax}%
\providecommand \BibitemShut  [1]{\csname bibitem#1\endcsname}%
\let\auto@bib@innerbib\@empty
\bibitem [{\citenamefont {Ivlev}\ \emph {et~al.}(2015)\citenamefont {Ivlev},
  \citenamefont {Bartnick}, \citenamefont {Heinen}, \citenamefont {Du},
  \citenamefont {Nosenko},\ and\ \citenamefont {L\"owen}}]{Ivlev2015PRX}%
  \BibitemOpen
  \bibfield  {author} {\bibinfo {author} {\bibfnamefont {A.~V.}\ \bibnamefont
  {Ivlev}}, \bibinfo {author} {\bibfnamefont {J.}~\bibnamefont {Bartnick}},
  \bibinfo {author} {\bibfnamefont {M.}~\bibnamefont {Heinen}}, \bibinfo
  {author} {\bibfnamefont {C.-R.}\ \bibnamefont {Du}}, \bibinfo {author}
  {\bibfnamefont {V.}~\bibnamefont {Nosenko}},\ and\ \bibinfo {author}
  {\bibfnamefont {H.}~\bibnamefont {L\"owen}},\ }\bibfield  {title} {\bibinfo
  {title} {Statistical mechanics where {N}ewton's third law is broken},\ }\href
  {https://doi.org/10.1103/PhysRevX.5.011035} {\bibfield  {journal} {\bibinfo
  {journal} {Phys. Rev. X}\ }\textbf {\bibinfo {volume} {5}},\ \bibinfo {pages}
  {011035} (\bibinfo {year} {2015})}\BibitemShut {NoStop}%
\bibitem [{\citenamefont {Brandenbourger}\ \emph {et~al.}(2019)\citenamefont
  {Brandenbourger}, \citenamefont {Locsin}, \citenamefont {Lerner},\ and\
  \citenamefont {Coulais}}]{Brandenbourger2019NatComm}%
  \BibitemOpen
  \bibfield  {author} {\bibinfo {author} {\bibfnamefont {M.}~\bibnamefont
  {Brandenbourger}}, \bibinfo {author} {\bibfnamefont {X.}~\bibnamefont
  {Locsin}}, \bibinfo {author} {\bibfnamefont {E.}~\bibnamefont {Lerner}},\
  and\ \bibinfo {author} {\bibfnamefont {C.}~\bibnamefont {Coulais}},\
  }\bibfield  {title} {\bibinfo {title} {Non-reciprocal robotic
  metamaterials},\ }\href {https://doi.org/10.1038/s41467-019-12599-3}
  {\bibfield  {journal} {\bibinfo  {journal} {Nat. Commun.}\ }\textbf {\bibinfo
  {volume} {10}},\ \bibinfo {pages} {4608} (\bibinfo {year}
  {2019})}\BibitemShut {NoStop}%
\bibitem [{\citenamefont {Nagulu}\ \emph {et~al.}(2020)\citenamefont {Nagulu},
  \citenamefont {Reiskarimian},\ and\ \citenamefont
  {Krishnaswamy}}]{Nagulu2020NatElec}%
  \BibitemOpen
  \bibfield  {author} {\bibinfo {author} {\bibfnamefont {A.}~\bibnamefont
  {Nagulu}}, \bibinfo {author} {\bibfnamefont {N.}~\bibnamefont
  {Reiskarimian}},\ and\ \bibinfo {author} {\bibfnamefont {H.}~\bibnamefont
  {Krishnaswamy}},\ }\bibfield  {title} {\bibinfo {title} {Non-reciprocal
  electronics based on temporal modulation},\ }\href
  {https://doi.org/10.1038/s41928-020-0400-5} {\bibfield  {journal} {\bibinfo
  {journal} {Nat. Electronics}\ }\textbf {\bibinfo {volume} {3}},\ \bibinfo
  {pages} {241} (\bibinfo {year} {2020})}\BibitemShut {NoStop}%
\bibitem [{\citenamefont {Baconnier}\ \emph {et~al.}(2022)\citenamefont
  {Baconnier}, \citenamefont {Shohat}, \citenamefont {L{\'o}pez}, \citenamefont
  {Coulais}, \citenamefont {D{\'e}mery}, \citenamefont {D{\"u}ring},\ and\
  \citenamefont {Dauchot}}]{Baconnier2022NatPhy}%
  \BibitemOpen
  \bibfield  {author} {\bibinfo {author} {\bibfnamefont {P.}~\bibnamefont
  {Baconnier}}, \bibinfo {author} {\bibfnamefont {D.}~\bibnamefont {Shohat}},
  \bibinfo {author} {\bibfnamefont {C.~H.}\ \bibnamefont {L{\'o}pez}}, \bibinfo
  {author} {\bibfnamefont {C.}~\bibnamefont {Coulais}}, \bibinfo {author}
  {\bibfnamefont {V.}~\bibnamefont {D{\'e}mery}}, \bibinfo {author}
  {\bibfnamefont {G.}~\bibnamefont {D{\"u}ring}},\ and\ \bibinfo {author}
  {\bibfnamefont {O.}~\bibnamefont {Dauchot}},\ }\bibfield  {title} {\bibinfo
  {title} {Selective and collective actuation in active solids},\ }\href
  {https://doi.org/10.1038/s41567-022-01704-x} {\bibfield  {journal} {\bibinfo
  {journal} {Nat. Phys.}\ }\textbf {\bibinfo {volume} {18}},\ \bibinfo {pages}
  {1234} (\bibinfo {year} {2022})}\BibitemShut {NoStop}%
\bibitem [{\citenamefont {Nassar}\ \emph {et~al.}(2020)\citenamefont {Nassar},
  \citenamefont {Yousefzadeh}, \citenamefont {Fleury}, \citenamefont {Ruzzene},
  \citenamefont {Al{\`u}}, \citenamefont {Daraio}, \citenamefont {Norris},
  \citenamefont {Huang},\ and\ \citenamefont {Haberman}}]{Nassar2020NatRevMat}%
  \BibitemOpen
  \bibfield  {author} {\bibinfo {author} {\bibfnamefont {H.}~\bibnamefont
  {Nassar}}, \bibinfo {author} {\bibfnamefont {B.}~\bibnamefont {Yousefzadeh}},
  \bibinfo {author} {\bibfnamefont {R.}~\bibnamefont {Fleury}}, \bibinfo
  {author} {\bibfnamefont {M.}~\bibnamefont {Ruzzene}}, \bibinfo {author}
  {\bibfnamefont {A.}~\bibnamefont {Al{\`u}}}, \bibinfo {author} {\bibfnamefont
  {C.}~\bibnamefont {Daraio}}, \bibinfo {author} {\bibfnamefont {A.~N.}\
  \bibnamefont {Norris}}, \bibinfo {author} {\bibfnamefont {G.}~\bibnamefont
  {Huang}},\ and\ \bibinfo {author} {\bibfnamefont {M.~R.}\ \bibnamefont
  {Haberman}},\ }\bibfield  {title} {\bibinfo {title} {Nonreciprocity in
  acoustic and elastic materials},\ }\href
  {https://doi.org/10.1038/s41578-020-0206-0} {\bibfield  {journal} {\bibinfo
  {journal} {Nat. Rev. Mater.}\ }\textbf {\bibinfo {volume} {5}},\ \bibinfo
  {pages} {667} (\bibinfo {year} {2020})}\BibitemShut {NoStop}%
\bibitem [{\citenamefont {Estep}\ \emph {et~al.}(2014)\citenamefont {Estep},
  \citenamefont {Sounas}, \citenamefont {Soric},\ and\ \citenamefont
  {Alu}}]{estep2014NatPhy}%
  \BibitemOpen
  \bibfield  {author} {\bibinfo {author} {\bibfnamefont {N.~A.}\ \bibnamefont
  {Estep}}, \bibinfo {author} {\bibfnamefont {D.~L.}\ \bibnamefont {Sounas}},
  \bibinfo {author} {\bibfnamefont {J.}~\bibnamefont {Soric}},\ and\ \bibinfo
  {author} {\bibfnamefont {A.}~\bibnamefont {Alu}},\ }\bibfield  {title}
  {\bibinfo {title} {Magnetic-free non-reciprocity and isolation based on
  parametrically modulated coupled-resonator loops},\ }\href
  {https://doi.org/10.1038/nphys3134} {\bibfield  {journal} {\bibinfo
  {journal} {Nat. Phys.}\ }\textbf {\bibinfo {volume} {10}},\ \bibinfo {pages}
  {923} (\bibinfo {year} {2014})}\BibitemShut {NoStop}%
\bibitem [{\citenamefont {Ruesink}\ \emph {et~al.}(2016)\citenamefont
  {Ruesink}, \citenamefont {Miri}, \citenamefont {Alu},\ and\ \citenamefont
  {Verhagen}}]{ruesink2016NatComm}%
  \BibitemOpen
  \bibfield  {author} {\bibinfo {author} {\bibfnamefont {F.}~\bibnamefont
  {Ruesink}}, \bibinfo {author} {\bibfnamefont {M.-A.}\ \bibnamefont {Miri}},
  \bibinfo {author} {\bibfnamefont {A.}~\bibnamefont {Alu}},\ and\ \bibinfo
  {author} {\bibfnamefont {E.}~\bibnamefont {Verhagen}},\ }\bibfield  {title}
  {\bibinfo {title} {Nonreciprocity and magnetic-free isolation based on
  optomechanical interactions},\ }\href {https://doi.org/10.1038/ncomms13662}
  {\bibfield  {journal} {\bibinfo  {journal} {Nat. Commun.}\ }\textbf {\bibinfo
  {volume} {7}},\ \bibinfo {pages} {13662} (\bibinfo {year}
  {2016})}\BibitemShut {NoStop}%
\bibitem [{\citenamefont {Fang}\ \emph {et~al.}(2017)\citenamefont {Fang},
  \citenamefont {Luo}, \citenamefont {Metelmann}, \citenamefont {Matheny},
  \citenamefont {Marquardt}, \citenamefont {Clerk},\ and\ \citenamefont
  {Painter}}]{fang2017NatPhy}%
  \BibitemOpen
  \bibfield  {author} {\bibinfo {author} {\bibfnamefont {K.}~\bibnamefont
  {Fang}}, \bibinfo {author} {\bibfnamefont {J.}~\bibnamefont {Luo}}, \bibinfo
  {author} {\bibfnamefont {A.}~\bibnamefont {Metelmann}}, \bibinfo {author}
  {\bibfnamefont {M.~H.}\ \bibnamefont {Matheny}}, \bibinfo {author}
  {\bibfnamefont {F.}~\bibnamefont {Marquardt}}, \bibinfo {author}
  {\bibfnamefont {A.~A.}\ \bibnamefont {Clerk}},\ and\ \bibinfo {author}
  {\bibfnamefont {O.}~\bibnamefont {Painter}},\ }\bibfield  {title} {\bibinfo
  {title} {Generalized non-reciprocity in an optomechanical circuit via
  synthetic magnetism and reservoir engineering},\ }\href
  {https://doi.org/10.1038/nphys4009} {\bibfield  {journal} {\bibinfo
  {journal} {Nat. Phys.}\ }\textbf {\bibinfo {volume} {13}},\ \bibinfo {pages}
  {465} (\bibinfo {year} {2017})}\BibitemShut {NoStop}%
\bibitem [{\citenamefont {Peterson}\ \emph {et~al.}(2017)\citenamefont
  {Peterson}, \citenamefont {Lecocq}, \citenamefont {Cicak}, \citenamefont
  {Simmonds}, \citenamefont {Aumentado},\ and\ \citenamefont
  {Teufel}}]{Peterson2017PRX}%
  \BibitemOpen
  \bibfield  {author} {\bibinfo {author} {\bibfnamefont {G.~A.}\ \bibnamefont
  {Peterson}}, \bibinfo {author} {\bibfnamefont {F.}~\bibnamefont {Lecocq}},
  \bibinfo {author} {\bibfnamefont {K.}~\bibnamefont {Cicak}}, \bibinfo
  {author} {\bibfnamefont {R.~W.}\ \bibnamefont {Simmonds}}, \bibinfo {author}
  {\bibfnamefont {J.}~\bibnamefont {Aumentado}},\ and\ \bibinfo {author}
  {\bibfnamefont {J.~D.}\ \bibnamefont {Teufel}},\ }\bibfield  {title}
  {\bibinfo {title} {Demonstration of efficient nonreciprocity in a microwave
  optomechanical circuit},\ }\href {https://doi.org/10.1103/PhysRevX.7.031001}
  {\bibfield  {journal} {\bibinfo  {journal} {Phys. Rev. X}\ }\textbf {\bibinfo
  {volume} {7}},\ \bibinfo {pages} {031001} (\bibinfo {year}
  {2017})}\BibitemShut {NoStop}%
\bibitem [{\citenamefont {Xu}\ \emph {et~al.}(2019)\citenamefont {Xu},
  \citenamefont {Jiang}, \citenamefont {Clerk},\ and\ \citenamefont
  {Harris}}]{xu2019Nature}%
  \BibitemOpen
  \bibfield  {author} {\bibinfo {author} {\bibfnamefont {H.}~\bibnamefont
  {Xu}}, \bibinfo {author} {\bibfnamefont {L.}~\bibnamefont {Jiang}}, \bibinfo
  {author} {\bibfnamefont {A.}~\bibnamefont {Clerk}},\ and\ \bibinfo {author}
  {\bibfnamefont {J.}~\bibnamefont {Harris}},\ }\bibfield  {title} {\bibinfo
  {title} {Nonreciprocal control and cooling of phonon modes in an
  optomechanical system},\ }\href {https://doi.org/10.1038/s41586-019-1061-2}
  {\bibfield  {journal} {\bibinfo  {journal} {Nature (London)}\ }\textbf
  {\bibinfo {volume} {568}},\ \bibinfo {pages} {65} (\bibinfo {year}
  {2019})}\BibitemShut {NoStop}%
\bibitem [{\citenamefont {Wanjura}\ \emph {et~al.}(2023)\citenamefont
  {Wanjura}, \citenamefont {Slim}, \citenamefont {del Pino}, \citenamefont
  {Brunelli}, \citenamefont {Verhagen},\ and\ \citenamefont
  {Nunnenkamp}}]{wanjura2023NatPhy}%
  \BibitemOpen
  \bibfield  {author} {\bibinfo {author} {\bibfnamefont {C.~C.}\ \bibnamefont
  {Wanjura}}, \bibinfo {author} {\bibfnamefont {J.~J.}\ \bibnamefont {Slim}},
  \bibinfo {author} {\bibfnamefont {J.}~\bibnamefont {del Pino}}, \bibinfo
  {author} {\bibfnamefont {M.}~\bibnamefont {Brunelli}}, \bibinfo {author}
  {\bibfnamefont {E.}~\bibnamefont {Verhagen}},\ and\ \bibinfo {author}
  {\bibfnamefont {A.}~\bibnamefont {Nunnenkamp}},\ }\bibfield  {title}
  {\bibinfo {title} {Quadrature nonreciprocity in bosonic networks without
  breaking time-reversal symmetry},\ }\href
  {https://doi.org/10.1038/s41567-023-02128-x} {\bibfield  {journal} {\bibinfo
  {journal} {Nat. Phys.}\ }\textbf {\bibinfo {volume} {19}},\ \bibinfo {pages}
  {1429} (\bibinfo {year} {2023})}\BibitemShut {NoStop}%
\bibitem [{\citenamefont {Reisenbauer}\ \emph {et~al.}(2024)\citenamefont
  {Reisenbauer}, \citenamefont {Rudolph}, \citenamefont {Egyed}, \citenamefont
  {Hornberger}, \citenamefont {Zasedatelev}, \citenamefont {Abuzarli},
  \citenamefont {Stickler},\ and\ \citenamefont
  {Deli{\'c}}}]{Reisenbauer2024NatPhy}%
  \BibitemOpen
  \bibfield  {author} {\bibinfo {author} {\bibfnamefont {M.}~\bibnamefont
  {Reisenbauer}}, \bibinfo {author} {\bibfnamefont {H.}~\bibnamefont
  {Rudolph}}, \bibinfo {author} {\bibfnamefont {L.}~\bibnamefont {Egyed}},
  \bibinfo {author} {\bibfnamefont {K.}~\bibnamefont {Hornberger}}, \bibinfo
  {author} {\bibfnamefont {A.~V.}\ \bibnamefont {Zasedatelev}}, \bibinfo
  {author} {\bibfnamefont {M.}~\bibnamefont {Abuzarli}}, \bibinfo {author}
  {\bibfnamefont {B.~A.}\ \bibnamefont {Stickler}},\ and\ \bibinfo {author}
  {\bibfnamefont {U.}~\bibnamefont {Deli{\'c}}},\ }\bibfield  {title} {\bibinfo
  {title} {Non-hermitian dynamics and non-reciprocity of optically coupled
  nanoparticles},\ }\href {https://doi.org/10.1038/s41567-024-02589-8}
  {\bibfield  {journal} {\bibinfo  {journal} {Nat. Phys.}\ }\textbf {\bibinfo
  {volume} {20}},\ \bibinfo {pages} {1629} (\bibinfo {year}
  {2024})}\BibitemShut {NoStop}%
\bibitem [{\citenamefont {Metelmann}\ and\ \citenamefont
  {Clerk}(2015)}]{Metelmann2015PRX}%
  \BibitemOpen
  \bibfield  {author} {\bibinfo {author} {\bibfnamefont {A.}~\bibnamefont
  {Metelmann}}\ and\ \bibinfo {author} {\bibfnamefont {A.~A.}\ \bibnamefont
  {Clerk}},\ }\bibfield  {title} {\bibinfo {title} {Nonreciprocal photon
  transmission and amplification via reservoir engineering},\ }\href
  {https://doi.org/10.1103/PhysRevX.5.021025} {\bibfield  {journal} {\bibinfo
  {journal} {Phys. Rev. X}\ }\textbf {\bibinfo {volume} {5}},\ \bibinfo {pages}
  {021025} (\bibinfo {year} {2015})}\BibitemShut {NoStop}%
\bibitem [{\citenamefont {Metelmann}\ and\ \citenamefont
  {Clerk}(2017)}]{Metelmann2017PRA}%
  \BibitemOpen
  \bibfield  {author} {\bibinfo {author} {\bibfnamefont {A.}~\bibnamefont
  {Metelmann}}\ and\ \bibinfo {author} {\bibfnamefont {A.~A.}\ \bibnamefont
  {Clerk}},\ }\bibfield  {title} {\bibinfo {title} {Nonreciprocal quantum
  interactions and devices via autonomous feedforward},\ }\href
  {https://doi.org/10.1103/PhysRevA.95.013837} {\bibfield  {journal} {\bibinfo
  {journal} {Phys. Rev. A}\ }\textbf {\bibinfo {volume} {95}},\ \bibinfo
  {pages} {013837} (\bibinfo {year} {2017})}\BibitemShut {NoStop}%
\bibitem [{\citenamefont {Huang}\ \emph {et~al.}(2021)\citenamefont {Huang},
  \citenamefont {Lu}, \citenamefont {Liang}, \citenamefont {Tao},\ and\
  \citenamefont {Liu}}]{Huang2021Light}%
  \BibitemOpen
  \bibfield  {author} {\bibinfo {author} {\bibfnamefont {X.}~\bibnamefont
  {Huang}}, \bibinfo {author} {\bibfnamefont {C.}~\bibnamefont {Lu}}, \bibinfo
  {author} {\bibfnamefont {C.}~\bibnamefont {Liang}}, \bibinfo {author}
  {\bibfnamefont {H.}~\bibnamefont {Tao}},\ and\ \bibinfo {author}
  {\bibfnamefont {Y.-C.}\ \bibnamefont {Liu}},\ }\bibfield  {title} {\bibinfo
  {title} {Loss-induced nonreciprocity},\ }\href
  {https://doi.org/10.1038/s41377-021-00464-2} {\bibfield  {journal} {\bibinfo
  {journal} {Light Sci. Appl.}\ }\textbf {\bibinfo {volume} {10}},\ \bibinfo
  {pages} {30} (\bibinfo {year} {2021})}\BibitemShut {NoStop}%
\bibitem [{\citenamefont {Wang}\ \emph {et~al.}(2023)\citenamefont {Wang},
  \citenamefont {Wang},\ and\ \citenamefont {Clerk}}]{Wang2023PRXQuantum}%
  \BibitemOpen
  \bibfield  {author} {\bibinfo {author} {\bibfnamefont {Y.-X.}\ \bibnamefont
  {Wang}}, \bibinfo {author} {\bibfnamefont {C.}~\bibnamefont {Wang}},\ and\
  \bibinfo {author} {\bibfnamefont {A.~A.}\ \bibnamefont {Clerk}},\ }\bibfield
  {title} {\bibinfo {title} {Quantum nonreciprocal interactions via dissipative
  gauge symmetry},\ }\href {https://doi.org/10.1103/PRXQuantum.4.010306}
  {\bibfield  {journal} {\bibinfo  {journal} {PRX Quantum}\ }\textbf {\bibinfo
  {volume} {4}},\ \bibinfo {pages} {010306} (\bibinfo {year}
  {2023})}\BibitemShut {NoStop}%
\bibitem [{\citenamefont {Manipatruni}\ \emph {et~al.}(2009)\citenamefont
  {Manipatruni}, \citenamefont {Robinson},\ and\ \citenamefont
  {Lipson}}]{Manipatruni2009PRL}%
  \BibitemOpen
  \bibfield  {author} {\bibinfo {author} {\bibfnamefont {S.}~\bibnamefont
  {Manipatruni}}, \bibinfo {author} {\bibfnamefont {J.~T.}\ \bibnamefont
  {Robinson}},\ and\ \bibinfo {author} {\bibfnamefont {M.}~\bibnamefont
  {Lipson}},\ }\bibfield  {title} {\bibinfo {title} {Optical nonreciprocity in
  optomechanical structures},\ }\href
  {https://doi.org/10.1103/PhysRevLett.102.213903} {\bibfield  {journal}
  {\bibinfo  {journal} {Phys. Rev. Lett.}\ }\textbf {\bibinfo {volume} {102}},\
  \bibinfo {pages} {213903} (\bibinfo {year} {2009})}\BibitemShut {NoStop}%
\bibitem [{\citenamefont {Lin}\ \emph {et~al.}(2011)\citenamefont {Lin},
  \citenamefont {Ramezani}, \citenamefont {Eichelkraut}, \citenamefont
  {Kottos}, \citenamefont {Cao},\ and\ \citenamefont
  {Christodoulides}}]{Lin2011PRL}%
  \BibitemOpen
  \bibfield  {author} {\bibinfo {author} {\bibfnamefont {Z.}~\bibnamefont
  {Lin}}, \bibinfo {author} {\bibfnamefont {H.}~\bibnamefont {Ramezani}},
  \bibinfo {author} {\bibfnamefont {T.}~\bibnamefont {Eichelkraut}}, \bibinfo
  {author} {\bibfnamefont {T.}~\bibnamefont {Kottos}}, \bibinfo {author}
  {\bibfnamefont {H.}~\bibnamefont {Cao}},\ and\ \bibinfo {author}
  {\bibfnamefont {D.~N.}\ \bibnamefont {Christodoulides}},\ }\bibfield  {title}
  {\bibinfo {title} {Unidirectional invisibility induced by
  $\mathcal{P}\mathcal{T}$-symmetric periodic structures},\ }\href
  {https://doi.org/10.1103/PhysRevLett.106.213901} {\bibfield  {journal}
  {\bibinfo  {journal} {Phys. Rev. Lett.}\ }\textbf {\bibinfo {volume} {106}},\
  \bibinfo {pages} {213901} (\bibinfo {year} {2011})}\BibitemShut {NoStop}%
\bibitem [{\citenamefont {Fan}\ \emph {et~al.}(2012)\citenamefont {Fan},
  \citenamefont {Wang}, \citenamefont {Varghese}, \citenamefont {Shen},
  \citenamefont {Niu}, \citenamefont {Xuan}, \citenamefont {Weiner},\ and\
  \citenamefont {Qi}}]{fan2012Science}%
  \BibitemOpen
  \bibfield  {author} {\bibinfo {author} {\bibfnamefont {L.}~\bibnamefont
  {Fan}}, \bibinfo {author} {\bibfnamefont {J.}~\bibnamefont {Wang}}, \bibinfo
  {author} {\bibfnamefont {L.~T.}\ \bibnamefont {Varghese}}, \bibinfo {author}
  {\bibfnamefont {H.}~\bibnamefont {Shen}}, \bibinfo {author} {\bibfnamefont
  {B.}~\bibnamefont {Niu}}, \bibinfo {author} {\bibfnamefont {Y.}~\bibnamefont
  {Xuan}}, \bibinfo {author} {\bibfnamefont {A.~M.}\ \bibnamefont {Weiner}},\
  and\ \bibinfo {author} {\bibfnamefont {M.}~\bibnamefont {Qi}},\ }\bibfield
  {title} {\bibinfo {title} {An all-silicon passive optical diode},\ }\href
  {https://www.science.org/doi/abs/10.1126/science.1214383} {\bibfield
  {journal} {\bibinfo  {journal} {Science}\ }\textbf {\bibinfo {volume}
  {335}},\ \bibinfo {pages} {447} (\bibinfo {year} {2012})}\BibitemShut
  {NoStop}%
\bibitem [{\citenamefont {Wang}\ \emph {et~al.}(2013)\citenamefont {Wang},
  \citenamefont {Zhou}, \citenamefont {Guo}, \citenamefont {Zhang},
  \citenamefont {Evers},\ and\ \citenamefont {Zhu}}]{Wang2013PRL}%
  \BibitemOpen
  \bibfield  {author} {\bibinfo {author} {\bibfnamefont {D.-W.}\ \bibnamefont
  {Wang}}, \bibinfo {author} {\bibfnamefont {H.-T.}\ \bibnamefont {Zhou}},
  \bibinfo {author} {\bibfnamefont {M.-J.}\ \bibnamefont {Guo}}, \bibinfo
  {author} {\bibfnamefont {J.-X.}\ \bibnamefont {Zhang}}, \bibinfo {author}
  {\bibfnamefont {J.}~\bibnamefont {Evers}},\ and\ \bibinfo {author}
  {\bibfnamefont {S.-Y.}\ \bibnamefont {Zhu}},\ }\bibfield  {title} {\bibinfo
  {title} {Optical diode made from a moving photonic crystal},\ }\href
  {https://doi.org/10.1103/PhysRevLett.110.093901} {\bibfield  {journal}
  {\bibinfo  {journal} {Phys. Rev. Lett.}\ }\textbf {\bibinfo {volume} {110}},\
  \bibinfo {pages} {093901} (\bibinfo {year} {2013})}\BibitemShut {NoStop}%
\bibitem [{\citenamefont {Shen}\ \emph {et~al.}(2016)\citenamefont {Shen},
  \citenamefont {Zhang}, \citenamefont {Chen}, \citenamefont {Zou},
  \citenamefont {Xiao}, \citenamefont {Zou}, \citenamefont {Sun}, \citenamefont
  {Guo},\ and\ \citenamefont {Dong}}]{shen2016NatPhot}%
  \BibitemOpen
  \bibfield  {author} {\bibinfo {author} {\bibfnamefont {Z.}~\bibnamefont
  {Shen}}, \bibinfo {author} {\bibfnamefont {Y.-L.}\ \bibnamefont {Zhang}},
  \bibinfo {author} {\bibfnamefont {Y.}~\bibnamefont {Chen}}, \bibinfo {author}
  {\bibfnamefont {C.-L.}\ \bibnamefont {Zou}}, \bibinfo {author} {\bibfnamefont
  {Y.-F.}\ \bibnamefont {Xiao}}, \bibinfo {author} {\bibfnamefont {X.-B.}\
  \bibnamefont {Zou}}, \bibinfo {author} {\bibfnamefont {F.-W.}\ \bibnamefont
  {Sun}}, \bibinfo {author} {\bibfnamefont {G.-C.}\ \bibnamefont {Guo}},\ and\
  \bibinfo {author} {\bibfnamefont {C.-H.}\ \bibnamefont {Dong}},\ }\bibfield
  {title} {\bibinfo {title} {Experimental realization of optomechanically
  induced non-reciprocity},\ }\href {https://doi.org/10.1038/nphoton.2016.161}
  {\bibfield  {journal} {\bibinfo  {journal} {Nat. Photonics}\ }\textbf
  {\bibinfo {volume} {10}},\ \bibinfo {pages} {657} (\bibinfo {year}
  {2016})}\BibitemShut {NoStop}%
\bibitem [{\citenamefont {Cao}\ \emph {et~al.}(2017)\citenamefont {Cao},
  \citenamefont {Wang}, \citenamefont {Dong}, \citenamefont {Jing},
  \citenamefont {Liu}, \citenamefont {Chen}, \citenamefont {Ge}, \citenamefont
  {Gong},\ and\ \citenamefont {Xiao}}]{Cao2017PRL}%
  \BibitemOpen
  \bibfield  {author} {\bibinfo {author} {\bibfnamefont {Q.-T.}\ \bibnamefont
  {Cao}}, \bibinfo {author} {\bibfnamefont {H.}~\bibnamefont {Wang}}, \bibinfo
  {author} {\bibfnamefont {C.-H.}\ \bibnamefont {Dong}}, \bibinfo {author}
  {\bibfnamefont {H.}~\bibnamefont {Jing}}, \bibinfo {author} {\bibfnamefont
  {R.-S.}\ \bibnamefont {Liu}}, \bibinfo {author} {\bibfnamefont
  {X.}~\bibnamefont {Chen}}, \bibinfo {author} {\bibfnamefont {L.}~\bibnamefont
  {Ge}}, \bibinfo {author} {\bibfnamefont {Q.}~\bibnamefont {Gong}},\ and\
  \bibinfo {author} {\bibfnamefont {Y.-F.}\ \bibnamefont {Xiao}},\ }\bibfield
  {title} {\bibinfo {title} {Experimental demonstration of spontaneous
  chirality in a nonlinear microresonator},\ }\href
  {https://doi.org/10.1103/PhysRevLett.118.033901} {\bibfield  {journal}
  {\bibinfo  {journal} {Phys. Rev. Lett.}\ }\textbf {\bibinfo {volume} {118}},\
  \bibinfo {pages} {033901} (\bibinfo {year} {2017})}\BibitemShut {NoStop}%
\bibitem [{\citenamefont {Xia}\ \emph {et~al.}(2018)\citenamefont {Xia},
  \citenamefont {Nori},\ and\ \citenamefont {Xiao}}]{Xia2018PRL}%
  \BibitemOpen
  \bibfield  {author} {\bibinfo {author} {\bibfnamefont {K.}~\bibnamefont
  {Xia}}, \bibinfo {author} {\bibfnamefont {F.}~\bibnamefont {Nori}},\ and\
  \bibinfo {author} {\bibfnamefont {M.}~\bibnamefont {Xiao}},\ }\bibfield
  {title} {\bibinfo {title} {Cavity-free optical isolators and circulators
  using a chiral cross-{Kerr} nonlinearity},\ }\href
  {https://doi.org/10.1103/PhysRevLett.121.203602} {\bibfield  {journal}
  {\bibinfo  {journal} {Phys. Rev. Lett.}\ }\textbf {\bibinfo {volume} {121}},\
  \bibinfo {pages} {203602} (\bibinfo {year} {2018})}\BibitemShut {NoStop}%
\bibitem [{\citenamefont {Tang}\ \emph {et~al.}(2022)\citenamefont {Tang},
  \citenamefont {Tang}, \citenamefont {Chen}, \citenamefont {Nori},
  \citenamefont {Xiao},\ and\ \citenamefont {Xia}}]{Tang2022PRL}%
  \BibitemOpen
  \bibfield  {author} {\bibinfo {author} {\bibfnamefont {L.}~\bibnamefont
  {Tang}}, \bibinfo {author} {\bibfnamefont {J.}~\bibnamefont {Tang}}, \bibinfo
  {author} {\bibfnamefont {M.}~\bibnamefont {Chen}}, \bibinfo {author}
  {\bibfnamefont {F.}~\bibnamefont {Nori}}, \bibinfo {author} {\bibfnamefont
  {M.}~\bibnamefont {Xiao}},\ and\ \bibinfo {author} {\bibfnamefont
  {K.}~\bibnamefont {Xia}},\ }\bibfield  {title} {\bibinfo {title} {Quantum
  squeezing induced optical nonreciprocity},\ }\href
  {https://doi.org/10.1103/PhysRevLett.128.083604} {\bibfield  {journal}
  {\bibinfo  {journal} {Phys. Rev. Lett.}\ }\textbf {\bibinfo {volume} {128}},\
  \bibinfo {pages} {083604} (\bibinfo {year} {2022})}\BibitemShut {NoStop}%
\bibitem [{\citenamefont {Fruchart}\ \emph {et~al.}(2021)\citenamefont
  {Fruchart}, \citenamefont {Hanai}, \citenamefont {Littlewood},\ and\
  \citenamefont {Vitelli}}]{Fruchart2021Nature}%
  \BibitemOpen
  \bibfield  {author} {\bibinfo {author} {\bibfnamefont {M.}~\bibnamefont
  {Fruchart}}, \bibinfo {author} {\bibfnamefont {R.}~\bibnamefont {Hanai}},
  \bibinfo {author} {\bibfnamefont {P.~B.}\ \bibnamefont {Littlewood}},\ and\
  \bibinfo {author} {\bibfnamefont {V.}~\bibnamefont {Vitelli}},\ }\bibfield
  {title} {\bibinfo {title} {Non-reciprocal phase transitions},\ }\href
  {https://doi.org/10.1038/s41586-021-03375-9} {\bibfield  {journal} {\bibinfo
  {journal} {Nature (London)}\ }\textbf {\bibinfo {volume} {592}},\ \bibinfo
  {pages} {363} (\bibinfo {year} {2021})}\BibitemShut {NoStop}%
\bibitem [{\citenamefont {Hanai}(2024)}]{Hanai2024}%
  \BibitemOpen
  \bibfield  {author} {\bibinfo {author} {\bibfnamefont {R.}~\bibnamefont
  {Hanai}},\ }\bibfield  {title} {\bibinfo {title} {Nonreciprocal frustration:
  Time crystalline order-by-disorder phenomenon and a spin-glass-like state},\
  }\href {https://doi.org/10.1103/physrevx.14.011029} {\bibfield  {journal}
  {\bibinfo  {journal} {Phys. Rev. X}\ }\textbf {\bibinfo {volume} {14}},\
  \bibinfo {pages} {011029} (\bibinfo {year} {2024})}\BibitemShut {NoStop}%
\bibitem [{\citenamefont {Begg}\ and\ \citenamefont
  {Hanai}(2024)}]{Hanai2024PRL}%
  \BibitemOpen
  \bibfield  {author} {\bibinfo {author} {\bibfnamefont {S.~E.}\ \bibnamefont
  {Begg}}\ and\ \bibinfo {author} {\bibfnamefont {R.}~\bibnamefont {Hanai}},\
  }\bibfield  {title} {\bibinfo {title} {Quantum criticality in open quantum
  spin chains with nonreciprocity},\ }\href
  {https://doi.org/10.1103/PhysRevLett.132.120401} {\bibfield  {journal}
  {\bibinfo  {journal} {Phys. Rev. Lett.}\ }\textbf {\bibinfo {volume} {132}},\
  \bibinfo {pages} {120401} (\bibinfo {year} {2024})}\BibitemShut {NoStop}%
\bibitem [{\citenamefont {Avni}\ \emph
  {et~al.}(2025{\natexlab{a}})\citenamefont {Avni}, \citenamefont {Fruchart},
  \citenamefont {Martin}, \citenamefont {Seara},\ and\ \citenamefont
  {Vitelli}}]{Avni2025PRL}%
  \BibitemOpen
  \bibfield  {author} {\bibinfo {author} {\bibfnamefont {Y.}~\bibnamefont
  {Avni}}, \bibinfo {author} {\bibfnamefont {M.}~\bibnamefont {Fruchart}},
  \bibinfo {author} {\bibfnamefont {D.}~\bibnamefont {Martin}}, \bibinfo
  {author} {\bibfnamefont {D.}~\bibnamefont {Seara}},\ and\ \bibinfo {author}
  {\bibfnamefont {V.}~\bibnamefont {Vitelli}},\ }\bibfield  {title} {\bibinfo
  {title} {Nonreciprocal {I}sing model},\ }\href
  {https://doi.org/10.1103/PhysRevLett.134.117103} {\bibfield  {journal}
  {\bibinfo  {journal} {Phys. Rev. Lett.}\ }\textbf {\bibinfo {volume} {134}},\
  \bibinfo {pages} {117103} (\bibinfo {year} {2025}{\natexlab{a}})}\BibitemShut
  {NoStop}%
\bibitem [{\citenamefont {Avni}\ \emph
  {et~al.}(2025{\natexlab{b}})\citenamefont {Avni}, \citenamefont {Fruchart},
  \citenamefont {Martin}, \citenamefont {Seara},\ and\ \citenamefont
  {Vitelli}}]{Avni2025PRE}%
  \BibitemOpen
  \bibfield  {author} {\bibinfo {author} {\bibfnamefont {Y.}~\bibnamefont
  {Avni}}, \bibinfo {author} {\bibfnamefont {M.}~\bibnamefont {Fruchart}},
  \bibinfo {author} {\bibfnamefont {D.}~\bibnamefont {Martin}}, \bibinfo
  {author} {\bibfnamefont {D.}~\bibnamefont {Seara}},\ and\ \bibinfo {author}
  {\bibfnamefont {V.}~\bibnamefont {Vitelli}},\ }\bibfield  {title} {\bibinfo
  {title} {Dynamical phase transitions in the nonreciprocal {I}sing model},\
  }\href {https://doi.org/10.1103/PhysRevE.111.034124} {\bibfield  {journal}
  {\bibinfo  {journal} {Phys. Rev. E}\ }\textbf {\bibinfo {volume} {111}},\
  \bibinfo {pages} {034124} (\bibinfo {year} {2025}{\natexlab{b}})}\BibitemShut
  {NoStop}%
\bibitem [{\citenamefont {Zhu}\ \emph {et~al.}(2024)\citenamefont {Zhu},
  \citenamefont {Hu}, \citenamefont {Wang}, \citenamefont {Qin}, \citenamefont
  {L\"u},\ and\ \citenamefont {Nori}}]{Zhu2024PRL}%
  \BibitemOpen
  \bibfield  {author} {\bibinfo {author} {\bibfnamefont {G.-L.}\ \bibnamefont
  {Zhu}}, \bibinfo {author} {\bibfnamefont {C.-S.}\ \bibnamefont {Hu}},
  \bibinfo {author} {\bibfnamefont {H.}~\bibnamefont {Wang}}, \bibinfo {author}
  {\bibfnamefont {W.}~\bibnamefont {Qin}}, \bibinfo {author} {\bibfnamefont
  {X.-Y.}\ \bibnamefont {L\"u}},\ and\ \bibinfo {author} {\bibfnamefont
  {F.}~\bibnamefont {Nori}},\ }\bibfield  {title} {\bibinfo {title}
  {Nonreciprocal superradiant phase transitions and multicriticality in a
  cavity {QED} system},\ }\href
  {https://doi.org/10.1103/PhysRevLett.132.193602} {\bibfield  {journal}
  {\bibinfo  {journal} {Phys. Rev. Lett.}\ }\textbf {\bibinfo {volume} {132}},\
  \bibinfo {pages} {193602} (\bibinfo {year} {2024})}\BibitemShut {NoStop}%
\bibitem [{\citenamefont {Weiderpass}\ \emph {et~al.}(2025)\citenamefont
  {Weiderpass}, \citenamefont {Sharma},\ and\ \citenamefont
  {Sethi}}]{Weiderpass2025PE}%
  \BibitemOpen
  \bibfield  {author} {\bibinfo {author} {\bibfnamefont {G.~A.}\ \bibnamefont
  {Weiderpass}}, \bibinfo {author} {\bibfnamefont {M.}~\bibnamefont {Sharma}},\
  and\ \bibinfo {author} {\bibfnamefont {S.}~\bibnamefont {Sethi}},\ }\bibfield
   {title} {\bibinfo {title} {Solving the kinetic {I}sing model with
  nonreciprocity},\ }\href {https://doi.org/10.1103/PhysRevE.111.024107}
  {\bibfield  {journal} {\bibinfo  {journal} {Phys. Rev. E}\ }\textbf {\bibinfo
  {volume} {111}},\ \bibinfo {pages} {024107} (\bibinfo {year}
  {2025})}\BibitemShut {NoStop}%
\bibitem [{\citenamefont {Brauns}\ and\ \citenamefont
  {Marchetti}(2024)}]{Brauns2024PRX}%
  \BibitemOpen
  \bibfield  {author} {\bibinfo {author} {\bibfnamefont {F.}~\bibnamefont
  {Brauns}}\ and\ \bibinfo {author} {\bibfnamefont {M.~C.}\ \bibnamefont
  {Marchetti}},\ }\bibfield  {title} {\bibinfo {title} {Nonreciprocal pattern
  formation of conserved fields},\ }\href
  {https://doi.org/10.1103/PhysRevX.14.021014} {\bibfield  {journal} {\bibinfo
  {journal} {Phys. Rev. X}\ }\textbf {\bibinfo {volume} {14}},\ \bibinfo
  {pages} {021014} (\bibinfo {year} {2024})}\BibitemShut {NoStop}%
\bibitem [{\citenamefont {Nadolny}\ \emph {et~al.}(2025)\citenamefont
  {Nadolny}, \citenamefont {Bruder},\ and\ \citenamefont
  {Brunelli}}]{Nadolny2025PRX}%
  \BibitemOpen
  \bibfield  {author} {\bibinfo {author} {\bibfnamefont {T.}~\bibnamefont
  {Nadolny}}, \bibinfo {author} {\bibfnamefont {C.}~\bibnamefont {Bruder}},\
  and\ \bibinfo {author} {\bibfnamefont {M.}~\bibnamefont {Brunelli}},\
  }\bibfield  {title} {\bibinfo {title} {Nonreciprocal synchronization of
  active quantum spins},\ }\href {https://doi.org/10.1103/PhysRevX.15.011010}
  {\bibfield  {journal} {\bibinfo  {journal} {Phys. Rev. X}\ }\textbf {\bibinfo
  {volume} {15}},\ \bibinfo {pages} {011010} (\bibinfo {year}
  {2025})}\BibitemShut {NoStop}%
\bibitem [{\citenamefont {Gu}\ \emph {et~al.}(2025)\citenamefont {Gu},
  \citenamefont {Guiselin}, \citenamefont {Bain}, \citenamefont {Zuriguel},\
  and\ \citenamefont {Bartolo}}]{Gu2025Nature}%
  \BibitemOpen
  \bibfield  {author} {\bibinfo {author} {\bibfnamefont {F.}~\bibnamefont
  {Gu}}, \bibinfo {author} {\bibfnamefont {B.}~\bibnamefont {Guiselin}},
  \bibinfo {author} {\bibfnamefont {N.}~\bibnamefont {Bain}}, \bibinfo {author}
  {\bibfnamefont {I.}~\bibnamefont {Zuriguel}},\ and\ \bibinfo {author}
  {\bibfnamefont {D.}~\bibnamefont {Bartolo}},\ }\bibfield  {title} {\bibinfo
  {title} {Emergence of collective oscillations in massive human crowds},\
  }\href {https://doi.org/10.1038/s41586-024-08514-6} {\bibfield  {journal}
  {\bibinfo  {journal} {Nature (London)}\ }\textbf {\bibinfo {volume} {638}},\
  \bibinfo {pages} {112} (\bibinfo {year} {2025})}\BibitemShut {NoStop}%
\bibitem [{\citenamefont {Chiacchio}\ and\ \citenamefont
  {Nunnenkamp}(2019)}]{Nunnenkamp2019PRL}%
  \BibitemOpen
  \bibfield  {author} {\bibinfo {author} {\bibfnamefont {E.~I.~R.}\
  \bibnamefont {Chiacchio}}\ and\ \bibinfo {author} {\bibfnamefont
  {A.}~\bibnamefont {Nunnenkamp}},\ }\bibfield  {title} {\bibinfo {title}
  {Dissipation-induced instabilities of a spinor {B}ose-{E}instein condensate
  inside an optical cavity},\ }\href
  {https://doi.org/10.1103/PhysRevLett.122.193605} {\bibfield  {journal}
  {\bibinfo  {journal} {Phys. Rev. Lett.}\ }\textbf {\bibinfo {volume} {122}},\
  \bibinfo {pages} {193605} (\bibinfo {year} {2019})}\BibitemShut {NoStop}%
\bibitem [{\citenamefont {Chiacchio}\ \emph {et~al.}(2023)\citenamefont
  {Chiacchio}, \citenamefont {Nunnenkamp},\ and\ \citenamefont
  {Brunelli}}]{Nunnenkamp2023PRL}%
  \BibitemOpen
  \bibfield  {author} {\bibinfo {author} {\bibfnamefont {E.~I.~R.}\
  \bibnamefont {Chiacchio}}, \bibinfo {author} {\bibfnamefont {A.}~\bibnamefont
  {Nunnenkamp}},\ and\ \bibinfo {author} {\bibfnamefont {M.}~\bibnamefont
  {Brunelli}},\ }\bibfield  {title} {\bibinfo {title} {Nonreciprocal {D}icke
  model},\ }\href {https://doi.org/10.1103/PhysRevLett.131.113602} {\bibfield
  {journal} {\bibinfo  {journal} {Phys. Rev. Lett.}\ }\textbf {\bibinfo
  {volume} {131}},\ \bibinfo {pages} {113602} (\bibinfo {year}
  {2023})}\BibitemShut {NoStop}%
\bibitem [{\citenamefont {Jachinowski}\ and\ \citenamefont
  {Littlewood}(2025)}]{Jachinowski2025}%
  \BibitemOpen
  \bibfield  {author} {\bibinfo {author} {\bibfnamefont {J.}~\bibnamefont
  {Jachinowski}}\ and\ \bibinfo {author} {\bibfnamefont {P.~B.}\ \bibnamefont
  {Littlewood}},\ }\bibfield  {title} {\bibinfo {title} {Spin-only dynamics of
  the multi-species nonreciprocal {D}icke model},\ }\href
  {https://arxiv.org/abs/2507.07960} {\bibfield  {journal} {\bibinfo  {journal}
  {arXiv:2507.07960}\ } (\bibinfo {year} {2025})}\BibitemShut {NoStop}%
\bibitem [{\citenamefont {Baumann}\ \emph {et~al.}(2010)\citenamefont
  {Baumann}, \citenamefont {Guerlin}, \citenamefont {Brennecke},\ and\
  \citenamefont {Esslinger}}]{Baumann2010nature}%
  \BibitemOpen
  \bibfield  {author} {\bibinfo {author} {\bibfnamefont {K.}~\bibnamefont
  {Baumann}}, \bibinfo {author} {\bibfnamefont {C.}~\bibnamefont {Guerlin}},
  \bibinfo {author} {\bibfnamefont {F.}~\bibnamefont {Brennecke}},\ and\
  \bibinfo {author} {\bibfnamefont {T.}~\bibnamefont {Esslinger}},\ }\bibfield
  {title} {\bibinfo {title} {Dicke quantum phase transition with a superfluid
  gas in an optical cavity},\ }\href {https://doi.org/10.1038/nature09009}
  {\bibfield  {journal} {\bibinfo  {journal} {Nature (London)}\ }\textbf
  {\bibinfo {volume} {464}},\ \bibinfo {pages} {1301} (\bibinfo {year}
  {2010})}\BibitemShut {NoStop}%
\bibitem [{\citenamefont {Baumann}\ \emph {et~al.}(2011)\citenamefont
  {Baumann}, \citenamefont {Mottl}, \citenamefont {Brennecke},\ and\
  \citenamefont {Esslinger}}]{Baumann2011PRL}%
  \BibitemOpen
  \bibfield  {author} {\bibinfo {author} {\bibfnamefont {K.}~\bibnamefont
  {Baumann}}, \bibinfo {author} {\bibfnamefont {R.}~\bibnamefont {Mottl}},
  \bibinfo {author} {\bibfnamefont {F.}~\bibnamefont {Brennecke}},\ and\
  \bibinfo {author} {\bibfnamefont {T.}~\bibnamefont {Esslinger}},\ }\bibfield
  {title} {\bibinfo {title} {Exploring symmetry breaking at the {D}icke quantum
  phase transition},\ }\href {https://doi.org/10.1103/PhysRevLett.107.140402}
  {\bibfield  {journal} {\bibinfo  {journal} {Phys. Rev. Lett.}\ }\textbf
  {\bibinfo {volume} {107}},\ \bibinfo {pages} {140402} (\bibinfo {year}
  {2011})}\BibitemShut {NoStop}%
\bibitem [{\citenamefont {Kirton}\ \emph {et~al.}(2019)\citenamefont {Kirton},
  \citenamefont {Roses}, \citenamefont {Keeling},\ and\ \citenamefont
  {Dalla~Torre}}]{Kirton2018AdvQuanTech}%
  \BibitemOpen
  \bibfield  {author} {\bibinfo {author} {\bibfnamefont {P.}~\bibnamefont
  {Kirton}}, \bibinfo {author} {\bibfnamefont {M.~M.}\ \bibnamefont {Roses}},
  \bibinfo {author} {\bibfnamefont {J.}~\bibnamefont {Keeling}},\ and\ \bibinfo
  {author} {\bibfnamefont {E.~G.}\ \bibnamefont {Dalla~Torre}},\ }\bibfield
  {title} {\bibinfo {title} {Introduction to the {D}icke model: From
  equilibrium to nonequilibrium, and vice versa},\ }\href
  {https://doi.org/https://doi.org/10.1002/qute.201800043} {\bibfield
  {journal} {\bibinfo  {journal} {Adv. Quantum Technol.}\ }\textbf {\bibinfo
  {volume} {2}},\ \bibinfo {pages} {1800043} (\bibinfo {year}
  {2019})}\BibitemShut {NoStop}%
\bibitem [{\citenamefont {Farokh~Mivehvar}\ and\ \citenamefont
  {Ritsch}(2021)}]{Mivehvar2021AdvPhy}%
  \BibitemOpen
  \bibfield  {author} {\bibinfo {author} {\bibfnamefont {T.~D.}\ \bibnamefont
  {Farokh~Mivehvar}, \bibfnamefont {Francesco~Piazza}}\ and\ \bibinfo {author}
  {\bibfnamefont {H.}~\bibnamefont {Ritsch}},\ }\bibfield  {title} {\bibinfo
  {title} {Cavity {QED} with quantum gases: new paradigms in many-body
  physics},\ }\href {https://doi.org/10.1080/00018732.2021.1969727} {\bibfield
  {journal} {\bibinfo  {journal} {Adv. Phys.}\ }\textbf {\bibinfo {volume}
  {70}},\ \bibinfo {pages} {1} (\bibinfo {year} {2021})}\BibitemShut {NoStop}%
\bibitem [{\citenamefont {Klinder}\ \emph
  {et~al.}(2015{\natexlab{a}})\citenamefont {Klinder}, \citenamefont {Keßler},
  \citenamefont {Wolke}, \citenamefont {Mathey},\ and\ \citenamefont
  {Hemmerich}}]{Klinder2015PNAS}%
  \BibitemOpen
  \bibfield  {author} {\bibinfo {author} {\bibfnamefont {J.}~\bibnamefont
  {Klinder}}, \bibinfo {author} {\bibfnamefont {H.}~\bibnamefont {Keßler}},
  \bibinfo {author} {\bibfnamefont {M.}~\bibnamefont {Wolke}}, \bibinfo
  {author} {\bibfnamefont {L.}~\bibnamefont {Mathey}},\ and\ \bibinfo {author}
  {\bibfnamefont {A.}~\bibnamefont {Hemmerich}},\ }\bibfield  {title} {\bibinfo
  {title} {Dynamical phase transition in the open {D}icke model},\ }\href
  {https://doi.org/10.1073/pnas.1417132112} {\bibfield  {journal} {\bibinfo
  {journal} {Proc. Natl. Acad. Sci. U.S.A.}\ }\textbf {\bibinfo {volume}
  {112}},\ \bibinfo {pages} {3290} (\bibinfo {year}
  {2015}{\natexlab{a}})}\BibitemShut {NoStop}%
\bibitem [{\citenamefont {Klinder}\ \emph
  {et~al.}(2015{\natexlab{b}})\citenamefont {Klinder}, \citenamefont
  {Ke\ss{}ler}, \citenamefont {Bakhtiari}, \citenamefont {Thorwart},\ and\
  \citenamefont {Hemmerich}}]{Klinder2015PRL}%
  \BibitemOpen
  \bibfield  {author} {\bibinfo {author} {\bibfnamefont {J.}~\bibnamefont
  {Klinder}}, \bibinfo {author} {\bibfnamefont {H.}~\bibnamefont {Ke\ss{}ler}},
  \bibinfo {author} {\bibfnamefont {M.~R.}\ \bibnamefont {Bakhtiari}}, \bibinfo
  {author} {\bibfnamefont {M.}~\bibnamefont {Thorwart}},\ and\ \bibinfo
  {author} {\bibfnamefont {A.}~\bibnamefont {Hemmerich}},\ }\bibfield  {title}
  {\bibinfo {title} {Observation of a superradiant {M}ott insulator in the
  {D}icke-{H}ubbard model},\ }\href
  {https://doi.org/10.1103/PhysRevLett.115.230403} {\bibfield  {journal}
  {\bibinfo  {journal} {Phys. Rev. Lett.}\ }\textbf {\bibinfo {volume} {115}},\
  \bibinfo {pages} {230403} (\bibinfo {year} {2015}{\natexlab{b}})}\BibitemShut
  {NoStop}%
\bibitem [{\citenamefont {L{\'e}onard}\ \emph {et~al.}(2017)\citenamefont
  {L{\'e}onard}, \citenamefont {Morales}, \citenamefont {Zupancic},
  \citenamefont {Esslinger},\ and\ \citenamefont {Donner}}]{Donner2017Nature}%
  \BibitemOpen
  \bibfield  {author} {\bibinfo {author} {\bibfnamefont {J.}~\bibnamefont
  {L{\'e}onard}}, \bibinfo {author} {\bibfnamefont {A.}~\bibnamefont
  {Morales}}, \bibinfo {author} {\bibfnamefont {P.}~\bibnamefont {Zupancic}},
  \bibinfo {author} {\bibfnamefont {T.}~\bibnamefont {Esslinger}},\ and\
  \bibinfo {author} {\bibfnamefont {T.}~\bibnamefont {Donner}},\ }\bibfield
  {title} {\bibinfo {title} {Supersolid formation in a quantum gas breaking a
  continuous translational symmetry},\ }\href
  {https://doi.org/10.1038/nature21067} {\bibfield  {journal} {\bibinfo
  {journal} {Nature (London)}\ }\textbf {\bibinfo {volume} {543}},\ \bibinfo
  {pages} {87} (\bibinfo {year} {2017})}\BibitemShut {NoStop}%
\bibitem [{\citenamefont {Kroeze}\ \emph {et~al.}(2018)\citenamefont {Kroeze},
  \citenamefont {Guo}, \citenamefont {Vaidya}, \citenamefont {Keeling},\ and\
  \citenamefont {Lev}}]{Benjamin2018PRL}%
  \BibitemOpen
  \bibfield  {author} {\bibinfo {author} {\bibfnamefont {R.~M.}\ \bibnamefont
  {Kroeze}}, \bibinfo {author} {\bibfnamefont {Y.}~\bibnamefont {Guo}},
  \bibinfo {author} {\bibfnamefont {V.~D.}\ \bibnamefont {Vaidya}}, \bibinfo
  {author} {\bibfnamefont {J.}~\bibnamefont {Keeling}},\ and\ \bibinfo {author}
  {\bibfnamefont {B.~L.}\ \bibnamefont {Lev}},\ }\bibfield  {title} {\bibinfo
  {title} {Spinor self-ordering of a quantum gas in a cavity},\ }\href
  {https://doi.org/10.1103/PhysRevLett.121.163601} {\bibfield  {journal}
  {\bibinfo  {journal} {Phys. Rev. Lett.}\ }\textbf {\bibinfo {volume} {121}},\
  \bibinfo {pages} {163601} (\bibinfo {year} {2018})}\BibitemShut {NoStop}%
\bibitem [{\citenamefont {Guo}\ \emph {et~al.}(2021)\citenamefont {Guo},
  \citenamefont {Kroeze}, \citenamefont {Marsh}, \citenamefont
  {Gopalakrishnan}, \citenamefont {Keeling},\ and\ \citenamefont
  {Lev}}]{Guo2021Nature}%
  \BibitemOpen
  \bibfield  {author} {\bibinfo {author} {\bibfnamefont {Y.}~\bibnamefont
  {Guo}}, \bibinfo {author} {\bibfnamefont {R.~M.}\ \bibnamefont {Kroeze}},
  \bibinfo {author} {\bibfnamefont {B.~P.}\ \bibnamefont {Marsh}}, \bibinfo
  {author} {\bibfnamefont {S.}~\bibnamefont {Gopalakrishnan}}, \bibinfo
  {author} {\bibfnamefont {J.}~\bibnamefont {Keeling}},\ and\ \bibinfo {author}
  {\bibfnamefont {B.~L.}\ \bibnamefont {Lev}},\ }\bibfield  {title} {\bibinfo
  {title} {An optical lattice with sound},\ }\href
  {https://doi.org/10.1038/s41586-021-03945-x} {\bibfield  {journal} {\bibinfo
  {journal} {Nature (London)}\ }\textbf {\bibinfo {volume} {599}},\ \bibinfo
  {pages} {211} (\bibinfo {year} {2021})}\BibitemShut {NoStop}%
\bibitem [{\citenamefont {Diep}(2005)}]{Diep2005book}%
  \BibitemOpen
  \bibfield  {author} {\bibinfo {author} {\bibfnamefont {H.~T.}\ \bibnamefont
  {Diep}},\ }\href {https://doi.org/10.1142/5697} {\emph {\bibinfo {title}
  {Frustrated Spin Systems}}}\ (\bibinfo  {publisher} {World Sientific},\
  \bibinfo {year} {2005})\BibitemShut {NoStop}%
\bibitem [{\citenamefont {Moessner}\ and\ \citenamefont
  {Ramirez}(2006)}]{Moessner2006PhysicsToday}%
  \BibitemOpen
  \bibfield  {author} {\bibinfo {author} {\bibfnamefont {R.}~\bibnamefont
  {Moessner}}\ and\ \bibinfo {author} {\bibfnamefont {A.~P.}\ \bibnamefont
  {Ramirez}},\ }\bibfield  {title} {\bibinfo {title} {Geometrical
  frustration},\ }\href {https://doi.org/10.1063/1.2186278} {\bibfield
  {journal} {\bibinfo  {journal} {Phys. Today}\ }\textbf {\bibinfo {volume}
  {59}},\ \bibinfo {pages} {24} (\bibinfo {year} {2006})}\BibitemShut {NoStop}%
\bibitem [{\citenamefont {Bramwell}\ and\ \citenamefont
  {Gingras}(2001)}]{Steven2001Science}%
  \BibitemOpen
  \bibfield  {author} {\bibinfo {author} {\bibfnamefont {S.~T.}\ \bibnamefont
  {Bramwell}}\ and\ \bibinfo {author} {\bibfnamefont {M.~J.~P.}\ \bibnamefont
  {Gingras}},\ }\bibfield  {title} {\bibinfo {title} {Spin ice state in
  frustrated magnetic pyrochlore materials},\ }\href
  {https://doi.org/10.1126/science.1064761} {\bibfield  {journal} {\bibinfo
  {journal} {Science}\ }\textbf {\bibinfo {volume} {294}},\ \bibinfo {pages}
  {1495} (\bibinfo {year} {2001})}\BibitemShut {NoStop}%
\bibitem [{\citenamefont {Ross}\ \emph {et~al.}(2011)\citenamefont {Ross},
  \citenamefont {Savary}, \citenamefont {Gaulin},\ and\ \citenamefont
  {Balents}}]{Ross2011PRL}%
  \BibitemOpen
  \bibfield  {author} {\bibinfo {author} {\bibfnamefont {K.~A.}\ \bibnamefont
  {Ross}}, \bibinfo {author} {\bibfnamefont {L.}~\bibnamefont {Savary}},
  \bibinfo {author} {\bibfnamefont {B.~D.}\ \bibnamefont {Gaulin}},\ and\
  \bibinfo {author} {\bibfnamefont {L.}~\bibnamefont {Balents}},\ }\bibfield
  {title} {\bibinfo {title} {Quantum excitations in quantum spin ice},\ }\href
  {https://doi.org/10.1103/PhysRevX.1.021002} {\bibfield  {journal} {\bibinfo
  {journal} {Phys. Rev. X}\ }\textbf {\bibinfo {volume} {1}},\ \bibinfo {pages}
  {021002} (\bibinfo {year} {2011})}\BibitemShut {NoStop}%
\bibitem [{\citenamefont {Lee}\ \emph {et~al.}(2012)\citenamefont {Lee},
  \citenamefont {Onoda},\ and\ \citenamefont {Balents}}]{Lee2012PRB}%
  \BibitemOpen
  \bibfield  {author} {\bibinfo {author} {\bibfnamefont {S.}~\bibnamefont
  {Lee}}, \bibinfo {author} {\bibfnamefont {S.}~\bibnamefont {Onoda}},\ and\
  \bibinfo {author} {\bibfnamefont {L.}~\bibnamefont {Balents}},\ }\bibfield
  {title} {\bibinfo {title} {Generic quantum spin ice},\ }\href
  {https://doi.org/10.1103/PhysRevB.86.104412} {\bibfield  {journal} {\bibinfo
  {journal} {Phys. Rev. B}\ }\textbf {\bibinfo {volume} {86}},\ \bibinfo
  {pages} {104412} (\bibinfo {year} {2012})}\BibitemShut {NoStop}%
\bibitem [{\citenamefont {Wang}\ \emph {et~al.}(2006)\citenamefont {Wang},
  \citenamefont {Nisoli}, \citenamefont {Freitas}, \citenamefont {Li},
  \citenamefont {McConville}, \citenamefont {Cooley}, \citenamefont {Lund},
  \citenamefont {Samarth}, \citenamefont {Leighton}, \citenamefont {Crespi},\
  and\ \citenamefont {Schiffer}}]{Wang2006Nature}%
  \BibitemOpen
  \bibfield  {author} {\bibinfo {author} {\bibfnamefont {R.~F.}\ \bibnamefont
  {Wang}}, \bibinfo {author} {\bibfnamefont {C.}~\bibnamefont {Nisoli}},
  \bibinfo {author} {\bibfnamefont {R.~S.~d.}\ \bibnamefont {Freitas}},
  \bibinfo {author} {\bibfnamefont {J.}~\bibnamefont {Li}}, \bibinfo {author}
  {\bibfnamefont {W.}~\bibnamefont {McConville}}, \bibinfo {author}
  {\bibfnamefont {B.~J.}\ \bibnamefont {Cooley}}, \bibinfo {author}
  {\bibfnamefont {M.~S.}\ \bibnamefont {Lund}}, \bibinfo {author}
  {\bibfnamefont {N.}~\bibnamefont {Samarth}}, \bibinfo {author} {\bibfnamefont
  {C.}~\bibnamefont {Leighton}}, \bibinfo {author} {\bibfnamefont {V.~H.}\
  \bibnamefont {Crespi}},\ and\ \bibinfo {author} {\bibfnamefont
  {P.}~\bibnamefont {Schiffer}},\ }\bibfield  {title} {\bibinfo {title}
  {Artificial ‘spin ice’ in a geometrically frustrated lattice of nanoscale
  ferromagnetic islands},\ }\href {https://doi.org/10.1038/nature04447}
  {\bibfield  {journal} {\bibinfo  {journal} {Nature (London)}\ }\textbf
  {\bibinfo {volume} {439}},\ \bibinfo {pages} {303} (\bibinfo {year}
  {2006})}\BibitemShut {NoStop}%
\bibitem [{\citenamefont {Balents}(2010)}]{Balents2010nature}%
  \BibitemOpen
  \bibfield  {author} {\bibinfo {author} {\bibfnamefont {L.}~\bibnamefont
  {Balents}},\ }\bibfield  {title} {\bibinfo {title} {Spin liquids in
  frustrated magnets},\ }\href {https://doi.org/10.1063/1.2186278} {\bibfield
  {journal} {\bibinfo  {journal} {Nature (London)}\ }\textbf {\bibinfo {volume}
  {464}},\ \bibinfo {pages} {199} (\bibinfo {year} {2010})}\BibitemShut
  {NoStop}%
\bibitem [{\citenamefont {Savary}\ and\ \citenamefont
  {Balents}(2016)}]{Savary2016RPP}%
  \BibitemOpen
  \bibfield  {author} {\bibinfo {author} {\bibfnamefont {L.}~\bibnamefont
  {Savary}}\ and\ \bibinfo {author} {\bibfnamefont {L.}~\bibnamefont
  {Balents}},\ }\bibfield  {title} {\bibinfo {title} {Quantum spin liquids: a
  review},\ }\href {https://doi.org/10.1088/0034-4885/80/1/016502} {\bibfield
  {journal} {\bibinfo  {journal} {Rep. Prog. Phys.}\ }\textbf {\bibinfo
  {volume} {80}},\ \bibinfo {pages} {016502} (\bibinfo {year}
  {2016})}\BibitemShut {NoStop}%
\bibitem [{\citenamefont {Broholm}\ \emph {et~al.}(2020)\citenamefont
  {Broholm}, \citenamefont {Cava}, \citenamefont {Kivelson}, \citenamefont
  {Nocera}, \citenamefont {Norman},\ and\ \citenamefont
  {Senthil}}]{Broholm2020science}%
  \BibitemOpen
  \bibfield  {author} {\bibinfo {author} {\bibfnamefont {C.}~\bibnamefont
  {Broholm}}, \bibinfo {author} {\bibfnamefont {R.~J.}\ \bibnamefont {Cava}},
  \bibinfo {author} {\bibfnamefont {S.~A.}\ \bibnamefont {Kivelson}}, \bibinfo
  {author} {\bibfnamefont {D.~G.}\ \bibnamefont {Nocera}}, \bibinfo {author}
  {\bibfnamefont {M.~R.}\ \bibnamefont {Norman}},\ and\ \bibinfo {author}
  {\bibfnamefont {T.}~\bibnamefont {Senthil}},\ }\bibfield  {title} {\bibinfo
  {title} {Quantum spin liquids},\ }\href
  {https://doi.org/10.1126/science.aay0668} {\bibfield  {journal} {\bibinfo
  {journal} {Science}\ }\textbf {\bibinfo {volume} {367}},\ \bibinfo {pages}
  {0668} (\bibinfo {year} {2020})}\BibitemShut {NoStop}%
\bibitem [{\citenamefont {Stein}\ and\ \citenamefont
  {Newman}(2013)}]{Stein2013book}%
  \BibitemOpen
  \bibfield  {author} {\bibinfo {author} {\bibfnamefont {D.~L.}\ \bibnamefont
  {Stein}}\ and\ \bibinfo {author} {\bibfnamefont {C.~M.}\ \bibnamefont
  {Newman}},\ }\href {http://www.jstor.org/stable/j.ctt12f4hf} {\emph {\bibinfo
  {title} {Spin Glasses and Complexity}}}\ (\bibinfo  {publisher} {Princeton
  University Press},\ \bibinfo {year} {2013})\BibitemShut {NoStop}%
\bibitem [{\citenamefont {Sherrington}\ and\ \citenamefont
  {Kirkpatrick}(1975)}]{Sherrington1975PRL}%
  \BibitemOpen
  \bibfield  {author} {\bibinfo {author} {\bibfnamefont {D.}~\bibnamefont
  {Sherrington}}\ and\ \bibinfo {author} {\bibfnamefont {S.}~\bibnamefont
  {Kirkpatrick}},\ }\bibfield  {title} {\bibinfo {title} {Solvable model of a
  spin-glass},\ }\href {https://doi.org/10.1103/PhysRevLett.35.1792} {\bibfield
   {journal} {\bibinfo  {journal} {Phys. Rev. Lett.}\ }\textbf {\bibinfo
  {volume} {35}},\ \bibinfo {pages} {1792} (\bibinfo {year}
  {1975})}\BibitemShut {NoStop}%
\bibitem [{\citenamefont {Kroeze}\ \emph {et~al.}(2023)\citenamefont {Kroeze},
  \citenamefont {Marsh}, \citenamefont {Atri~Schuller}, \citenamefont {Hunt},
  \citenamefont {Bourzutschky}, \citenamefont {Winer}, \citenamefont
  {Gopalakrishnan}, \citenamefont {Keeling},\ and\ \citenamefont
  {Lev}}]{Benjamin2023arXiv}%
  \BibitemOpen
  \bibfield  {author} {\bibinfo {author} {\bibfnamefont {R.~M.}\ \bibnamefont
  {Kroeze}}, \bibinfo {author} {\bibfnamefont {B.~P.}\ \bibnamefont {Marsh}},
  \bibinfo {author} {\bibfnamefont {D.}~\bibnamefont {Atri~Schuller}}, \bibinfo
  {author} {\bibfnamefont {H.~S.}\ \bibnamefont {Hunt}}, \bibinfo {author}
  {\bibfnamefont {A.~N.}\ \bibnamefont {Bourzutschky}}, \bibinfo {author}
  {\bibfnamefont {M.}~\bibnamefont {Winer}}, \bibinfo {author} {\bibfnamefont
  {S.}~\bibnamefont {Gopalakrishnan}}, \bibinfo {author} {\bibfnamefont
  {J.}~\bibnamefont {Keeling}},\ and\ \bibinfo {author} {\bibfnamefont {B.~L.}\
  \bibnamefont {Lev}},\ }\bibfield  {title} {\bibinfo {title} {Directly
  observing replica symmetry breaking in a vector quantum-optical spin glass},\
  }\href {https://doi.org/10.48550/arXiv.2311.04216} {\bibfield  {journal}
  {\bibinfo  {journal} {arXiv.2311.04216}\ } (\bibinfo {year}
  {2023})}\BibitemShut {NoStop}%
\bibitem [{\citenamefont {Marsh}\ \emph {et~al.}(2024)\citenamefont {Marsh},
  \citenamefont {Kroeze}, \citenamefont {Ganguli}, \citenamefont
  {Gopalakrishnan}, \citenamefont {Keeling},\ and\ \citenamefont
  {Lev}}]{Benjamin2024PRX}%
  \BibitemOpen
  \bibfield  {author} {\bibinfo {author} {\bibfnamefont {B.~P.}\ \bibnamefont
  {Marsh}}, \bibinfo {author} {\bibfnamefont {R.~M.}\ \bibnamefont {Kroeze}},
  \bibinfo {author} {\bibfnamefont {S.}~\bibnamefont {Ganguli}}, \bibinfo
  {author} {\bibfnamefont {S.}~\bibnamefont {Gopalakrishnan}}, \bibinfo
  {author} {\bibfnamefont {J.}~\bibnamefont {Keeling}},\ and\ \bibinfo {author}
  {\bibfnamefont {B.~L.}\ \bibnamefont {Lev}},\ }\bibfield  {title} {\bibinfo
  {title} {Entanglement and replica symmetry breaking in a driven-dissipative
  quantum spin glass},\ }\href {https://doi.org/10.1103/PhysRevX.14.011026}
  {\bibfield  {journal} {\bibinfo  {journal} {Phys. Rev. X}\ }\textbf {\bibinfo
  {volume} {14}},\ \bibinfo {pages} {011026} (\bibinfo {year}
  {2024})}\BibitemShut {NoStop}%
\bibitem [{\citenamefont {Chiocchetta}\ \emph {et~al.}(2021)\citenamefont
  {Chiocchetta}, \citenamefont {Kiese}, \citenamefont {Zelle}, \citenamefont
  {Piazza},\ and\ \citenamefont {Diehl}}]{Chiocchetta2021NatComm}%
  \BibitemOpen
  \bibfield  {author} {\bibinfo {author} {\bibfnamefont {A.}~\bibnamefont
  {Chiocchetta}}, \bibinfo {author} {\bibfnamefont {D.}~\bibnamefont {Kiese}},
  \bibinfo {author} {\bibfnamefont {C.~P.}\ \bibnamefont {Zelle}}, \bibinfo
  {author} {\bibfnamefont {F.}~\bibnamefont {Piazza}},\ and\ \bibinfo {author}
  {\bibfnamefont {S.}~\bibnamefont {Diehl}},\ }\bibfield  {title} {\bibinfo
  {title} {Cavity-induced quantum spin liquids},\ }\href
  {https://doi.org/10.1038/s41467-021-26076-3} {\bibfield  {journal} {\bibinfo
  {journal} {Nat. Commun.}\ }\textbf {\bibinfo {volume} {12}},\ \bibinfo
  {pages} {5901} (\bibinfo {year} {2021})}\BibitemShut {NoStop}%
\bibitem [{\citenamefont {Kim}\ \emph {et~al.}(2010)\citenamefont {Kim},
  \citenamefont {Chang}, \citenamefont {Korenblit}, \citenamefont {Islam},
  \citenamefont {Edwards}, \citenamefont {Freericks}, \citenamefont {Lin},
  \citenamefont {Duan},\ and\ \citenamefont {Monroe}}]{Kim2010Nature}%
  \BibitemOpen
  \bibfield  {author} {\bibinfo {author} {\bibfnamefont {K.}~\bibnamefont
  {Kim}}, \bibinfo {author} {\bibfnamefont {M.-S.}\ \bibnamefont {Chang}},
  \bibinfo {author} {\bibfnamefont {S.}~\bibnamefont {Korenblit}}, \bibinfo
  {author} {\bibfnamefont {R.}~\bibnamefont {Islam}}, \bibinfo {author}
  {\bibfnamefont {E.~E.}\ \bibnamefont {Edwards}}, \bibinfo {author}
  {\bibfnamefont {J.~K.}\ \bibnamefont {Freericks}}, \bibinfo {author}
  {\bibfnamefont {G.-D.}\ \bibnamefont {Lin}}, \bibinfo {author} {\bibfnamefont
  {L.-M.}\ \bibnamefont {Duan}},\ and\ \bibinfo {author} {\bibfnamefont
  {C.}~\bibnamefont {Monroe}},\ }\bibfield  {title} {\bibinfo {title} {Quantum
  simulation of frustrated {I}sing spins with trapped ions},\ }\href
  {https://doi.org/10.1038/nature09071} {\bibfield  {journal} {\bibinfo
  {journal} {Nature (London)}\ }\textbf {\bibinfo {volume} {465}},\ \bibinfo
  {pages} {590} (\bibinfo {year} {2010})}\BibitemShut {NoStop}%
\bibitem [{\citenamefont {Gopalakrishnan}\ \emph {et~al.}(2009)\citenamefont
  {Gopalakrishnan}, \citenamefont {Lev},\ and\ \citenamefont
  {Goldbart}}]{Benjamin2009NatPhy}%
  \BibitemOpen
  \bibfield  {author} {\bibinfo {author} {\bibfnamefont {S.}~\bibnamefont
  {Gopalakrishnan}}, \bibinfo {author} {\bibfnamefont {B.~L.}\ \bibnamefont
  {Lev}},\ and\ \bibinfo {author} {\bibfnamefont {P.~M.}\ \bibnamefont
  {Goldbart}},\ }\bibfield  {title} {\bibinfo {title} {Emergent crystallinity
  and frustration with {B}ose--{E}instein condensates in multimode cavities},\
  }\href {https://doi.org/10.1038/nphys1403} {\bibfield  {journal} {\bibinfo
  {journal} {Nat. Phys.}\ }\textbf {\bibinfo {volume} {5}},\ \bibinfo {pages}
  {845} (\bibinfo {year} {2009})}\BibitemShut {NoStop}%
\bibitem [{\citenamefont {Marsh}\ \emph {et~al.}(2021)\citenamefont {Marsh},
  \citenamefont {Guo}, \citenamefont {Kroeze}, \citenamefont {Gopalakrishnan},
  \citenamefont {Ganguli}, \citenamefont {Keeling},\ and\ \citenamefont
  {Lev}}]{Benjamin2021PRX}%
  \BibitemOpen
  \bibfield  {author} {\bibinfo {author} {\bibfnamefont {B.~P.}\ \bibnamefont
  {Marsh}}, \bibinfo {author} {\bibfnamefont {Y.}~\bibnamefont {Guo}}, \bibinfo
  {author} {\bibfnamefont {R.~M.}\ \bibnamefont {Kroeze}}, \bibinfo {author}
  {\bibfnamefont {S.}~\bibnamefont {Gopalakrishnan}}, \bibinfo {author}
  {\bibfnamefont {S.}~\bibnamefont {Ganguli}}, \bibinfo {author} {\bibfnamefont
  {J.}~\bibnamefont {Keeling}},\ and\ \bibinfo {author} {\bibfnamefont {B.~L.}\
  \bibnamefont {Lev}},\ }\bibfield  {title} {\bibinfo {title} {Enhancing
  associative memory recall and storage capacity using confocal cavity {QED}},\
  }\href {https://doi.org/10.1103/PhysRevX.11.021048} {\bibfield  {journal}
  {\bibinfo  {journal} {Phys. Rev. X}\ }\textbf {\bibinfo {volume} {11}},\
  \bibinfo {pages} {021048} (\bibinfo {year} {2021})}\BibitemShut {NoStop}%
\bibitem [{\citenamefont {Zhao}\ and\ \citenamefont
  {Hwang}(2022)}]{Zhao2022PRL}%
  \BibitemOpen
  \bibfield  {author} {\bibinfo {author} {\bibfnamefont {J.}~\bibnamefont
  {Zhao}}\ and\ \bibinfo {author} {\bibfnamefont {M.-J.}\ \bibnamefont
  {Hwang}},\ }\bibfield  {title} {\bibinfo {title} {Frustrated superradiant
  phase transition},\ }\href {https://doi.org/10.1103/PhysRevLett.128.163601}
  {\bibfield  {journal} {\bibinfo  {journal} {Phys. Rev. Lett.}\ }\textbf
  {\bibinfo {volume} {128}},\ \bibinfo {pages} {163601} (\bibinfo {year}
  {2022})}\BibitemShut {NoStop}%
\bibitem [{\citenamefont {Zhao}\ and\ \citenamefont
  {Hwang}(2023)}]{Zhao2023PRL}%
  \BibitemOpen
  \bibfield  {author} {\bibinfo {author} {\bibfnamefont {J.}~\bibnamefont
  {Zhao}}\ and\ \bibinfo {author} {\bibfnamefont {M.-J.}\ \bibnamefont
  {Hwang}},\ }\bibfield  {title} {\bibinfo {title} {Anomalous criticality with
  bounded fluctuations and long-range frustration induced by broken
  time-reversal symmetry},\ }\href
  {https://doi.org/10.1103/PhysRevResearch.5.L042016} {\bibfield  {journal}
  {\bibinfo  {journal} {Phys. Rev. Res.}\ }\textbf {\bibinfo {volume} {5}},\
  \bibinfo {pages} {L042016} (\bibinfo {year} {2023})}\BibitemShut {NoStop}%
\bibitem [{\citenamefont {Fallas~Padilla}\ \emph {et~al.}(2022)\citenamefont
  {Fallas~Padilla}, \citenamefont {Pu}, \citenamefont {Cheng},\ and\
  \citenamefont {Zhang}}]{Zhang2022PRL}%
  \BibitemOpen
  \bibfield  {author} {\bibinfo {author} {\bibfnamefont {D.}~\bibnamefont
  {Fallas~Padilla}}, \bibinfo {author} {\bibfnamefont {H.}~\bibnamefont {Pu}},
  \bibinfo {author} {\bibfnamefont {G.-J.}\ \bibnamefont {Cheng}},\ and\
  \bibinfo {author} {\bibfnamefont {Y.-Y.}\ \bibnamefont {Zhang}},\ }\bibfield
  {title} {\bibinfo {title} {Understanding the quantum {R}abi ring using
  analogies to quantum magnetism},\ }\href
  {https://doi.org/10.1103/PhysRevLett.129.183602} {\bibfield  {journal}
  {\bibinfo  {journal} {Phys. Rev. Lett.}\ }\textbf {\bibinfo {volume} {129}},\
  \bibinfo {pages} {183602} (\bibinfo {year} {2022})}\BibitemShut {NoStop}%
\bibitem [{\citenamefont {Zhang}\ \emph {et~al.}(2024)\citenamefont {Zhang},
  \citenamefont {Liang}, \citenamefont {Lambert},\ and\ \citenamefont
  {Cirio}}]{Zhang2024PRR}%
  \BibitemOpen
  \bibfield  {author} {\bibinfo {author} {\bibfnamefont {C.}~\bibnamefont
  {Zhang}}, \bibinfo {author} {\bibfnamefont {P.}~\bibnamefont {Liang}},
  \bibinfo {author} {\bibfnamefont {N.}~\bibnamefont {Lambert}},\ and\ \bibinfo
  {author} {\bibfnamefont {M.}~\bibnamefont {Cirio}},\ }\bibfield  {title}
  {\bibinfo {title} {Closed and open unbalanced {D}icke trimer model: Critical
  properties and nonlinear semiclassical dynamics},\ }\href
  {https://doi.org/10.1103/PhysRevResearch.6.023012} {\bibfield  {journal}
  {\bibinfo  {journal} {Phys. Rev. Res.}\ }\textbf {\bibinfo {volume} {6}},\
  \bibinfo {pages} {023012} (\bibinfo {year} {2024})}\BibitemShut {NoStop}%
\bibitem [{\citenamefont {Luo}\ \emph {et~al.}(2025)\citenamefont {Luo},
  \citenamefont {Wang},\ and\ \citenamefont {Xiang}}]{Luo2025}%
  \BibitemOpen
  \bibfield  {author} {\bibinfo {author} {\bibfnamefont {J.-W.}\ \bibnamefont
  {Luo}}, \bibinfo {author} {\bibfnamefont {B.}~\bibnamefont {Wang}},\ and\
  \bibinfo {author} {\bibfnamefont {Z.-L.}\ \bibnamefont {Xiang}},\ }\bibfield
  {title} {\bibinfo {title} {Quantum phase transitions in a {D}icke trimer with
  both photon and atom hoppings},\ }\href {https://arxiv.org/abs/2502.10839}
  {\bibfield  {journal} {\bibinfo  {journal} {arXiv:2502.10839}\ } (\bibinfo
  {year} {2025})}\BibitemShut {NoStop}%
\bibitem [{\citenamefont {Kongkhambut}\ \emph {et~al.}(2022)\citenamefont
  {Kongkhambut}, \citenamefont {Skulte}, \citenamefont {Mathey}, \citenamefont
  {Cosme}, \citenamefont {Hemmerich},\ and\ \citenamefont
  {Keßler}}]{Kongkhambut2022Science}%
  \BibitemOpen
  \bibfield  {author} {\bibinfo {author} {\bibfnamefont {P.}~\bibnamefont
  {Kongkhambut}}, \bibinfo {author} {\bibfnamefont {J.}~\bibnamefont {Skulte}},
  \bibinfo {author} {\bibfnamefont {L.}~\bibnamefont {Mathey}}, \bibinfo
  {author} {\bibfnamefont {J.~G.}\ \bibnamefont {Cosme}}, \bibinfo {author}
  {\bibfnamefont {A.}~\bibnamefont {Hemmerich}},\ and\ \bibinfo {author}
  {\bibfnamefont {H.}~\bibnamefont {Keßler}},\ }\bibfield  {title} {\bibinfo
  {title} {Observation of a continuous time crystal},\ }\href
  {https://doi.org/10.1126/science.abo3382} {\bibfield  {journal} {\bibinfo
  {journal} {Science}\ }\textbf {\bibinfo {volume} {377}},\ \bibinfo {pages}
  {670} (\bibinfo {year} {2022})}\BibitemShut {NoStop}%
\bibitem [{\citenamefont {Iemini}\ \emph {et~al.}(2018)\citenamefont {Iemini},
  \citenamefont {Russomanno}, \citenamefont {Keeling}, \citenamefont
  {Schir\`o}, \citenamefont {Dalmonte},\ and\ \citenamefont
  {Fazio}}]{Iemini2018PRL}%
  \BibitemOpen
  \bibfield  {author} {\bibinfo {author} {\bibfnamefont {F.}~\bibnamefont
  {Iemini}}, \bibinfo {author} {\bibfnamefont {A.}~\bibnamefont {Russomanno}},
  \bibinfo {author} {\bibfnamefont {J.}~\bibnamefont {Keeling}}, \bibinfo
  {author} {\bibfnamefont {M.}~\bibnamefont {Schir\`o}}, \bibinfo {author}
  {\bibfnamefont {M.}~\bibnamefont {Dalmonte}},\ and\ \bibinfo {author}
  {\bibfnamefont {R.}~\bibnamefont {Fazio}},\ }\bibfield  {title} {\bibinfo
  {title} {Boundary time crystals},\ }\href
  {https://doi.org/10.1103/PhysRevLett.121.035301} {\bibfield  {journal}
  {\bibinfo  {journal} {Phys. Rev. Lett.}\ }\textbf {\bibinfo {volume} {121}},\
  \bibinfo {pages} {035301} (\bibinfo {year} {2018})}\BibitemShut {NoStop}%
\bibitem [{\citenamefont {Sacha}\ and\ \citenamefont
  {Zakrzewski}(2017)}]{Sacha2018Reports}%
  \BibitemOpen
  \bibfield  {author} {\bibinfo {author} {\bibfnamefont {K.}~\bibnamefont
  {Sacha}}\ and\ \bibinfo {author} {\bibfnamefont {J.}~\bibnamefont
  {Zakrzewski}},\ }\bibfield  {title} {\bibinfo {title} {Time crystals: a
  review},\ }\href {https://doi.org/10.1088/1361-6633/aa8b38} {\bibfield
  {journal} {\bibinfo  {journal} {Rep. Prog. Phys.}\ }\textbf {\bibinfo
  {volume} {81}},\ \bibinfo {pages} {016401} (\bibinfo {year}
  {2017})}\BibitemShut {NoStop}%
\bibitem [{\citenamefont {Dogra}\ \emph {et~al.}(2019)\citenamefont {Dogra},
  \citenamefont {Landini}, \citenamefont {Kroeger}, \citenamefont {Hruby},
  \citenamefont {Donner},\ and\ \citenamefont {Esslinger}}]{Donner2019Science}%
  \BibitemOpen
  \bibfield  {author} {\bibinfo {author} {\bibfnamefont {N.}~\bibnamefont
  {Dogra}}, \bibinfo {author} {\bibfnamefont {M.}~\bibnamefont {Landini}},
  \bibinfo {author} {\bibfnamefont {K.}~\bibnamefont {Kroeger}}, \bibinfo
  {author} {\bibfnamefont {L.}~\bibnamefont {Hruby}}, \bibinfo {author}
  {\bibfnamefont {T.}~\bibnamefont {Donner}},\ and\ \bibinfo {author}
  {\bibfnamefont {T.}~\bibnamefont {Esslinger}},\ }\bibfield  {title} {\bibinfo
  {title} {Dissipation-induced structural instability and chiral dynamics in a
  quantum gas},\ }\href
  {https://www.science.org/doi/abs/10.1126/science.aaw4465} {\bibfield
  {journal} {\bibinfo  {journal} {Science}\ }\textbf {\bibinfo {volume}
  {366}},\ \bibinfo {pages} {1496} (\bibinfo {year} {2019})}\BibitemShut
  {NoStop}%
\bibitem [{\citenamefont {Landini}\ \emph {et~al.}(2018)\citenamefont
  {Landini}, \citenamefont {Dogra}, \citenamefont {Kroeger}, \citenamefont
  {Hruby}, \citenamefont {Donner},\ and\ \citenamefont
  {Esslinger}}]{Donner2018PRL}%
  \BibitemOpen
  \bibfield  {author} {\bibinfo {author} {\bibfnamefont {M.}~\bibnamefont
  {Landini}}, \bibinfo {author} {\bibfnamefont {N.}~\bibnamefont {Dogra}},
  \bibinfo {author} {\bibfnamefont {K.}~\bibnamefont {Kroeger}}, \bibinfo
  {author} {\bibfnamefont {L.}~\bibnamefont {Hruby}}, \bibinfo {author}
  {\bibfnamefont {T.}~\bibnamefont {Donner}},\ and\ \bibinfo {author}
  {\bibfnamefont {T.}~\bibnamefont {Esslinger}},\ }\bibfield  {title} {\bibinfo
  {title} {Formation of a spin texture in a quantum gas coupled to a cavity},\
  }\href {https://doi.org/10.1103/PhysRevLett.120.223602} {\bibfield  {journal}
  {\bibinfo  {journal} {Phys. Rev. Lett.}\ }\textbf {\bibinfo {volume} {120}},\
  \bibinfo {pages} {223602} (\bibinfo {year} {2018})}\BibitemShut {NoStop}%
\bibitem [{SM_()}]{SM_link}%
  \BibitemOpen
  \href@noop {} {\bibinfo  {journal} {See Supplemental Material for details.}\
  }\BibitemShut {NoStop}%
\bibitem [{\citenamefont {Torre}\ \emph {et~al.}(2013)\citenamefont {Torre},
  \citenamefont {Diehl}, \citenamefont {Lukin}, \citenamefont {Sachdev},\ and\
  \citenamefont {Strack}}]{TorrePRA2013}%
  \BibitemOpen
\bibfield  {journal} {  }\bibfield  {author} {\bibinfo {author} {\bibfnamefont
  {E.~G.~D.}\ \bibnamefont {Torre}}, \bibinfo {author} {\bibfnamefont
  {S.}~\bibnamefont {Diehl}}, \bibinfo {author} {\bibfnamefont {M.~D.}\
  \bibnamefont {Lukin}}, \bibinfo {author} {\bibfnamefont {S.}~\bibnamefont
  {Sachdev}},\ and\ \bibinfo {author} {\bibfnamefont {P.}~\bibnamefont
  {Strack}},\ }\bibfield  {title} {\bibinfo {title} {Keldysh approach for
  nonequilibrium phase transitions in quantum optics: Beyond the {D}icke model
  in optical cavities},\ }\href {https://doi.org/10.1103/PhysRevA.87.023831}
  {\bibfield  {journal} {\bibinfo  {journal} {Phys. Rev. A}\ }\textbf {\bibinfo
  {volume} {87}},\ \bibinfo {pages} {023831} (\bibinfo {year}
  {2013})}\BibitemShut {NoStop}%
\bibitem [{\citenamefont {Hwang}\ \emph {et~al.}(2018)\citenamefont {Hwang},
  \citenamefont {Rabl},\ and\ \citenamefont {Plenio}}]{Hwang2018PRA}%
  \BibitemOpen
  \bibfield  {author} {\bibinfo {author} {\bibfnamefont {M.-J.}\ \bibnamefont
  {Hwang}}, \bibinfo {author} {\bibfnamefont {P.}~\bibnamefont {Rabl}},\ and\
  \bibinfo {author} {\bibfnamefont {M.~B.}\ \bibnamefont {Plenio}},\ }\bibfield
   {title} {\bibinfo {title} {Dissipative phase transition in the open quantum
  {R}abi model},\ }\href {https://doi.org/10.1103/PhysRevA.97.013825}
  {\bibfield  {journal} {\bibinfo  {journal} {Phys. Rev. A}\ }\textbf {\bibinfo
  {volume} {97}},\ \bibinfo {pages} {013825} (\bibinfo {year}
  {2018})}\BibitemShut {NoStop}%
\end{thebibliography}%


\begin{thebibliography}{9}%
\makeatletter
\providecommand \@ifxundefined [1]{%
 \@ifx{#1\undefined}
}%
\providecommand \@ifnum [1]{%
 \ifnum #1\expandafter \@firstoftwo
 \else \expandafter \@secondoftwo
 \fi
}%
\providecommand \@ifx [1]{%
 \ifx #1\expandafter \@firstoftwo
 \else \expandafter \@secondoftwo
 \fi
}%
\providecommand \natexlab [1]{#1}%
\providecommand \enquote  [1]{``#1''}%
\providecommand \bibnamefont  [1]{#1}%
\providecommand \bibfnamefont [1]{#1}%
\providecommand \citenamefont [1]{#1}%
\providecommand \href@noop [0]{\@secondoftwo}%
\providecommand \href [0]{\begingroup \@sanitize@url \@href}%
\providecommand \@href[1]{\@@startlink{#1}\@@href}%
\providecommand \@@href[1]{\endgroup#1\@@endlink}%
\providecommand \@sanitize@url [0]{\catcode `\\12\catcode `\$12\catcode
  `\&12\catcode `\#12\catcode `\^12\catcode `\_12\catcode `\%12\relax}%
\providecommand \@@startlink[1]{}%
\providecommand \@@endlink[0]{}%
\providecommand \url  [0]{\begingroup\@sanitize@url \@url }%
\providecommand \@url [1]{\endgroup\@href {#1}{\urlprefix }}%
\providecommand \urlprefix  [0]{URL }%
\providecommand \Eprint [0]{\href }%
\providecommand \doibase [0]{https://doi.org/}%
\providecommand \selectlanguage [0]{\@gobble}%
\providecommand \bibinfo  [0]{\@secondoftwo}%
\providecommand \bibfield  [0]{\@secondoftwo}%
\providecommand \translation [1]{[#1]}%
\providecommand \BibitemOpen [0]{}%
\providecommand \bibitemStop [0]{}%
\providecommand \bibitemNoStop [0]{.\EOS\space}%
\providecommand \EOS [0]{\spacefactor3000\relax}%
\providecommand \BibitemShut  [1]{\csname bibitem#1\endcsname}%
\let\auto@bib@innerbib\@empty
\bibitem [{\citenamefont {Chiacchio}\ and\ \citenamefont
  {Nunnenkamp}(2019)}]{Nunnenkamp2019PRL}%
  \BibitemOpen
  \bibfield  {author} {\bibinfo {author} {\bibfnamefont {E.~I.~R.}\
  \bibnamefont {Chiacchio}}\ and\ \bibinfo {author} {\bibfnamefont
  {A.}~\bibnamefont {Nunnenkamp}},\ }\bibfield  {title} {\bibinfo {title}
  {Dissipation-induced instabilities of a spinor {B}ose-{E}instein condensate
  inside an optical cavity},\ }\href
  {https://doi.org/10.1103/PhysRevLett.122.193605} {\bibfield  {journal}
  {\bibinfo  {journal} {Phys. Rev. Lett.}\ }\textbf {\bibinfo {volume} {122}},\
  \bibinfo {pages} {193605} (\bibinfo {year} {2019})}\BibitemShut {NoStop}%
\bibitem [{\citenamefont {Fruchart}\ \emph {et~al.}(2021)\citenamefont
  {Fruchart}, \citenamefont {Hanai}, \citenamefont {Littlewood},\ and\
  \citenamefont {Vitelli}}]{Fruchart2021Nature}%
  \BibitemOpen
  \bibfield  {author} {\bibinfo {author} {\bibfnamefont {M.}~\bibnamefont
  {Fruchart}}, \bibinfo {author} {\bibfnamefont {R.}~\bibnamefont {Hanai}},
  \bibinfo {author} {\bibfnamefont {P.~B.}\ \bibnamefont {Littlewood}},\ and\
  \bibinfo {author} {\bibfnamefont {V.}~\bibnamefont {Vitelli}},\ }\bibfield
  {title} {\bibinfo {title} {Non-reciprocal phase transitions},\ }\href
  {https://doi.org/10.1038/s41586-021-03375-9} {\bibfield  {journal} {\bibinfo
  {journal} {Nature (London)}\ }\textbf {\bibinfo {volume} {592}},\ \bibinfo
  {pages} {363} (\bibinfo {year} {2021})}\BibitemShut {NoStop}%
\bibitem [{\citenamefont {Chiacchio}\ \emph {et~al.}(2023)\citenamefont
  {Chiacchio}, \citenamefont {Nunnenkamp},\ and\ \citenamefont
  {Brunelli}}]{Nunnenkamp2023PRL}%
  \BibitemOpen
  \bibfield  {author} {\bibinfo {author} {\bibfnamefont {E.~I.~R.}\
  \bibnamefont {Chiacchio}}, \bibinfo {author} {\bibfnamefont {A.}~\bibnamefont
  {Nunnenkamp}},\ and\ \bibinfo {author} {\bibfnamefont {M.}~\bibnamefont
  {Brunelli}},\ }\bibfield  {title} {\bibinfo {title} {Nonreciprocal {D}icke
  model},\ }\href {https://doi.org/10.1103/PhysRevLett.131.113602} {\bibfield
  {journal} {\bibinfo  {journal} {Phys. Rev. Lett.}\ }\textbf {\bibinfo
  {volume} {131}},\ \bibinfo {pages} {113602} (\bibinfo {year}
  {2023})}\BibitemShut {NoStop}%
\bibitem [{\citenamefont {Stitely}\ \emph {et~al.}(2020)\citenamefont
  {Stitely}, \citenamefont {Masson}, \citenamefont {Giraldo}, \citenamefont
  {Krauskopf},\ and\ \citenamefont {Parkins}}]{Stitely2020PRA}%
  \BibitemOpen
  \bibfield  {author} {\bibinfo {author} {\bibfnamefont {K.~C.}\ \bibnamefont
  {Stitely}}, \bibinfo {author} {\bibfnamefont {S.~J.}\ \bibnamefont {Masson}},
  \bibinfo {author} {\bibfnamefont {A.}~\bibnamefont {Giraldo}}, \bibinfo
  {author} {\bibfnamefont {B.}~\bibnamefont {Krauskopf}},\ and\ \bibinfo
  {author} {\bibfnamefont {S.}~\bibnamefont {Parkins}},\ }\bibfield  {title}
  {\bibinfo {title} {{Superradiant switching, quantum hysteresis, and
  oscillations in a generalized {D}icke model}},\ }\href
  {https://doi.org/10.1103/PhysRevA.102.063702} {\bibfield  {journal} {\bibinfo
   {journal} {Phys. Rev. A}\ }\textbf {\bibinfo {volume} {102}},\ \bibinfo
  {pages} {063702} (\bibinfo {year} {2020})}\BibitemShut {NoStop}%
\bibitem [{\citenamefont {Ferri}\ \emph {et~al.}(2021)\citenamefont {Ferri},
  \citenamefont {Rosa-Medina}, \citenamefont {Finger}, \citenamefont {Dogra},
  \citenamefont {Soriente}, \citenamefont {Zilberberg}, \citenamefont
  {Donner},\ and\ \citenamefont {Esslinger}}]{ETH2021PRX}%
  \BibitemOpen
  \bibfield  {author} {\bibinfo {author} {\bibfnamefont {F.}~\bibnamefont
  {Ferri}}, \bibinfo {author} {\bibfnamefont {R.}~\bibnamefont {Rosa-Medina}},
  \bibinfo {author} {\bibfnamefont {F.}~\bibnamefont {Finger}}, \bibinfo
  {author} {\bibfnamefont {N.}~\bibnamefont {Dogra}}, \bibinfo {author}
  {\bibfnamefont {M.}~\bibnamefont {Soriente}}, \bibinfo {author}
  {\bibfnamefont {O.}~\bibnamefont {Zilberberg}}, \bibinfo {author}
  {\bibfnamefont {T.}~\bibnamefont {Donner}},\ and\ \bibinfo {author}
  {\bibfnamefont {T.}~\bibnamefont {Esslinger}},\ }\bibfield  {title} {\bibinfo
  {title} {Emerging dissipative phases in a superradiant quantum gas with
  tunable decay},\ }\href {https://doi.org/10.1103/PhysRevX.11.041046}
  {\bibfield  {journal} {\bibinfo  {journal} {Phys. Rev. X}\ }\textbf {\bibinfo
  {volume} {11}},\ \bibinfo {pages} {041046} (\bibinfo {year}
  {2021})}\BibitemShut {NoStop}%
\bibitem [{\citenamefont {Lyu}\ \emph {et~al.}(2024)\citenamefont {Lyu},
  \citenamefont {Kottmann}, \citenamefont {Plenio},\ and\ \citenamefont
  {Hwang}}]{Lyu2024PRR}%
  \BibitemOpen
  \bibfield  {author} {\bibinfo {author} {\bibfnamefont {G.}~\bibnamefont
  {Lyu}}, \bibinfo {author} {\bibfnamefont {K.}~\bibnamefont {Kottmann}},
  \bibinfo {author} {\bibfnamefont {M.~B.}\ \bibnamefont {Plenio}},\ and\
  \bibinfo {author} {\bibfnamefont {M.-J.}\ \bibnamefont {Hwang}},\ }\bibfield
  {title} {\bibinfo {title} {Multicritical dissipative phase transitions in the
  anisotropic open quantum {R}abi model},\ }\href
  {https://doi.org/10.1103/PhysRevResearch.6.033075} {\bibfield  {journal}
  {\bibinfo  {journal} {Phys. Rev. Res.}\ }\textbf {\bibinfo {volume} {6}},\
  \bibinfo {pages} {033075} (\bibinfo {year} {2024})}\BibitemShut {NoStop}%
\bibitem [{\citenamefont {Rodriguez}\ \emph {et~al.}(2017)\citenamefont
  {Rodriguez}, \citenamefont {Casteels}, \citenamefont {Storme}, \citenamefont
  {Carlon~Zambon}, \citenamefont {Sagnes}, \citenamefont {Le~Gratiet},
  \citenamefont {Galopin}, \citenamefont {Lema\^{\i}tre}, \citenamefont {Amo},
  \citenamefont {Ciuti},\ and\ \citenamefont {Bloch}}]{Rodriguez2017PRL}%
  \BibitemOpen
  \bibfield  {author} {\bibinfo {author} {\bibfnamefont {S.~R.~K.}\
  \bibnamefont {Rodriguez}}, \bibinfo {author} {\bibfnamefont {W.}~\bibnamefont
  {Casteels}}, \bibinfo {author} {\bibfnamefont {F.}~\bibnamefont {Storme}},
  \bibinfo {author} {\bibfnamefont {N.}~\bibnamefont {Carlon~Zambon}}, \bibinfo
  {author} {\bibfnamefont {I.}~\bibnamefont {Sagnes}}, \bibinfo {author}
  {\bibfnamefont {L.}~\bibnamefont {Le~Gratiet}}, \bibinfo {author}
  {\bibfnamefont {E.}~\bibnamefont {Galopin}}, \bibinfo {author} {\bibfnamefont
  {A.}~\bibnamefont {Lema\^{\i}tre}}, \bibinfo {author} {\bibfnamefont
  {A.}~\bibnamefont {Amo}}, \bibinfo {author} {\bibfnamefont {C.}~\bibnamefont
  {Ciuti}},\ and\ \bibinfo {author} {\bibfnamefont {J.}~\bibnamefont {Bloch}},\
  }\bibfield  {title} {\bibinfo {title} {Probing a dissipative phase transition
  via dynamical optical hysteresis},\ }\href
  {https://doi.org/10.1103/PhysRevLett.118.247402} {\bibfield  {journal}
  {\bibinfo  {journal} {Phys. Rev. Lett.}\ }\textbf {\bibinfo {volume} {118}},\
  \bibinfo {pages} {247402} (\bibinfo {year} {2017})}\BibitemShut {NoStop}%
\bibitem [{\citenamefont {Jachinowski}\ and\ \citenamefont
  {Littlewood}(2025)}]{Jachinowski2025}%
  \BibitemOpen
  \bibfield  {author} {\bibinfo {author} {\bibfnamefont {J.}~\bibnamefont
  {Jachinowski}}\ and\ \bibinfo {author} {\bibfnamefont {P.~B.}\ \bibnamefont
  {Littlewood}},\ }\bibfield  {title} {\bibinfo {title} {Spin-only dynamics of
  the multi-species nonreciprocal {D}icke model},\ }\href
  {https://arxiv.org/abs/2507.07960} {\bibfield  {journal} {\bibinfo  {journal}
  {arXiv:2507.07960}\ } (\bibinfo {year} {2025})}\BibitemShut {NoStop}%
\bibitem [{\citenamefont {Landini}\ \emph {et~al.}(2018)\citenamefont
  {Landini}, \citenamefont {Dogra}, \citenamefont {Kroeger}, \citenamefont
  {Hruby}, \citenamefont {Donner},\ and\ \citenamefont
  {Esslinger}}]{Donner2018PRL}%
  \BibitemOpen
  \bibfield  {author} {\bibinfo {author} {\bibfnamefont {M.}~\bibnamefont
  {Landini}}, \bibinfo {author} {\bibfnamefont {N.}~\bibnamefont {Dogra}},
  \bibinfo {author} {\bibfnamefont {K.}~\bibnamefont {Kroeger}}, \bibinfo
  {author} {\bibfnamefont {L.}~\bibnamefont {Hruby}}, \bibinfo {author}
  {\bibfnamefont {T.}~\bibnamefont {Donner}},\ and\ \bibinfo {author}
  {\bibfnamefont {T.}~\bibnamefont {Esslinger}},\ }\bibfield  {title} {\bibinfo
  {title} {Formation of a spin texture in a quantum gas coupled to a cavity},\
  }\href {https://doi.org/10.1103/PhysRevLett.120.223602} {\bibfield  {journal}
  {\bibinfo  {journal} {Phys. Rev. Lett.}\ }\textbf {\bibinfo {volume} {120}},\
  \bibinfo {pages} {223602} (\bibinfo {year} {2018})}\BibitemShut {NoStop}%
\end{thebibliography}%

\end{document}


\title{--- Supplemental Material ---\\
		Nonreciprocal and Geometric Frustration in Dissipative Quantum Spins
	}

\author{Guitao Lyu}
\affiliation{Division of Natural and Applied Sciences, Duke Kunshan University, Kunshan, Jiangsu 215300, China}

\author{Myung-Joong Hwang}	
\affiliation{Division of Natural and Applied Sciences, Duke Kunshan University, Kunshan, Jiangsu 215300, China}

	\date{\today}

    \maketitle



\twocolumngrid

\begin{center}

	\newcommand{\beginsupplement}{%
		\setcounter{table}{0}
		\renewcommand{\thetable}{S\arabic{table}}%
		\setcounter{figure}{0}
		\renewcommand{\thefigure}{S\arabic{figure}}%
	}

\end{center}
\newcommand{\beginsupplement}{%
	\setcounter{table}{0}
	\renewcommand{\thetable}{S\arabic{table}}%
	\setcounter{figure}{0}
	\renewcommand{\thefigure}{S\arabic{figure}}%
}

\setcounter{equation}{0}
\setcounter{figure}{0}
\setcounter{table}{0}
\setcounter{page}{1}
\makeatletter
\renewcommand{\theequation}{S\arabic{equation}}
\renewcommand{\thefigure}{S\arabic{figure}}
\renewcommand{\bibnumfmt}[1]{[S#1]}
\renewcommand{\citenumfont}[1]{S#1}
\vspace{0.8 in}



\vspace{-3.3cm}


\subsection{Spectrum of dynamical matrix} \label{SM: unstable phase}
By setting the time derivative of EOMs~(2) and (7) of the main text to be zero, and using the spin conservation $S_{x,m}^2+S_{y,m}^2 +S_{z,m}^2=N^2$, one can obtain stationary solutions. Considering small deviations from the stationary solutions, we obtain the linearized dynamical matrix $\mathbf{M}$, defined by $ \dot{\mathbf{\delta S}}=\mathbf{M} \mathbf{\delta S}$, where $ \mathbf{\delta S}=\left(\delta S_{x,1}, \delta S_{x,0},\delta S_{x,-1}, \delta S_{y,1}, \delta S_{y,0}, \delta S_{y,-1}\right)^{\mathsf{T}} $.
For the normal phase (NP), $S_{x,m}=S_{y,m}=0$ and $S_{z,m}=-N$, the dynamical matrix reads
\begin{align}
	\label{eq: M_NP}
	\mathbf{M_\text{NP}} =
	\begin{pmatrix}
		-\gamma \mathbf{I_3} & -\Omega \mathbf{I_3} \\
		\mathbf{B} +\Omega \mathbf{I_3}& -\gamma \mathbf{I_3}
	\end{pmatrix},
\end{align}
where $ \mathbf{I_3}$ is a $ 3 \times 3 $ identity matrix, and the submatrix
\begin{align}
	\mathbf{B} =
	- \frac{\lambda^2}{\omega_c^2 + \kappa^2}
	\begin{pmatrix}
		D[0] \ & D[\phi] & D[2\phi] \\
		\ D[-\phi] & D[0] \ & D[\phi] \\
		D[-2\phi] & D[-\phi] & D[0] \
	\end{pmatrix},
\end{align}
with $ D[\Phi] \equiv \omega _c\cos\Phi +\kappa  \sin\Phi $.
The diagonal block, $-\gamma \mathbf{I_3}$, and a part of the off-diagonal blocks, $\pm \Omega \mathbf{I_3}$, represent the decay rate and the precession frequency of each individual spin species, respectively. The submatrix $\mathbf{B}$ includes the cavity-mediated shifts of the precession frequency (diagonal part) and the cavity-mediated nonreciprocal coupling (off-diagonal part)~\cite{Nunnenkamp2019PRL}.

The boundary of the NP is determined by the eigenvalues of  $\mathbf{M_\text{NP}}$. In the following analyses of spectrum, we only consider $\gamma=0$ case, as the nonzero spin damping merely shifts the real part of eigenvalues by $-\gamma $. The six eigenvalues are given by
\begin{equation}
	\nu_\text{NP}=\pm i\Omega,
\end{equation}
which is decoupled from the cavity field and therefore independent of $\phi$, and
\begin{equation}
	\mu_\text{NP}=-\gamma \pm i\sqrt{\Omega ^2 - \lambda ^2 \Omega ( 3\omega _c \pm \sqrt{P})/[2 \left(\omega _c^2+\kappa ^2\right)] }
\end{equation}
where $P=\omega _c^2 (2\cos 2 \phi +1)^2-8 \kappa ^2 \sin^2\phi (\cos 2 \phi +2)$, which stems from cavity-mediated nonreciprocal couplings.

The presence of the decoupled spin modes with $\nu_\text{NP}=\pm i\Omega$ can be understood as following. The $x$ components of spins $\mathbf{\delta S_{x}}=\left(\delta S_{x,1}, \delta S_{x,0},\delta S_{x,-1}\right)^{\mathsf{T}}$ is governed by the second-order differential equations,
\begin{equation}
	\ddot{\mathbf{\delta S_{x}}}=-\Omega(\mathbf{B}+\Omega \mathbf{I_3})\mathbf{\delta S_{x}}.
\end{equation}
The same equation holds for the $y$ components of spins. Interestingly, there is a kernel vector of $\mathbf{B}$ that depends only on the phase of the coherent spin-cavity interaction, $\left(1, -2\cos\phi,1\right)^{\mathsf{T}}$. Therefore, there always exists a collective mode of spins that decouples from the cavity and oscillates at the bare spin frequency $\Omega$.

\begin{figure}[tpb!]
	\includegraphics[width=1\linewidth]{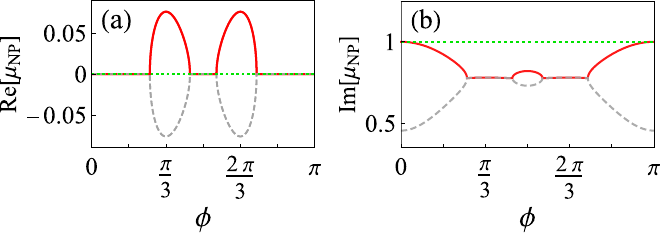}
	\caption{Eigenvalues of the dynamical matrix of  NP. The (a) real and (b) imaginary part of the eigenvalues $\mu_\textrm{NP}$ and $\nu_\textrm{NP}$ for $\lambda=12$ as a function of $\phi$ are presented, with identical parameters given in Fig.~1(a) of the main text. The real and imaginary part of an eigenvalue are shown by the same color and line style. The exceptional point emerges where the imaginary parts coalesce, which marks the nonreciprocal phase transition point.
	}
	\label{fig: DM_NP_eigenvalues}
\end{figure}

The dynamics induced by the nonreciprocal coupling is therefore determined only by $\mu_\text{NP}$, which features the exceptional point with level attractions, as shown in Fig.~\ref{fig: DM_NP_eigenvalues}, indicating the nonreciprocal phase transition~\cite{Fruchart2021Nature,Nunnenkamp2023PRL}. For $P<0$, $\sqrt{P}$ becomes a pure imaginary, giving rise to nonzero real part of $\mu_\text{NP}$, one of which has a positive sign. Therefore, $P=0$ determines the boundary of the DP shown by the red vertical lines in Fig.~1(a) of the main text, which is given by
\begin{equation}
	\phi_c = \arccos\left( \pm \frac{1}{2} \sqrt{1 \pm \frac{3\kappa}{\sqrt{\kappa^2 + \omega_c^2}}} \right).
\end{equation}
From the above expression, we note that the island of NP near $\pi/2$ only exists for $\kappa/\omega_c<\frac{1}{2\sqrt{2}}$, which corresponds to the choice of our parameters. Otherwise, there are only two solutions of $\phi_c$, leading to the single area of DP within the range $0\leq\phi\leq\pi$.

Finally, by setting $\mu_\text{NP}=0 $, we identify the critical points,
\begin{equation}
	\lambda_c =\sqrt{\Omega (-3 \omega_c + \sqrt{P})/(2 \cos 2 \phi +\cos 4 \phi -3)},
\end{equation}
at which a second-order phase transition to the self-ordered phases occurs. This phase boundary between the NP and SOP is shown as the black dashed curve in Fig.~1(a) of the main text.

The dynamical matrix for the SOPs can also be constructed by considering the stationary solutions in each phases. However, since these solutions are obtained numerically, the dynamical matrix and its spectrum are also calculated numerically using these solutions. The results are shown in the main text.

\subsection{First-order phase transitions}  \label{SM: phase transitions}
\begin{figure}[tb!]
	\includegraphics[width=0.95\linewidth]{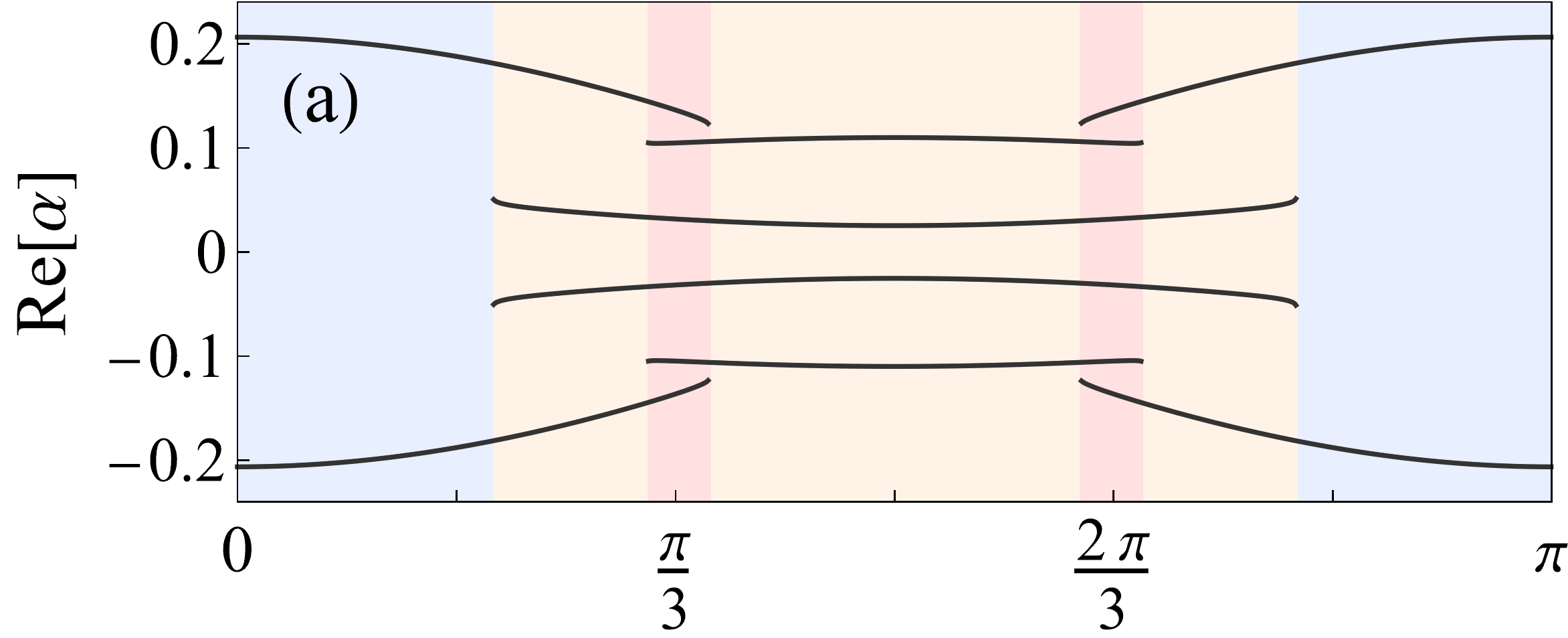}
	\includegraphics[width=0.95\linewidth]{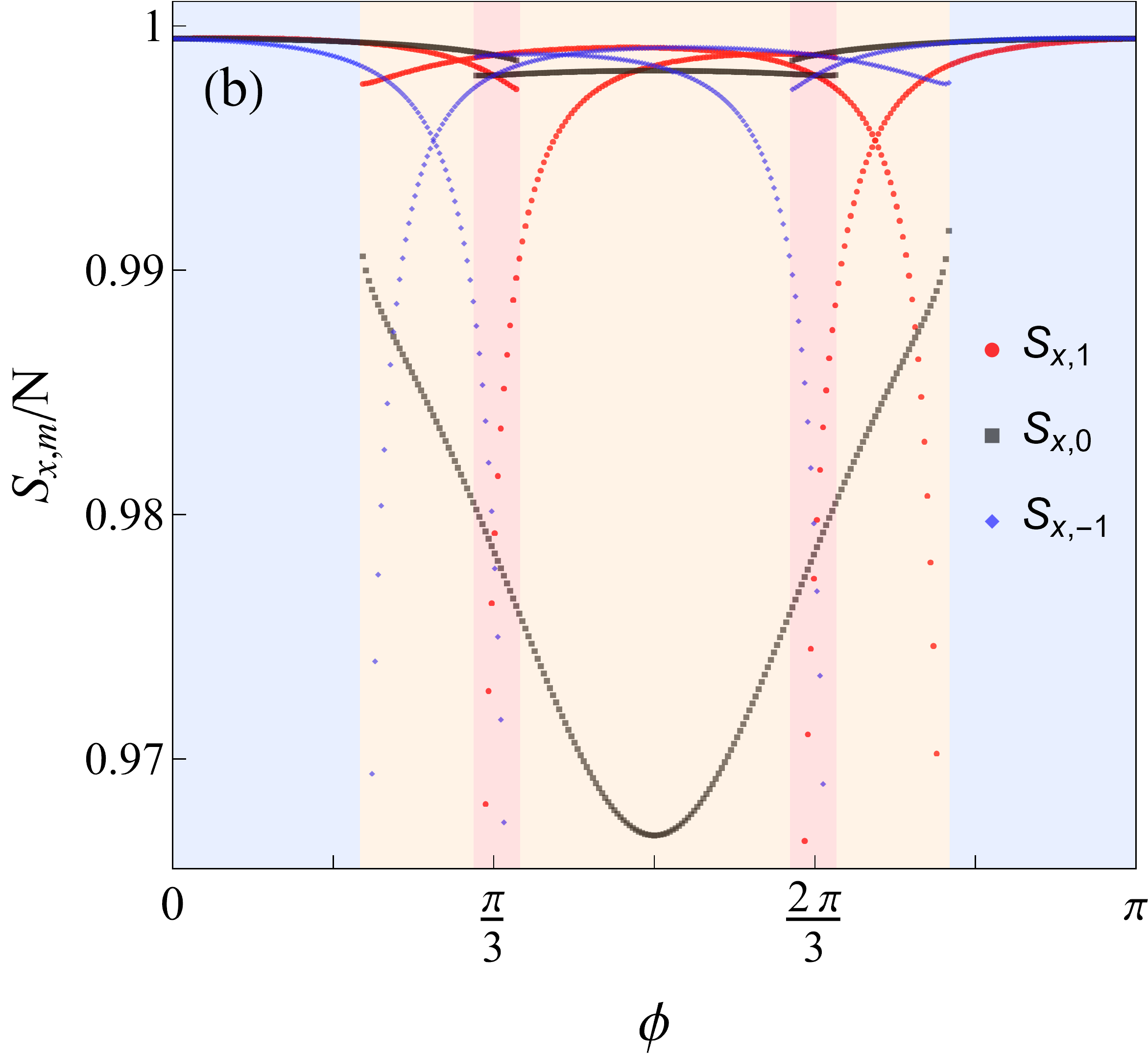}
	\caption{Solutions for the order parameter: (a) the cavity field $\text{Re}[\alpha]$ and (b) the spins $S_{x,m=(1,0,-1)}$, along the line $\lambda=75$ of Fig.~1(a) of the main text. For spins, only solutions with positive values are shown since the negative values are simply mirror reflections with opposite signs. The SOP, pFSOP, and FSOP are colored in blue, orange, and red, respectively.}
	\label{fig: phase transitions}
\end{figure}

We present the the order parameters for the cavity field and the $S_{x,m}$ as a function of $\phi$ for a fixed value of $\lambda\gg\lambda_c$ in Fig.~\ref{fig: phase transitions}. The results clearly show that the number of stable solutions increases or decreases by two at each transition point, and that the emergent order parameter exhibits discontinuous jumps at these points, confirming the first-order nature of transitions between the SOP and pFSOP, as well as between the pFSOP and FSOP. We note that the presence of multistable regions may be experimentally probed via hysteresis effects, as demonstrated in \cite{Stitely2020PRA, ETH2021PRX, Lyu2024PRR,Rodriguez2017PRL}.

\subsection{Ground-state phase diagram ($\kappa=\gamma=0$)} \label{SM: H_eff}
\begin{figure}[htpb!]
	\includegraphics[width=0.95\linewidth]{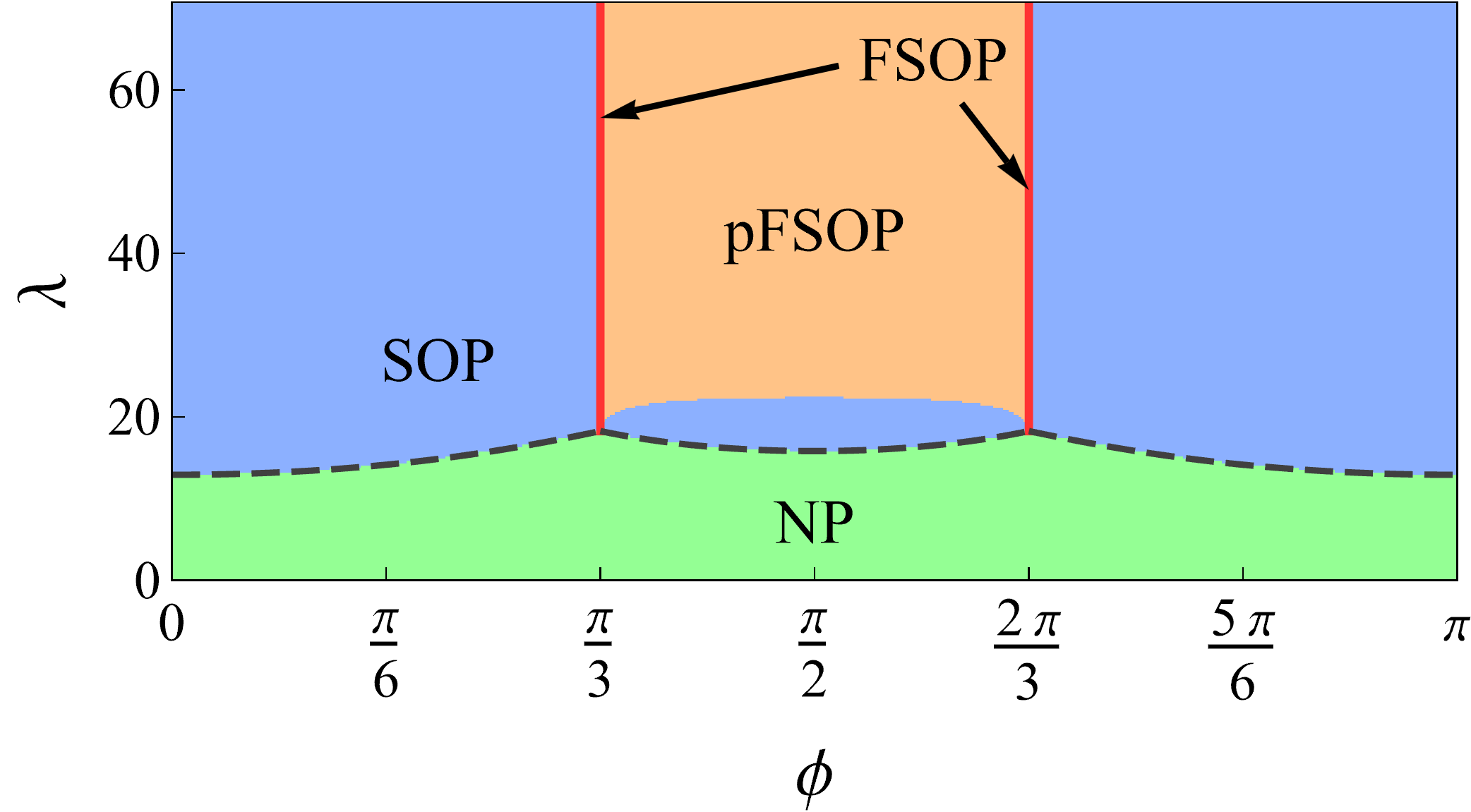}
	\caption{Ground-state phase diagram in the reciprocal limit $\kappa,\gamma=0$. The black dashed curve is the boundary of NP. In contrast to the steady-state phase diagram presented in the main text, the FSOP emerges only at the fine-tuned values of $\phi = \pi/3$ and $2\pi/3$, as indicated by the red lines.}
	\label{fig: ground state phase diagram}
\end{figure}

To obtain the mean-field ground-state phase diagram of the Hamiltonian presented in Eq. (1) of the main text, we derive the spin-only mean-field energy by integrating out the cavity field. This can be done as follows. We take the mean-field approximation where the operators are replaced by expectation values, and solve the extremum condition for the cavity field in terms of spin expectation values. By plugging it back into the mean-field energy, we obtain Eq. (8) in the main text, which we reproduce here:
\begin{align}
	\label{eq: E_eff_A}
	E_\text{eff}=& \sum_{m=-1}^1 -\frac{\Omega }{2} \sqrt{N^2 -S_{x,m}^2}
	-\frac{\lambda ^2}{4 N \omega_c} S_{x,m}^2 +E_\text{int},
\end{align}
with $E_\text{int}= -\frac{\lambda ^2 }{2N \omega_c} (S_{x,-1} S_{x,0} \cos\phi + S_{x,1} S_{x,0}\cos\phi  + S_{x,-1}S_{x,1}\cos2\phi ) $. Here, only $S_{z,m}<0$ is considered, which leads to the lowest mean-field energy. The ground-state phase diagram is obtained by minimizing Eq.~(\ref{eq: E_eff_A}), which is presented in Fig.~\ref{fig: ground state phase diagram}.

Here, we aim to provide qualitative explanation for the emergence of the frustrated phases. To this end, note first that the local spin energy, $-\frac{\Omega }{2} \sqrt{N^2 -S_{x,m}^2}
-\frac{\lambda ^2}{4 N \omega_c} S_{x,m}^2$, has $S_{x,m}=0$ as a single minimum for $\lambda<\sqrt{\Omega\omega_c}$ and two minima
\begin{equation}
	S_{x,m}=\pm N \sqrt{1-\frac{\Omega^2\omega_c^2}{2\lambda^2}}\quad\textrm{for}\quad\lambda>\sqrt{\Omega\omega_c}.
\end{equation}
In the limit of $\lambda\gg\sqrt{\Omega\omega_c}$, the spin can be approximated as fully aligned along the $x$ direction,  $S_{x,m}/N\approx\pm1$. Each local spin should choose the sign for $S_{x,m}$ so that all interaction energies among spins are simultaneously minimized, and frustration occurs when it is not possible.

Let us first consider the case where $|\cos\theta|=|\cos2\theta|$. For $\phi=2\pi/3$, all interactions become antiferromagnetic, which cannot be simultaneously satisfied on a triangular motif. This geometric frustration leads to a fully frustrated self-organized phase (FSOP), characterized by six degenerate solutions
\begin{equation}
	\mathbf{S}_x^\textrm{FSOP}/N\;\approx\;
	\pm(1,\, 1,\,-1),\pm(1,\,-1,\,1),
	\pm(-1,\,1,\,1)
\end{equation}
where $\mathbf{S}_x\equiv\left(S_{x,-1},\, S_{x,0},\, S_{x,1}\right)$. For $\phi=\pi/3$, two pairs of spins favor alignment while the remaining pair favors anti-alignment. This competing interaction pattern also cannot be satisfied on a triangle, resulting in the same FSOP with sixfold degeneracy. Second, we consider the case $|\cos2\theta|>|\cos\theta|$ where the interaction between spins $m=1$ and $-1$ is stronger than the others. In this regime, the system may alleviate frustration by preferentially satisfying the dominant coupling between these two spins. Since $\cos2\phi$ is negative throughout the interval $\pi/3<\phi < 2\pi/3$, the coupling between spins $m=1$ and $-1$ is antiferromagnetic. In this case, neither alignment nor anti-alignment of the spin $m=0$ is compatible with both neighbors, and both configurations result in the same energy. This gives rise to the partially frustrated self-organized phase (pFSOP), with four degenerate solutions:
\begin{equation}
	\mathbf{S}_x^\textrm{pFSOP}/N
	\;\approx\;
	\pm(1,\, 1,\,-1),
	\pm(-1,\,1,\,1).
\end{equation}
Finally, when $|\cos 2\theta| < |\cos \theta|$, the interactions between the pairs of spins $m = 1$ and $0$ as well $m = -1$ and $0$ are  dominant. In this case, the extra degeneracy arising from geometric frustration is fully lifted, as the system minimizes its energy by preferentially satisfying the stronger couplings between $m = \pm1$ and $m = 0$. Therefore, this leads to non-frustrated conventional SOP with two degenerate solutions,
\begin{align}
	\mathbf{S}_x^\textrm{SOP}/N &\;\approx\;\pm(1,\, 1,\,1)\quad\textrm{for}\quad0\leq \phi <\pi/3,\\\nonumber
	&\;\approx\;\pm(1,\, -1,\,1)\quad\textrm{for}\quad 2\pi/3< \phi \leq \pi.
\end{align}
The above analysis agrees very well with the ground-state phase diagram given in Fig.~\ref{fig: ground state phase diagram} for large $\lambda$.

Interestingly, we find an island of the conventional SOP with no frustration in the range $\pi/3<\phi < 2\pi/3$ near the phase boundary to NP. Note that the above analysis, based on the approximation of fully polarized spins $S_{x,m}/N \approx \pm 1$, is valid only in the large-$\lambda$ limit. For $\lambda$ near the phase boundary, the competition between local and interaction energies gives rise to more intricate ground-state solutions with inhomogeneous magnetization, which leads to a double transition from NP to SOP, followed by SOP to pFSOP, a feature that is also found in the steady-state phase diagram.

\subsection{Chaotic phase in between the chiral and swap phases}  \label{SM: other dynamical phases}
\begin{figure}[tpb!]
	\includegraphics[width=0.36\linewidth]{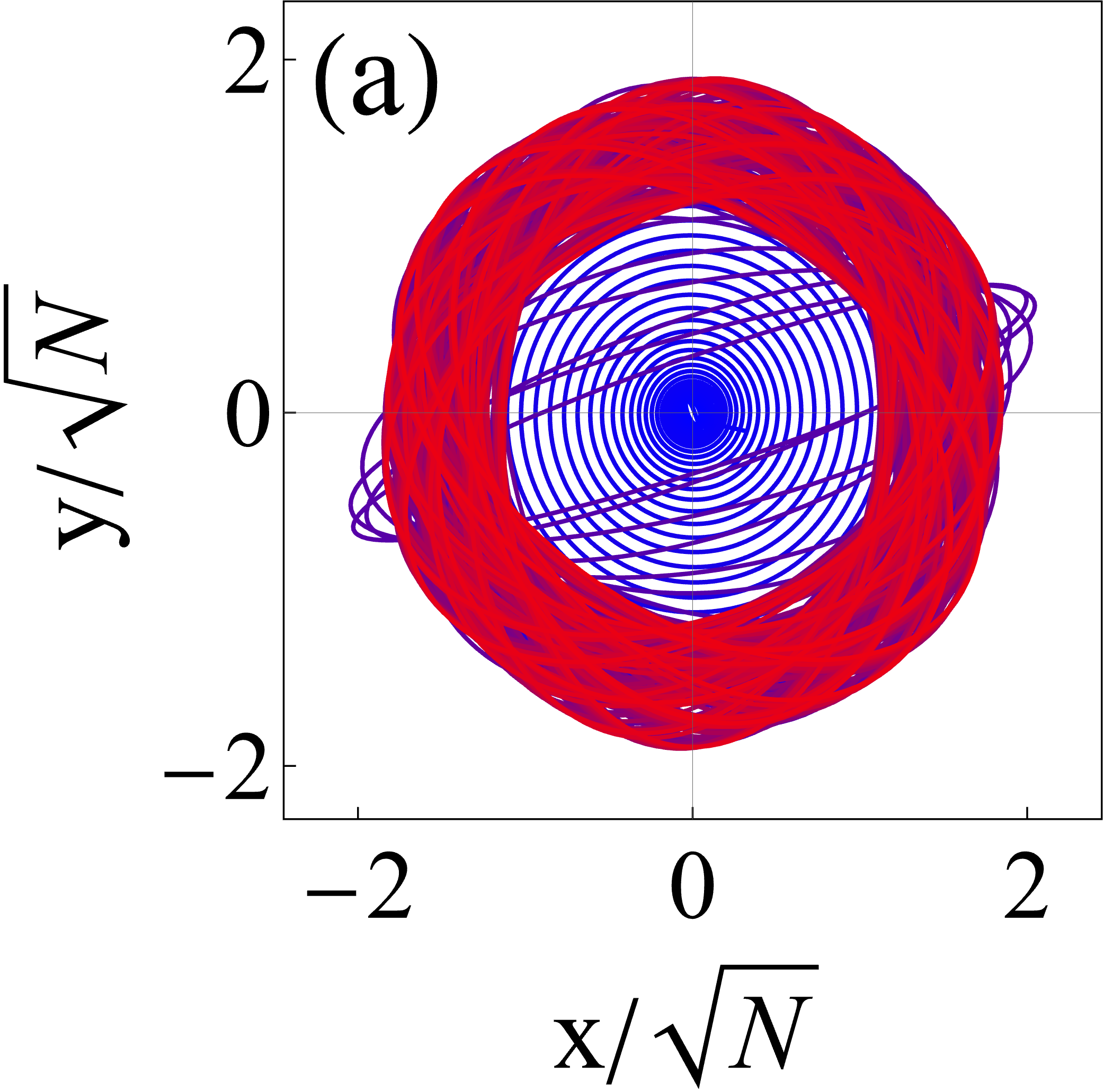}
	\includegraphics[width=0.31\linewidth]{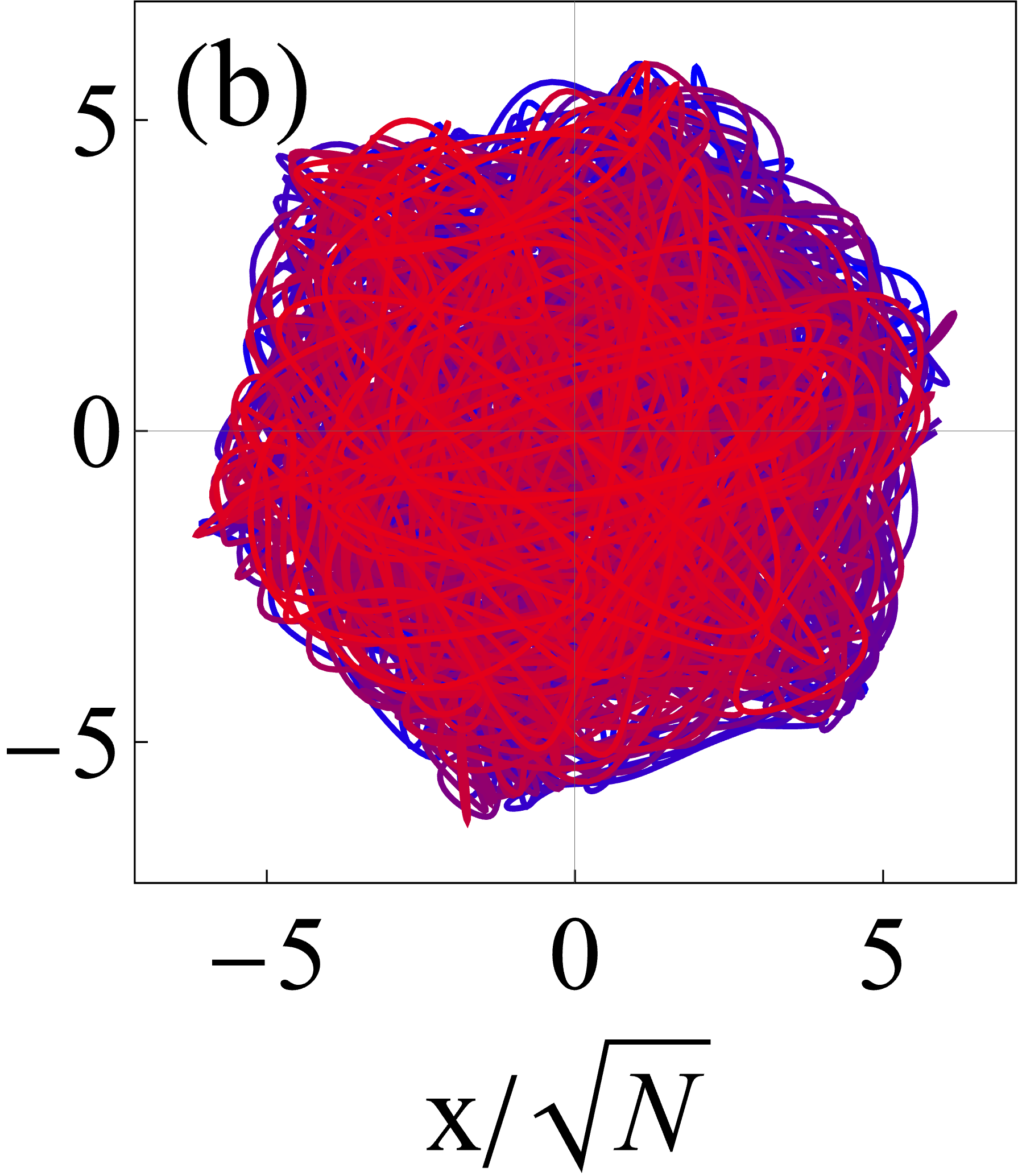}
	\includegraphics[width=0.31\linewidth]{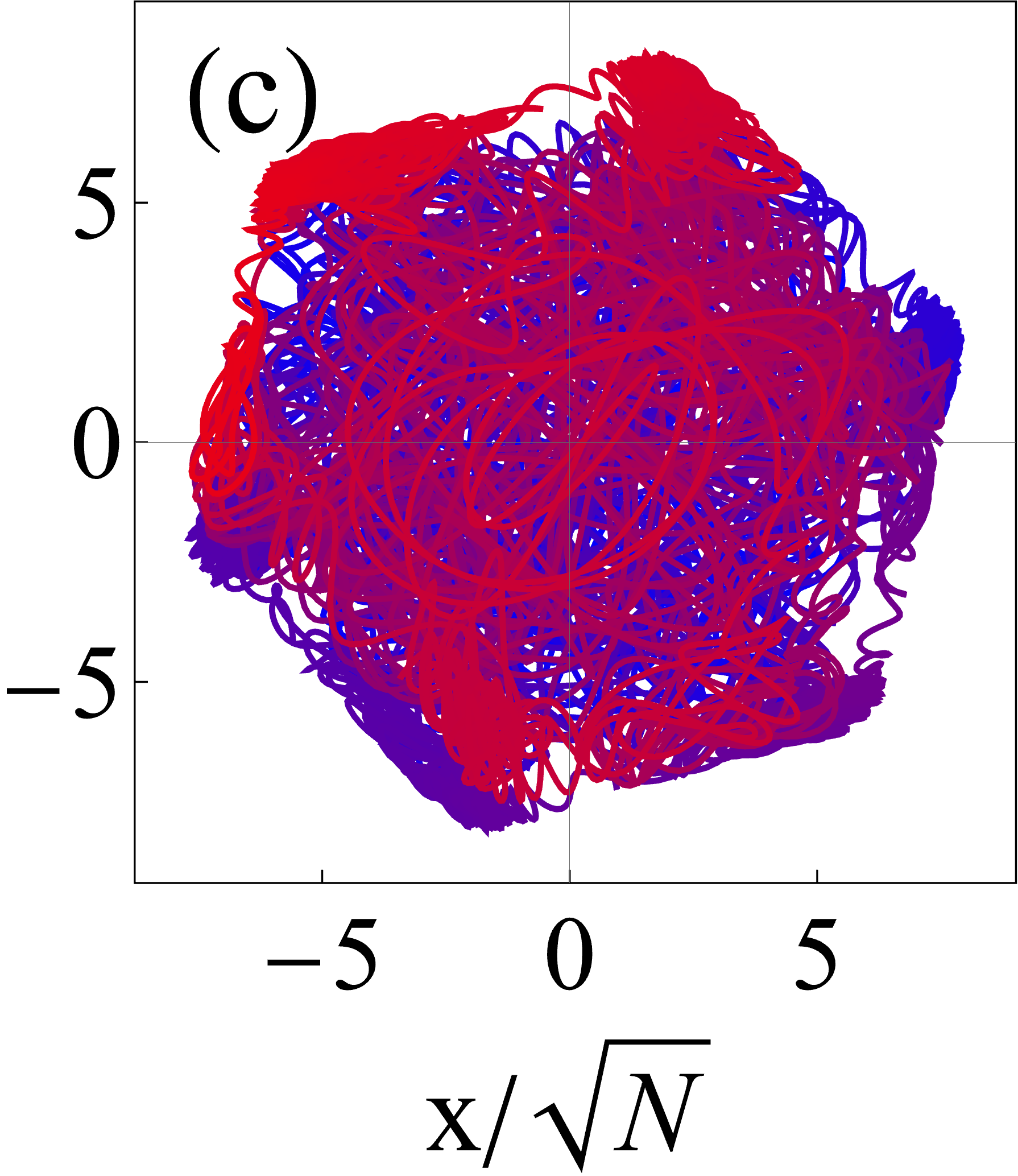}
	\caption{Dynamics of the cavity field ($\alpha\equiv x+iy$) at $\phi=2\pi/3$ for various coupling strength, (a) $\lambda=11.5$ , (b) $\lambda=35$, (c) $\lambda=45$, which demonstrates that the chiral phase gradually transitions into the chaotic phase, followed by the emergence of swap phase. Here, we take the other parameters as in Fig. 1(b) of the main text. The dynamics starts from the NP solution ($S_{x,m}=S_{y,m}=0$, $S_{z,m}=-1$, and $\alpha=0$) with a tiny deviation, and evolves up to time $t=2000$.
	}
	\label{fig: Dynamics}
\end{figure}

In the main text, we have demonstrated that the chiral phase in the weak coupling regime, characterized by a limit cycle on a circle with a single frequency and synchronized relative phases, is a fundamentally different order from the swap phase in the strong coupling regime, characterized by a limit cycle on a hexagon with an emergent frequency and its multiple harmonics. Here, we show that these two distinct phases are separated by a chaotic phase. Upon increasing the coupling strength $\lambda$ from the chiral phase, the circular limit cycle acquires finite width [Fig.~\ref{fig: Dynamics} (a)]. This is followed by a chaotic dynamics [Fig.~\ref{fig: Dynamics} (b)], confined to the hexagon defined by the FSOP. Upon further increasing $\lambda$, the dynamics remain chaotic, but the system increasingly dwells near the vertices of the hexagon, signaling the onset of the swap-phase dynamics.

\begin{figure}[tpb!]
	\includegraphics[width=0.47\linewidth]{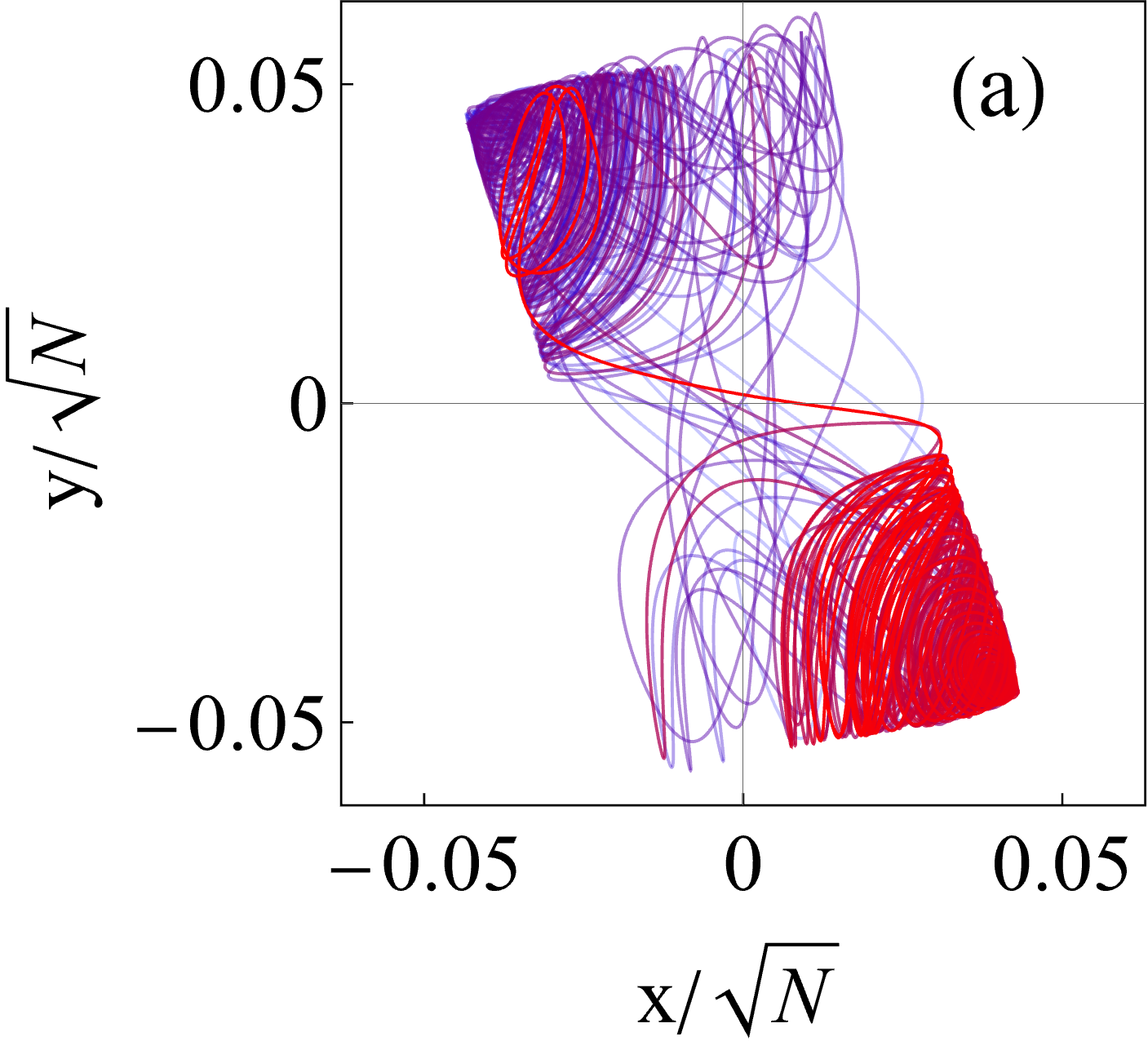}
	\includegraphics[width=0.52\linewidth]{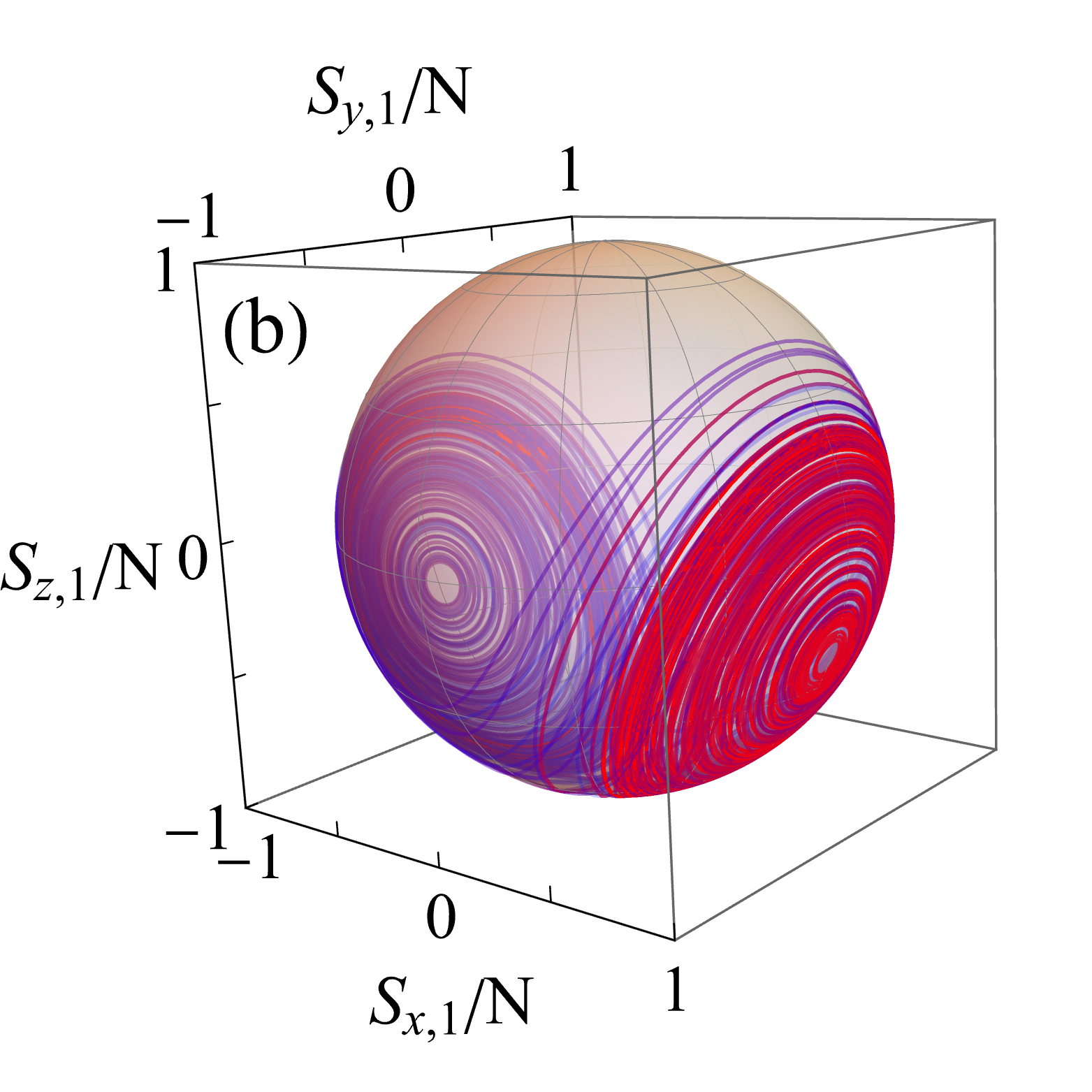}
	\caption{A representative swap phase dynamics in the DP embedded in the SOP. (a) the cavity and (b) individual spin dynamics (spin $m=1$) are presented for $\phi = 1.51$ and $\lambda = 30$, corresponding to the DP within the SOP shown in Fig.~1(b) of the main text. Trajectories are color-coded to indicate time evolution: starting in semi-transparent blue and gradually becoming opaque red from beginning to the end t=1000 (in the unit of $1/\Omega$). The cavity field, equivalently the collective spin coordinates, oscillates between two points in phase space, each corresponding to the stable solution in the neighboring SOP.}
	\label{fig: dynamics midisland} 
\end{figure}

We note that Ref.~\cite{Jachinowski2025} finds the emergence of a circular limit cycle in the two-species nonreciprocal Dicke model. While the authors didn't focus on the relative phases between two spins, but we have numerically confirmed that they are fixed by $\pi/2$ (not shown), which preserves the extended parity symmetry in two-species model at this point. Therefore, we identify that the circular limit cycle in the two-species case also corresponds to the chiral phase, as in the three-species case. Previous studies ~\cite{Nunnenkamp2023PRL,Jachinowski2025} have also reported that chaotic phase emerges upon increasing coupling strength. Our numerics further confirms that the swap phase, similar to the one shown in Fig.~\ref{fig: dynamics midisland}, emerges after the chaotic phase (not shown). Therefore, the chiral-chaos-swap transition appears to be a generic feature of nonreciprocal Dicke model, that would be present in more general multi-species models. A detailed investigation of the mechanisms and nature of these transitions remains an important subject for future work.


\subsection{Dynamical phase embedded in the self-organized phases}  \label{SM: mid-island} We observe a dynamical phase emerge inside the SOP phase, a feature that is absent in the two-species phase diagram. A representative behavior is shown in Fig.~\ref{fig: dynamics midisland}. The three spins precess around one of the two $Z_2$ symmetry-breaking configurations, $(1, -1, -1)$ and $(-1, 1, 1)$, and gradually transition toward the other. These two spin configurations correspond to two stable solutions in SOP near the phase boundary of DP and corresponds to two points in the phase space of the cavity. The resulting dynamics therefore realizes a swap phase, as discussed in the main text, where the metastable points form a line instead of a hexagon due to the absence of frustration.

\subsection{Compensation Scheme for Inhomogeneous Hyperfine Coupling Strengths} \label{SM: Implementation}
In this section, we examine the effect of inhomogeneous coupling strengths between the cavity and spin species and propose a compensation scheme to enable the observation of the predicted nonreciprocal dynamics of the frustrated self-organized states. Let us consider a general setting where, for each spin species $m$, the coupling strength $\lambda_m$ and the number of spins $N_m$ may differ. The spin-only equation of motions become
\begin{align}
	\label{eq: eoms1new}
	\dot S_{x,m}=-\Omega_m  S_{y,m}
\end{align}
and
\begin{align}
	\dot S_{y,m}&=\Omega_m  S_{x,m}+ \frac{\lambda_m  S_{z,m} \tilde F_m }{\sqrt{N_m}(\omega _c^2+\kappa ^2)},
\end{align}
where $\tilde F_m \equiv \sum_{m'=-1}^1 \frac{\lambda_{m'}}{\sqrt{N_{m'}}}D[(m-m')\phi] S_{x,m'}$
with $ D[\Phi] \equiv \omega _c\cos\Phi +\kappa  \sin\Phi $. Compared to $F_m$ defined in the main text, $\tilde F_m$ include inhomogeneous spin parameters. Since the magnitude of spins depends on $N_m$, let us define a normalized spin variables $s_{(x,y,z),m}=S_{(x,y,z),m}/N_m$. We propose to choose $\lambda_m$ and $N_m$ such that
\begin{align}
	\label{eq: lambda_sqrt_N}
	\lambda_{m} \sqrt{N_m}\equiv \Lambda,
\end{align}
which leads to the EOM for a renormalized spin variables
\begin{align}
	\label{eq: eoms3new}
	\dot s_{y,m}&=\Omega_m  s_{x,m}+ \frac{\Lambda^2  s_{z,m} F_m }{N_m(\omega _c^2+\kappa ^2)}.
\end{align}
We emphasize that $F_m$ is identical with the one defined in the main text for the homogeneous case, except that $S_{x,m}$ is replaced by $s_{x,m}$. For steady-states, choosing a renormalized coupling strength $\Lambda=\lambda_{m} \sqrt{N_m}$ to be $m$-independent is equivalent to renormalizing the spin frequencies by $\Omega_mN_m$. Therefore, by tuning each spin frequency so that $\Omega\equiv\Omega_mN_m$, one obtains identical phase diagram with the homogeneous case. Hence, in the presence of inherent inhomogeneous coupling strengths, as it is the case for the hyperfine BEC, it is possible to realize the phase diagrams obtained in the main text for the homogeneous case by choosing the spin population $N_m$ and the spin frequency $\Omega_m$ accordingly.

\begin{figure}[t]
	\includegraphics[width=0.65\linewidth]{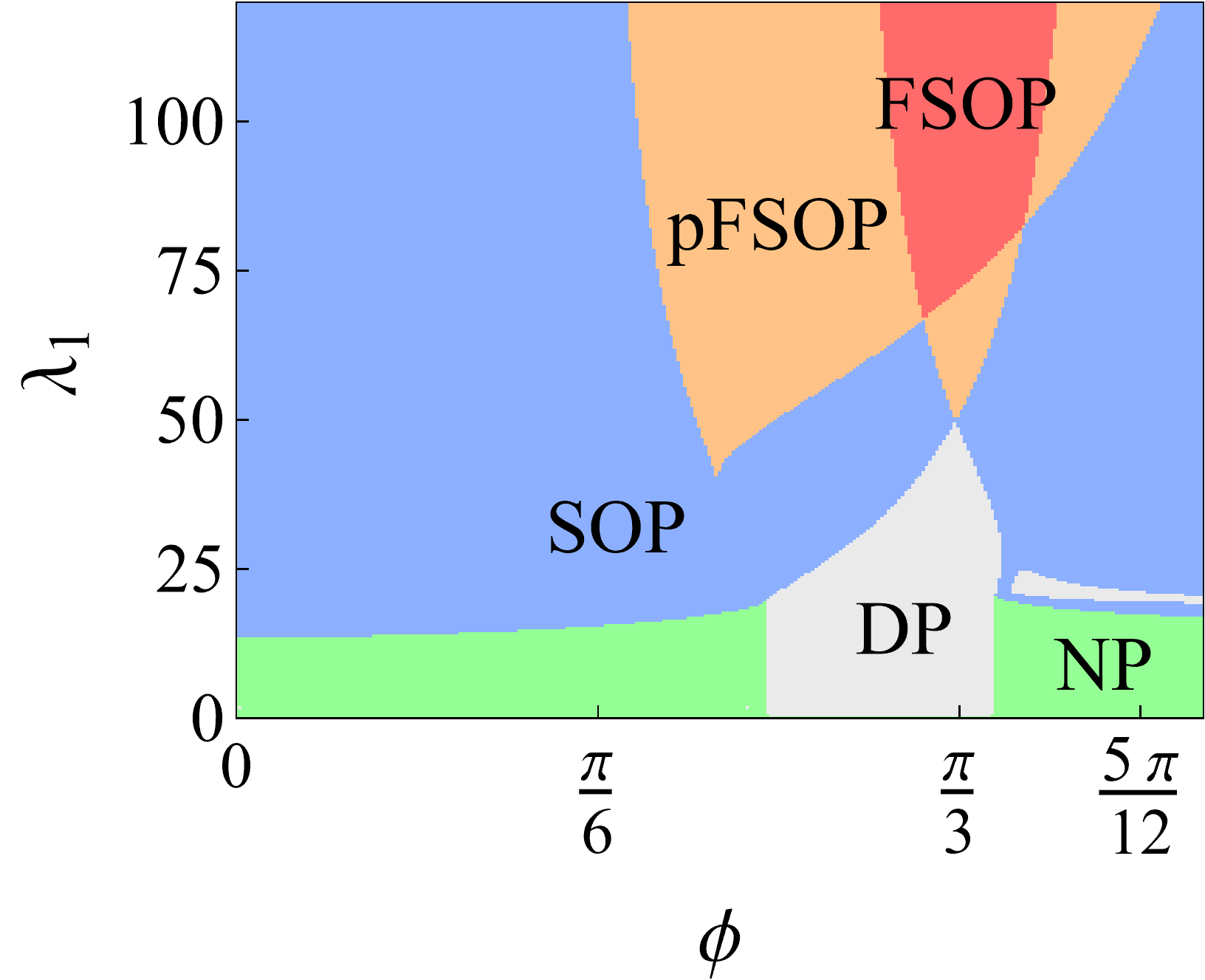}
	\caption{Phase diagram with inhomogeneous coupling. Here we set the phase dependent population imbalance, $ N_0/N_{\pm1}=\sqrt{1+\tan^2\phi}$. The parameters $\omega_c = 500,\ \kappa=150 $, $\Omega=1$, and $\gamma=0$ are chosen as in Fig.1(a) of the main text. }
	\label{fig: asymmetry phase}
\end{figure}

\begin{figure}[t]
	\includegraphics[width=0.49\linewidth]{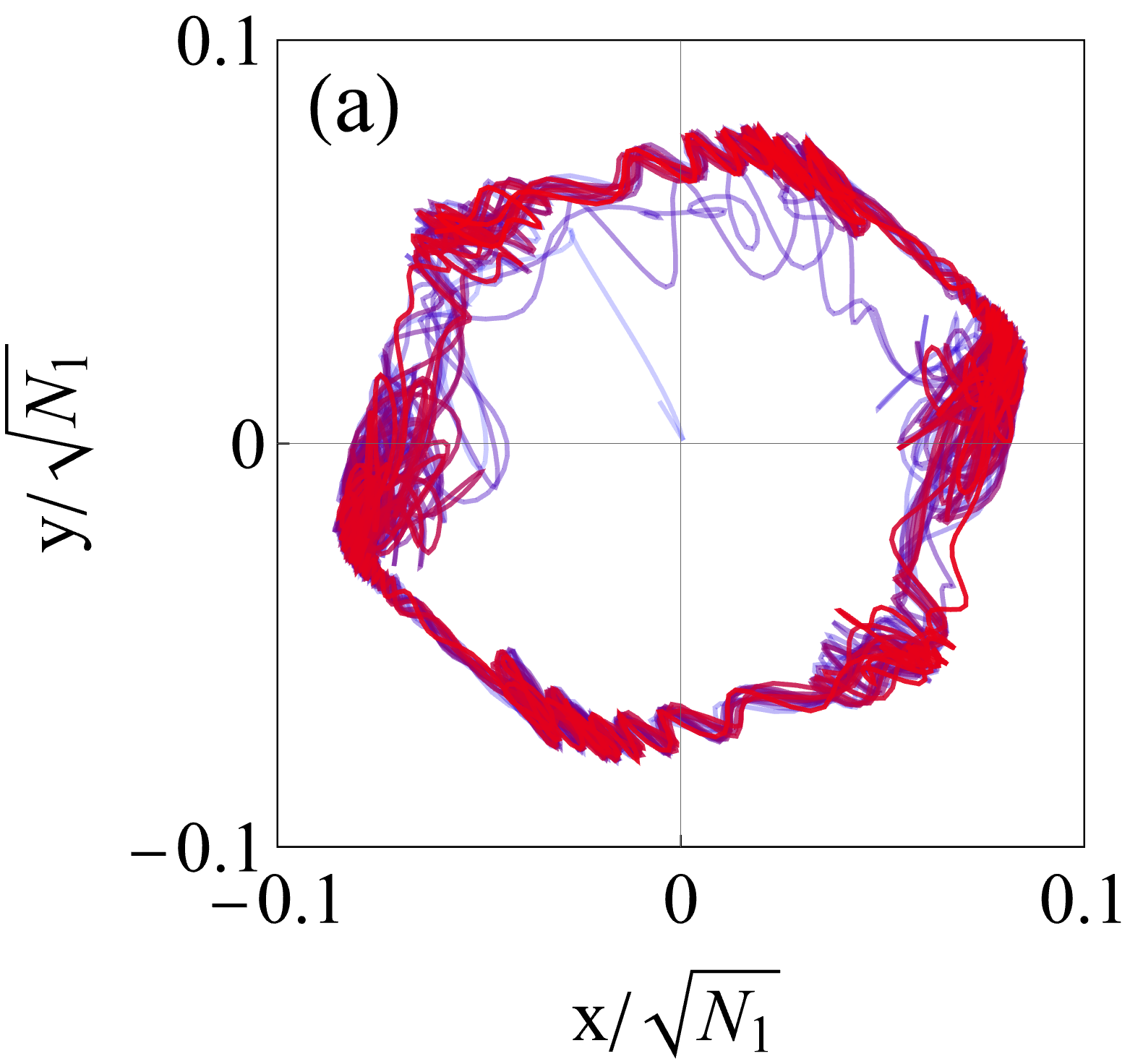}
	\includegraphics[width=0.49\linewidth]{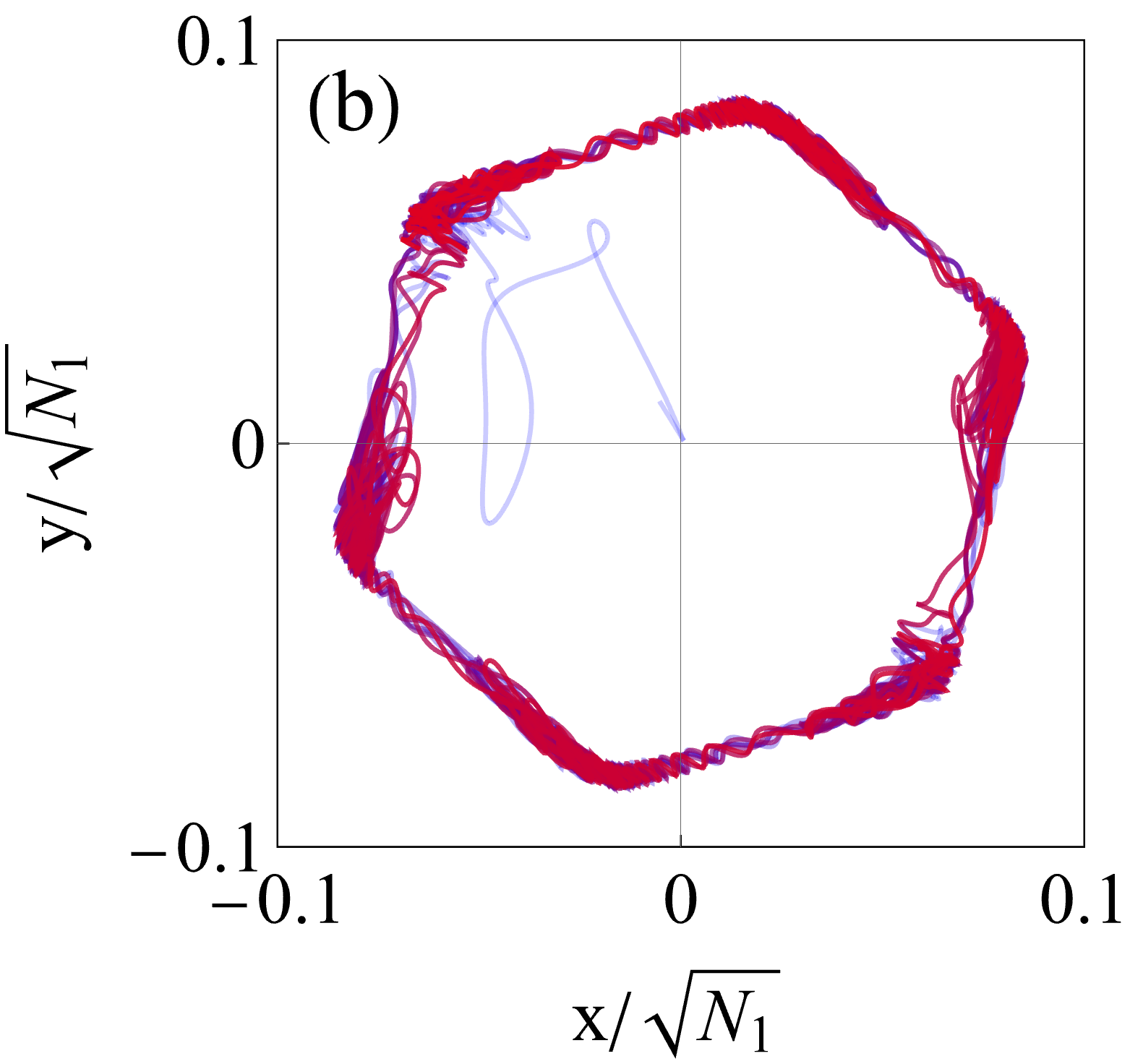}
	\caption{Dynamics of the cavity field ($\alpha\equiv x+iy$) at $\phi=\pi/3$ with population imbalance  $N_0=2N_{\pm1}$ and (a) identical spin frequency $\Omega_0=\Omega_{\pm}=1$ and (b) with spin frequency modulation $\Omega_0=\Omega_{\pm}/2=1/2$.  Here we take $\omega_c = 500, \kappa=150 $, $\Omega=1$, and $\gamma=0.3$. The dynamics starts from the NP solution with a tiny deviation. Trajectories are color-coded to indicate time evolution: starting in semi-transparent blue and gradually becoming opaque red from beginning to the end $t=1400$. }
	\label{fig: asymmetry hexagon}
\end{figure}

Ref.~\cite{Donner2018PRL} presents an experiment that realizes the spin-dependent interaction between the cavity and the multicomponent spinor BEC. In this setup, the coupling strength between the cavity and the hyperfine spins are given by
\begin{equation}
	\lambda_{m_F}= \sqrt{\lambda_s^2 \cos^2 \varphi + \lambda_v^2 m_F^2 \sin^2 \varphi},
\end{equation}
where $\varphi$ is a polarization angle of the optical pump, which is related to the spin-dependent phase as $\phi_{m_F}=m_F \phi= \arctan( \frac{\lambda_v m_F}{\lambda_s} \tan \varphi )$. Therefore, the coupling strength ratio is
\begin{equation}
	\frac{\lambda_{\pm1}}{\lambda_0}= \sqrt{\frac{N_{\pm1}}{N_0}}\sqrt{1+\frac{\alpha_v^2}{4\alpha_s^2}\tan\varphi}=\sqrt{\frac{N_{\pm1}}{N_0}}\sqrt{1+\tan\phi^2}
\end{equation}
where $\alpha_{s,v}$ are quantities determined by electronic structure of the atom and pump laser. In order to set the spin independent renormalized coupling strength $\Lambda$, one may choose
\begin{equation}
	\label{eq: imbalance}
	\frac{N_{0}}{N_{\pm1}}=\sqrt{1+\tan\phi^2}.
\end{equation}
To demonstrate the robustness of the phase diagram and nonreciprocal dynamics discussed in the main text, we present the steady-state phase diagram in Fig.~\ref{fig: asymmetry phase}, obtained under the spin population imbalance specified in Eq.~(\ref{eq: imbalance}). While the phase boundaries are shifted, all predicted phases including pFSOP, FSOP, and DP remain clearly identifiable. In Fig.~\ref{fig: asymmetry hexagon} (a), we present the cavity field dynamics in swap phase at $\phi = \pi/3$ with the population imbalance $N_0=2N_{\pm1}$. Despite slight distortions, the characteristic chiral dynamics along the hexagonal trajectory persists, highlighting the robustness of the nonreciprocal phase transition mechanism under both nonreciprocal and geometrically frustrated interactions. As discussed above, by modulating the spin frequencies $\Omega_{0}=\Omega_{\pm1}/2$, one obtains the identical steady-state phase diagram as in the homogeneous case. However, the nonstationary dynamics is in general not identical, since the time-derivative term does not vanishes, making impossible to make all effective parameters homogeneous. Nonetheless, we show that the dynamics in the swap phase with both population and frequency modulation become more closer to the homogeneous case shown in the main text [See Fig.~3(a)].

\bibliography{reference_of_SM}